\documentclass[
  journal=largetwo,
  manuscript=article-type,
  year=2026,
  volume=38,
]{cup-journal}

\usepackage{amssymb, amsmath}
\usepackage[nopatch]{microtype}
\usepackage{booktabs}
\usepackage{multirow}
\usepackage{longtable}
\usepackage{graphicx,times}             
\usepackage[breaklinks,colorlinks=true,citecolor=blue,linkcolor=Purple,urlcolor=Violet,bookmarks=true]{hyperref}
\usepackage{aas-macros}
\usepackage[dvipsnames]{xcolor} 

\makeatletter
\newcommand*{\citelinktext}[2]{%
	\hyper@@link[cite]{}{cite.0@#1}{#2}}
\makeatother

\title{Isochrone Fitting of Galactic Globular Clusters -- VII. NGC\,1904 (M79), NGC\,4372, and revision of NGC\,288, NGC\,362, NGC\,5904 (M5), NGC\,6205 (M13), and NGC\,6218 (M12).
}

\author{George A. Gontcharov}
\affiliation{Central (Pulkovo) Astronomical Observatory, Russian Academy of Sciences, Pulkovskoye chaussee 65/1, St. Petersburg 196140, Russia}
\email[G. A. Gontcharov]{georgegontcharov@yahoo.com}

\author{Sergey S. Savchenko}
\affiliation{Central (Pulkovo) Astronomical Observatory, Russian Academy of Sciences, Pulkovskoye chaussee 65/1, St. Petersburg 196140, Russia}
\alsoaffiliation{Saint Petersburg State University, 7/9 Universitetskaya nab., St. Petersburg, 199034, Russia}

\author{Olga S. Ryutina}
\affiliation{Saint Petersburg State University, 7/9 Universitetskaya nab., St. Petersburg, 199034, Russia}

\author{Charles J. Bonatto}
\affiliation{Departamento de Astronomia, Instituto de F\'isica, UFRGS, Av. Bento Gon\c{c}alves, 9500, Porto Alegre, RS, Brazil}

\author{Jae-Woo Lee}
\affiliation{Department of Physics and Astronomy, Sejong University, 209 Neungdo-ro, Gwangjin-Gu, Seoul 05006, Korea}

\author{Vladimir B. Il'in}
\affiliation{Central (Pulkovo) Astronomical Observatory, Russian Academy of Sciences, Pulkovskoye chaussee 65/1, St. Petersburg 196140, Russia}
\alsoaffiliation{Saint Petersburg State University, 7/9 Universitetskaya nab., St. Petersburg, 199034, Russia}

\author{Maxim Yu. Khovritchev}
\affiliation{Central (Pulkovo) Astronomical Observatory, Russian Academy of Sciences, Pulkovskoye chaussee 65/1, St. Petersburg 196140, Russia}
\alsoaffiliation{Saint Petersburg State University, 7/9 Universitetskaya nab., St. Petersburg, 199034, Russia}

\author{Alexander A. Marchuk}
\affiliation{Central (Pulkovo) Astronomical Observatory, Russian Academy of Sciences, Pulkovskoye chaussee 65/1, St. Petersburg 196140, Russia}
\alsoaffiliation{Saint Petersburg State University, 7/9 Universitetskaya nab., St. Petersburg, 199034, Russia}

\author{Aleksandr V. Mosenkov}
\affiliation{Astrophysical Research Consortium, c/o Department of Astronomy, University of Washington, Box 351580, Seattle, WA 98195, USA}

\author{Denis M. Poliakov}
\affiliation{Central (Pulkovo) Astronomical Observatory, Russian Academy of Sciences, Pulkovskoye chaussee 65/1, St. Petersburg 196140, Russia}

\author{Anton A. Smirnov}
\affiliation{Central (Pulkovo) Astronomical Observatory, Russian Academy of Sciences, Pulkovskoye chaussee 65/1, St. Petersburg 196140, Russia}


\begin{document}
\defcitealias{ngc5904}{Paper I}
\defcitealias{ngc6205}{Paper II}
\defcitealias{ngc288}{Paper III}
\defcitealias{ngc6362}{Paper IV}
\defcitealias{ngc6397}{Paper V}
\defcitealias{ngc5024}{Paper VI}
\defcitealias{nardiello2018}{NLP18}
\defcitealias{stetson2019}{SPZ19}
\defcitealias{grundahl1999}{GCL99}
\defcitealias{vasiliev2021}{VB21}
\defcitealias{baumgardt2021}{BV21}
\defcitealias{ccm89}{CCM89}
\defcitealias{bonatto2013}{BCK13}
\defcitealias{sfd98}{SFD98}
\defcitealias{gms2025}{GMS25}
\defcitealias{green2019}{GSZ19}
\defcitealias{ceccarelli2025}{CARMA}

\begin{abstract}
We estimate key parameters for the Galactic globular clusters NGC\,1904 (M79) and NGC\,4372.
Additionally, we update the parameters for NGC\,288, NGC\,362, NGC\,5904 (M5), NGC\,6205 (M13), and NGC\,6218 (M12), which were analysed in our previous papers, to incorporate significant advancements in data sets and isochrones in recent years. 
We fit various colour--magnitude diagrams (CMDs) of the clusters using isochrones from the Dartmouth Stellar Evolution Database and the Bag of Stellar Tracks and Isochrones, adopting an $\alpha$--enrichment value of $[\alpha/\text{Fe}] = +0.4$.
The CMDs are constructed from data sets provided by the {\it Hubble Space Telescope}, {\it Gaia}, SkyMapper Southern Sky Survey Data Release 4, 
a large compilation of ground-based observations by Stetson, and other sources, using multiple filters for each cluster.
Our cross-identification of almost all the data sets with those from {\it Gaia} or {\it Hubble Space Telescope} allows us to use their astrometry to precisely select cluster members in all the data sets.
We obtain the following estimates, along with their total uncertainties, for NGC\,288, NGC\,362, NGC\,1904, NGC\,4372, NGC\,5904, NGC\,6205 and NGC\,6218, respectively:

metallicities [Fe/H]$=-1.28\pm0.08$, $-1.26\pm0.07$, $-1.64\pm0.09$, $-2.28\pm0.09$, $-1.33\pm0.10$, $-1.56\pm0.09$, and $-1.27\pm0.10$ dex;
ages         $12.94\pm0.76$, $10.33\pm0.75$, $13.16\pm0.76$, $12.81\pm0.81$, $11.53\pm0.76$, $12.75\pm0.76$, and $13.03\pm0.81$ Gyr;
distances    $8.83\pm0.21$, $9.00\pm0.21$, $12.66\pm0.36$, $5.17\pm0.15$, $7.24\pm0.16$, $7.39\pm0.08$, and $4.92\pm0.13$ kpc;
reddenings  $E(B-V)=0.022\pm0.024$, $0.029\pm0.025$, $0.031\pm0.018$, $0.545\pm0.032$, $0.045\pm0.027$, $0.024\pm0.021$, and $0.210\pm0.028$\,mag;
extinctions $A_\mathrm{V}=0.09\pm0.06$, $0.09\pm0.06$, $0.11\pm0.06$, $1.58\pm0.06$, $0.13\pm0.06$, $0.09\pm0.06$, and $0.67\pm0.06$\,mag;
and extinction-to-reddening ratio $R_\mathrm{V}=3.9\pm0.7$, $3.0\pm0.5$, $3.8\pm0.5$, $2.9\pm0.4$, $2.9\pm0.2$, $3.6\pm0.7$, and $3.2\pm0.1$.
The $R_\mathrm{V}$ estimates are fairly accurate, as the cross-identification of data sets enables us to calculate extinction across all ultraviolet, optical, and infrared filters used, thereby allowing us to derive an empirical extinction law for each combination of cluster, data set, and model.
We confirm that the differences in horizontal branch morphology among the 16 Galactic globular clusters analysed in our studies can be explained by variations in their metallicity, age, mass-loss efficiency, and the loss of low-mass members during cluster evolution.
Accordingly, most clusters indicate a relatively high mass-loss efficiency, consistent with the Reimers mass-loss law with $\eta > 0.3$.
\end{abstract} 
%
%


\section{Introduction}
\label{intro}

Globular clusters are among the oldest stellar systems in the Universe, serving as natural laboratories for studying the evolution of stars and the dynamical processes shaping stellar populations. These dense assemblies of stars provide critical insights into the early formation history of the Milky Way and other galaxies \citep[][and references therein]{searle1978,mackey2019,monty2023}. By analysing the colour–magnitude diagrams (CMDs) of globular clusters, one can infer their key parameters such as metallicity [Fe/H], age, distance $R$ from the Sun, and interstellar extinction across multiple filters, which are essential for understanding the evolution of both individual stars and stellar populations \citep{marin2009,dotter2011,vandenberg2013,stetson2019,ying2025}, see also the recent CARMA project papers \citep[][hereafter CARMA]{massari2023,aguado2025,ceccarelli2025}. CMDs reveal the distribution of stars across different evolutionary stages, offering a direct way to test stellar evolution models against observational data \citep{dotter2007}. In addition, the study of CMDs enables the exploration of phenomena such as mass loss, helium enrichment, and the impact of multiple stellar populations, which are crucial for refining theories of stellar and cluster evolution \citep{carretta2010, milone2017}. 

In our previous papers, we have estimated the key parameters for 14 Galactic clusters by fitting their CMDs with theoretical isochrones derived from stellar evolution models.
Our approach and results are presented in 
\citet[][hereafter Paper I]{ngc5904}, \citet[][hereafter Paper II]{ngc6205}, \citet[][hereafter Paper III]{ngc288}, \citet[][hereafter Paper IV]{ngc6362}, 
\citet[][hereafter Paper V]{ngc6397}, and \citet[][hereafter Paper VI]{ngc5024}.
The CMDs in our papers are based on accurate selection of cluster members, using astrometry from the {\it Hubble Space Telescope (HST}; \citealt{libralato2022}) or {\it Gaia} \citep{gaiadr3},
and photometry of these members in ultraviolet (UV), optical, or infrared (IR) filters. 
We use isochrones from
Dartmouth Stellar Evolution Database (DSED, \citealt{dotter2007,dotter2008})\footnote{\url{http://stellar.dartmouth.edu/models/}} and
a Bag of Stellar Tracks and Isochrones (BaSTI, \citealt{newbasti,pietrinferni2021})\footnote{\url{http://basti-iac.oa-abruzzo.inaf.it/index.html}}, as they provide comprehensive and well-tested stellar evolution models with broad parameter coverage, including variations in metallicity, age, and $\alpha$--enhancement, making them ideal for accurately fitting the colour--magnitude diagrams of globular clusters.
Both the observations and isochrones successfully reproduce on the CMD the main stages of stellar evolution, 
including the main sequence (MS), turn-off (TO), subgiant branch (SGB), red giant branch (RGB), horizontal branch (HB), and asymptotic giant branch (AGB).

Our previous papers demonstrate that the use of a large number of filters, CMDs, and cross-identified data sets allows us to reduce statistical uncertainties and identify systematic differences between the data sets. 
Moreover, by calculating interstellar extinction for each filter used, the inclusion of a large number of filters enables precise drawing and analysis of empirical extinction laws, i.e., the dependence of extinction on wavelength, for the globular clusters under consideration. Consequently, in this study, we use as many filters and high-quality data sets as possible, performing a comprehensive cross-identification of these data sets.

The continuous advancements in models, isochrones, photometry, and astrometry compel us to revise all the results presented in 
\citetalias{ngc5904}, \citetalias{ngc6205}, and \citetalias{ngc288}, 
i.e. for NGC\,288, NGC\,362, NGC\,5904 (Messier 5, M5), NGC\,6205 (Messier 13, M13), and NGC\,6218 (Messier 12, M12).
In contrast, the results presented in \citetalias{ngc6362}--\citetalias{ngc5024} remain unchanged and require no revision.
The most significant developments and improvements are: 
(i) revisions of the BaSTI isochrones, including very significant revision in 2021,
(ii) our experience and studies of other authors (e.g. see \citealt{mucciarelli2020}) approve determination of [Fe/H] as a parameter in our isochrone fitting
instead of fixing [Fe/H] from the published results, as done in \citetalias{ngc5904}--\citetalias{ngc288},
(iii) {\it Gaia} Data Release 3 (DR3; \citealt{gaiadr3}) since \citetalias{ngc5904} and \citetalias{ngc6205},
(iv) ongoing updates of the $UBVRI$ photometry collection from various ground-based telescopes \citep[][hereafter SPZ19]{stetson2019},
\footnote{\url{http://cdsarc.u-strasbg.fr/viz-bin/cat/J/MNRAS/485/3042}, \url{https://www.canfar.net/storage/vault/list/STETSON/homogeneous/Latest_photometry_for_targets_with_at_least_BVI}}
(v) SkyMapper Southern Sky Survey Data Release 4 (SMSS, SMSS DR4; \citealt{onken2024})\footnote{\url{https://skymapper.anu.edu.au}}
(vi) VISTA Hemisphere Survey catalog Data Release 5 (its older version is presented by \citealt{vista}),
(vii) Dark Energy Survey Data Release 2 (DES DR2; \citealt{des}),\footnote{\url{https://cdsarc.cds.unistra.fr/viz-bin/cat/II/371, https://des.ncsa.illinois.edu/releases/dr2}.}
(viii) data sets of Str\"omgren photometry for many clusters obtained by Jae-Woo Lee with the 1-m telescope at Cerro Tololo Inter-American Observatory (CTIO) and the 0.9-m telescope at Kitt Peak National Observatory (KPNO), which were partially presented by \citet{lee2009} and \citet{lee2017,lee2021}.

Besides our revised analysis of NGC\,288, NGC\,362, NGC\,5904, NGC\,6205, and NGC\,6218, this study includes an application of our methodology to very interesting NGC\,4372 and NGC\,1904 (Messier79, M79).

As this is the seventh paper in the series, many details of our analysis are discussed in our previous works.
We refer the reader to those papers, especially to \citetalias{ngc6362}--\citetalias{ngc5024}, which are almost free from the described shortcomings of \citetalias{ngc5904}--\citetalias{ngc288}.

This paper is organised as follows.
In Sect.~\ref{clusterproperties} we consider some properties of the clusters under consideration.
In Sect.~\ref{datasets} we present the data sets used.
We provide some analysis of the data and models in Sect.~\ref{analysis}.
In Sect.~\ref{results} we present and discuss the results of our isochrone fitting.
We summarise our main findings and conclusions in Sect.~\ref{conclusions}.
Some details of our study and additional CMDs of the clusters are provided in Appendixes and Supplemental material.

\section{Properties of the clusters}
\label{clusterproperties}

\begin{table*}
\def\baselinestretch{1}\normalsize\scriptsize
\caption[]{Some properties of the clusters under consideration. \\
Coordinates are taken from \citet{goldsbury2010} or calculated by us as medians for cluster members,
$r_t/r_c$ is the ratio of tidal and core radii, 
$\Delta(V-I)$ is the median colour difference between the HB and RGB from \citet{dotter2010},
$R$ is the distance from the Sun,
$\delta Y_{2G,1G}$ is the average helium mass fraction difference between the second and first stellar generations,
$\delta Y_{max}$ is the maximum internal variation of helium mass fraction,
$\overline{\Delta E(B-V)}$ and $\Delta E(B-V)_\mathrm{max}$ are the mean and maximum differential reddening from \citetalias{bonatto2013}, respectively,
$\Delta E(B-V)$ is the difference between the 98th and the 2nd percentile of differential-reddening distributions from \citet{jang2022}, while
$dE(B-V)_\mathrm{max}$ is the total differential reddening from \citet{pancino2024}.
We use the $R$ and [Fe/H] estimates of \citet{arellano2024} for the RRc variable stars, with the [Fe/H] estimates being on the [Fe/H] scale of \citet{carretta2009}.
The \citealt{arellano2024} values without uncertainties are based on the only measurement.
}
\label{properties}
\[
\begin{tabular}{lccccccc}
\hline
\noalign{\smallskip}
Property                                              &    NGC\,288     &    NGC\,362     &    NGC\,1904    &   NGC\,4372     &    NGC\,5904    &    NGC\,6205    & NGC\,6218 \\
\hline
\noalign{\smallskip}
RA J2000 (h~m~s)                     & \hphantom{$-$}00 52 45 & \hphantom{$-$}01 03 14 & \hphantom{$-$}$05$ $24$ $11$ & \hphantom{$-$}12 25 49 & \hphantom{$-$}15 18 33  & \hphantom{$-$}16 41 41 & \hphantom{$-$}16 47 14 \\
Dec. J2000 ($^{\circ}$ \verb|'| \verb|"|)              &   $-$26 34 57   &   $-$70 50 56   & $-$24 31 35 &   $-$72 39 33   &   $+$02 04 52   &  $+$36 27 36    & $-$01 56 55   \\
Galactic longitude ($^{\circ}$)                          &   151.2815      &   301.5330      &  227.2313     &   300.9976      &     3.8586      &   59.0073       & 15.7151 \\
Galactic latitude ($^{\circ}$)                           &   $-89.3804$    &   $-46.2474$    &   $-29.3500$    &   $-9.8841$     &    $+46.7964$   &   $+40.9131$    & $+26.3134$ \\
\hline
\noalign{\smallskip}
Angular radius (arcmin) from \citet{bica2019}         &  15             &  15             &   18          &  24             &  25             &  33             & 19   \\
Tidal radius (arcmin) from \citetalias{baumgardt2021} &  36.4           &  33.9           &   17.4        &  44.6           &  37.4           &  60.4           & 30.7 \\
Truncation radius (arcmin) from this study            &  18.5           &  20.0           &   13.0        &  18.0           &  34.5           &  27.0           & 18.7 \\
\hline
\noalign{\smallskip}
Core density (solar mass per cubic pc) from \citetalias{baumgardt2021} &      $68$      & $37\,154$ & $36\,308$ & $132$ & $4677$ & $2239$ & $1445$ \\
$r_t/r_c$ from \citetalias{baumgardt2021}             &      27         &       202       &      201      &       21        &        72       &          74     & 39  \\
\hline
\noalign{\smallskip}
$\Delta(V-I)$ from \citet{dotter2010}                 & $1.022\pm0.025$ & $0.195\pm0.003$ &                 &                 & $0.874\pm0.024$ & $1.047\pm0.010$ & $0.986\pm0.009$ \\
$\tau_{HB}$ index from \citet{torelli2019}            &   $9.39\pm0.37$ &   $3.24\pm0.01$ &                 &                 &   $5.04\pm0.14$ &  $13.37\pm0.72$ & $9.05\pm0.32$  \\
HB type from \citet{torelli2019}                      &  $+0.98\pm0.12$ &  $-0.76\pm0.07$ &                 &                 &  $+0.42\pm0.03$ &  $+0.99\pm0.05$ & $1.00\pm0.10$ \\
HB type from \citet{arellano2024}                     &     $+0.98$     &                 &      $+0.74$    &                 &      $+0.31$    &     $+0.97$     &  \\
Mean HB type from this study    & $+0.94^{+0.02}_{-0.02}$ & $-0.82^{+0.05}_{-0.05}$ & $+0.90^{+0.06}_{-0.06}$ & $+1.00$ & $+0.48^{+0.04}_{-0.04}$ & $+0.98^{+0.01}_{-0.01}$ & $+0.98^{+0.01}_{-0.03}$  \\
\hline
\noalign{\smallskip}
$R$ (kpc) from \citet{harris}, 2010 revision\footnotemark\ &       8.9  &       8.6       & 12.9          &       5.8       &        7.5      &       7.1       & 4.8     \\
$R$ (kpc) from \citetalias{baumgardt2021}             &  $8.99\pm0.09$  &  $8.83\pm0.10$  &  $13.08\pm0.18$ &  $5.71\pm0.21$  &  $7.48\pm0.06$  &  $7.42\pm0.08$  & $5.11\pm0.05$  \\
$R$ (kpc) from \citet{arellano2024}                   &     $8.0$       &                 &      $12.9$     &                 &   $7.5\pm0.3$   &   $6.8\pm0.3$   &   \\
$R$ (kpc) from \citet{valcin2020}             & $9.77^{+0.20}_{-0.20}$ & $9.12^{+0.21}_{-0.21}$ &  &  & $7.53^{+0.11}_{-0.17}$ & $7.79^{+0.09}_{-0.12}$ & $5.27^{+0.12}_{-0.04}$ \\        
$R$ (kpc) from \citetalias{ceccarelli2025}   & $8.99\pm0.04$ & $9.08\pm0.04$ &  &  & & $7.38\pm0.03$ & \\
$R$ (kpc) from this study&  $8.83\pm0.21$  &  $9.00\pm0.21$  & $12.66\pm0.36$  &  $5.17\pm0.15$  &  $7.24\pm0.16$  &  $7.39\pm0.08$  & $4.92\pm0.13$ \\
\hline
\noalign{\smallskip}
$[$Fe$/$H$]$ from \citet{carretta2009}                &  $-1.32\pm0.02$ &  $-1.30\pm0.04$ &  $-1.58\pm0.02$ &  $-2.19\pm0.08$ &  $-1.33\pm0.02$ &  $-1.58\pm0.04$ & $-1.33\pm0.02$ \\
$[$Fe$/$H$]$ from \citet{meszaros2020}                &  $-1.18\pm0.11$ &  $-1.03\pm0.08$ &  $-1.47\pm0.09$ &                 &  $-1.18\pm0.10$ &  $-1.43\pm0.13$ & $-1.17\pm0.09$ \\
Spectroscopic $[$Fe$/$H$]$ from \citet{jurcsik2023}   &                 &  $-1.17\pm0.07$ &  $-1.55\pm0.02$ &                 &  $-1.26\pm0.03$ &  $-1.51\pm0.02$ &  \\
Photometric $[$Fe$/$H$]$ from \citet{jurcsik2023}     &                 &  $-1.35\pm0.04$ &  $-1.76\pm0.20$ &                 &  $-1.38\pm0.04$ &  $-1.88$        &  \\  
$[$Fe$/$H$]$ from \citet{arellano2024}                &     $-1.52$     &                 &  $-1.66$        &                 &  $-1.39\pm0.11$ &  $-1.63\pm0.20$ &   \\
$[$Fe$/$H$]$ from \citet{valcin2020} & $-1.43^{+0.18}_{-0.11}$ & $-1.30^{+0.14}_{-0.12}$ &  &  & $-1.30^{+0.10}_{-0.16}$ & $-1.48^{+0.08}_{-0.16}$ & $-1.51^{+0.13}_{-0.11}$ \\
$[$Fe$/$H$]$ from this study &  $-1.28\pm0.07$ &  $-1.26\pm0.07$ &  $-1.64\pm0.09$ &  $-2.28\pm0.09$ &  $-1.33\pm0.10$ &  $-1.56\pm0.09$ & $-1.27\pm0.10$ \\
\hline
\noalign{\smallskip}
$\delta Y_{2G,1G}$ from \citet{milone2018}            & $0.015\pm0.010$ & $0.008\pm0.006$ &                 &                 & $0.012\pm0.004$ & $0.020\pm0.004$ & $0.009\pm0.007$ \\
$\delta Y_{max}$ from \citet{milone2018}              & $0.016\pm0.012$ & $0.026\pm0.008$ &                 &                 & $0.037\pm0.007$ & $0.052\pm0.004$ & $0.011\pm0.011$ \\ 
\hline
\noalign{\smallskip}
Age (Gyr) from \citet{dotter2010}                     & $12.50\pm0.50$  & $11.50\pm0.50$  &                 &                 & $12.25\pm0.75$  & $13.00\pm0.50$  & $13.25\pm0.75$  \\
Age (Gyr) from \citet{vandenberg2013}                 & $11.50\pm0.38$  & $10.75\pm0.25$  &                 &                 & $11.50\pm0.25$  & $12.00\pm0.38$  & $13.00\pm0.50$ \\
Age (Gyr) from \citet{valcin2020}      & $11.20^{+0.67}_{-0.67}$ & $11.52^{+0.84}_{-0.84}$ &  &  & $12.75^{+0.80}_{-0.80}$ & $13.49^{+0.62}_{-0.45}$ & $14.64^{+0.29}_{-0.64}$ \\
Age (Gyr) from \citetalias{ceccarelli2025} & $13.75^{+0.28}_{-0.22}$ & $11.47^{+0.11}_{-0.10}$ &  &  & & $14.06^{+0.35}_{-0.41}$ & \\
Age (Gyr) from this study & $12.94\pm0.76$ &  $10.33\pm0.75$ &  $13.16\pm0.76$ & $12.81\pm0.81$  &  $11.53\pm0.76$ &  $12.75\pm0.76$ & $13.03\pm0.81$ \\
\hline
\noalign{\smallskip}
$\overline{\Delta E(B-V)}$ (mag) from \citetalias{bonatto2013}    & $0.047\pm0.018$ & $0.032\pm0.009$ &     &                 & $0.033\pm0.009$ & $0.026\pm0.009$ & $0.027\pm0.008$ \\
$\Delta E(B-V)_\mathrm{max}$ (mag) from \citetalias{bonatto2013}  &      0.091      &       0.056     &     &                 &       0.068     &       0.054     &      0.056   \\
$\Delta E(B-V)$ (mag) from \citet{jang2022}           & $0.010\pm0.001$ &                 & $0.020\pm0.007$ & $0.187\pm0.021$ & $0.014\pm0.002$ & $0.013\pm0.001$ & $0.029\pm0.003$ \\
$dE(B-V)_\mathrm{max}$ (mag) from \citet{pancino2024} & $0.015\pm0.015$ &                 & $0.007\pm0.011$ & $0.334\pm0.047$ & $0.017\pm0.020$ & $0.017\pm0.018$ & $0.044\pm0.024$ \\
\hline
\noalign{\smallskip}
$E(B-V)$ (mag) from \citet{harris}, 2010 revision     &      0.03       &       0.05      &      0.01     &      0.39       &       0.03      &       0.02      &      0.19  \\
$E(B-V)$ (mag) from \citetalias{ceccarelli2025}     &      0.02       &       0.03      &         &         &         &       0.01      &     \\
$E(B-V)$ (mag) from \citetalias{sfd98}                &      0.01       &       0.03      &      0.03     &      0.57       &       0.04      &       0.02      &      0.18 \\ 
$E(B-V)$ (mag) from \citet{schlaflyfinkbeiner2011}    &       0.01      &       0.03      &      0.03     &      0.50       &       0.03      &       0.02      &      0.16   \\ 
$E(B-V)$ (mag) from \citet{planck}                    &       0.02      &       0.03      &      0.04     &      0.54       &       0.03      &       0.01      &      0.19  \\
$E(B-V)$ (mag) from \citet{lallement2019}             &       0.01      &       0.02      &      0.02     &      0.32       &       0.04      &       0.03      &      0.14   \\
$E(B-V)$ (mag) from \citetalias{green2019}            &       0.03      &                 &      0.05     &                 &       0.09      &       0.09      &      0.22   \\
$E(B-V)$ (mag) from \citetalias{gms2025}              &       0.03      &       0.04      &      0.06     &      0.43       &       0.06      &       0.05      &      0.19    \\
$E(B-V)$ (mag) from this study & $0.022\pm0.024$ & $0.029\pm0.025$ & $0.031\pm0.018$ & $0.545\pm0.032$ & $0.045\pm0.027$ & $0.024\pm0.021$  & $0.210\pm0.028$   \\
\hline
\end{tabular}
\]
\end{table*}
\footnotetext{The commonly used database of clusters by \citet{harris} (\url{https://www.physics.mcmaster.ca/~harris/mwgc.dat}), 2010 revision.}

The key properties of the clusters under consideration are presented in Table~\ref{properties}.
The table highlights that some characteristics of these intriguing clusters remain uncertain and warrant further clarification.
This is particularly significant for the relatively distant NGC\,1904 and heterogeneously reddened NGC\,4372. 
The data sets obtained with the {\it HST} Wide Field Channel of the Advanced Camera for Surveys (ACS) and Wide Field Camera 3 (WFC3) \citep[][hereafter NLP18]{nardiello2018}\footnote{\url{http://groups.dfa.unipd.it/ESPG/treasury.php}} have been widely used by various authors to derive the cluster properties listed in Table~\ref{properties}. Such data sets are not available for NGC\,1904 and NGC\,4372.\footnote{Although {\it HST} Wide Field and Planetary Camera 2 (WFPC2) photometry is available for these clusters \citep{piotto2002}, it cannot be used due to significant photometric discrepancies between the CCD chips.}
As a result, our study provides one of the few available estimates of age, distance, and photometric [Fe/H] for NGC\,1904 and NGC\,4372.

Among the seven clusters analysed in this study, five (excluding NGC\,6218 and NGC\,4372) exhibit extremely low, nearly zero extinction. 
For NGC\,288 this is justified by the fact that this cluster is located very close to the South Galactic Pole.
Thus, extinction estimates from this study are valuable for establishing a lower limit on total Galactic extinction across the entire Galactic dust layer. 

Also, we investigate the parameters influencing the different HB morphology of globular clusters. Specifically, NGC\,1904, the currently revised NGC\,6205, and NGC\,5272 from \citetalias{ngc5024} share a similar metallicity of [Fe/H]$\approx-1.6$, yet exhibit distinct HB morphologies. Furthermore, the HB morphology of a metal-poor cluster NGC\,4372 should be compared with that of NGC\,5024, NGC\,5053, NGC\,5466, and NGC\,7099 from \citetalias{ngc5024}, as they have a similar [Fe/H]$\approx-2$.
Finally, the quartet of clusters --- NGC\,288, NGC\,362, NGC\,5904, and NGC\,6218 --- with similar metallicity [Fe/H]$\approx-1.3$ is well-known for their distinct HB morphology difference.
This difference suggests that one or more parameters beyond metallicity must be responsible for the HB morphology \citep{lee1999,dalessandro2013,lee2024}.
Various authors have introduced HB morphology indices to emphasise the difference.
These indices for the clusters under consideration are presented in Table~\ref{properties}:
$\Delta(V-I)$ defined by \citet{dotter2010}, $\tau_{HB}$ defined by \citet{torelli2019}, and the HB type\footnote{The HB type is defined as $(N_B-N_R)/(N_B+N_V+N_R)$, 
where $N_B$, $N_V$, and $N_R$ are the number of stars that lie blueward of the instability strip, the number of RR~Lyrae variable stars, and the number of stars that lie redward of the instability strip, respectively \citep{lee1994}.}
calculated by \citet{torelli2019} and \citet{arellano2024}.

Furthermore, NGC\,362 is interesting posing a challenge in distinguishing its members from the background Small Magellanic Cloud.

NGC\,288 and NGC\,6218 are interesting as some of the oldest Galactic globular clusters.
The important BaSTI revision in 2021 seems to decrease its predicted age by about 1.2 Gyr making the BaSTI age estimates consistent with those from DSED despite the different input physics in these models.
Therefore, this study has the potential to deliver some of the most precise estimates of the age of the oldest globular clusters relative to the age of the Universe.

The clusters under consideration have, at least, two stellar generations \citep{lee2009,lee2017,milone2017,lee2021,jang2025}, 
hereafter designated as 1G and 2G, respectively, with a similar enrichment by $\alpha$ elements [$\alpha$/Fe]$\approx0.4$ \citep{johnson2006,carretta2010,sanroman2015,masseron2019}, but a mild difference $\delta Y_{2G,1G}$ in helium mass fraction $Y$ between 2G and 1G, as seen from Table~\ref{properties}.

We choose the clusters with a rather small $\delta Y_{2G,1G}$ to fit an observed mix of the generations. We adopt the helium mass fraction $Y_\mathrm{mix}$ of the mix calculated as 
\begin{equation}
\label{ymixformula}
Y_\mathrm{mix}=Y_\mathrm{1G}\cdot N_\mathrm{1G}/N_\mathrm{TOT}+(Y_\mathrm{1G}+\delta Y_\mathrm{2G,1G})\cdot(1-N_\mathrm{1G}/N_\mathrm{TOT}),
\end{equation}
where $Y_\mathrm{1G}$ is the primordial $Y$ given by BaSTI for the cluster's $[$Fe$/$H$]$ and $[\alpha/$Fe$]$, $N_\mathrm{1G}/N_\mathrm{TOT}$ is the fraction of 1G stars adopted from \citet{jang2025,dondoglio2021} or adopted for NGC\,4372 as 0.3 based on the spectroscopic data from \citet{sanroman2015}, $\delta Y_\mathrm{2G,1G}$ is adopted from \citet{milone2018} or adopted for NGC\,1904 the same as for its metallicity and age analog NGC\,6205 and for NGC\,4372 $\delta Y_\mathrm{2G,1G}=0.006$ as the average between the estimates for its metallicity and age analogs NGC\,5053 and NGC\,7099 from \citetalias{ngc5024} (see Sect.~\ref{hb}). 
The uncertainty of $Y_\mathrm{mix}$ is calculated from the uncertainties of the input arguments.
For $Y_\mathrm{mix}$ of NGC\,1904 and NGC\,4372 we adopt a higher uncertainty 0.01.
All the quantities are presented in Table~\ref{ymix}.

\begin{table}
\def\baselinestretch{1}\normalsize\small
\caption[]{$Y_\mathrm{mix}$ calculated with formula~(\ref{ymixformula}) and adopted for the fitted mix of stellar generations.
$Y_\mathrm{1G}$ is the primordial $Y$ stated by BaSTI for the cluster's $[$Fe$/$H$]$,
$N_\mathrm{1G}/N_\mathrm{TOT}$ is the fraction of 1G stars,
$\delta Y_\mathrm{2G,1G}$ is the average difference between 2G and 1G stars (see the text).
}
\label{ymix}
\[
\begin{tabular}{lcccc}
\hline
\noalign{\smallskip}
Cluster    & $Y_\mathrm{1G}$ & $N_\mathrm{1G}/N_\mathrm{TOT}$ & $\delta Y_\mathrm{2G,1G}$ & $Y_\mathrm{mix}$ \\    
\hline
\noalign{\smallskip}
NGC\,288   & 0.249 & 0.56 & 0.015 & $0.256\pm0.005$ \\
NGC\,362   & 0.249 & 0.28 & 0.008 & $0.255\pm0.003$ \\
NGC\,1904  & 0.248 & 0.33 & 0.020 & $0.255\pm0.010$ \\
NGC\,4372  & 0.247 & 0.30 & 0.006 & $0.251\pm0.010$ \\
NGC\,5904  & 0.249 & 0.29 & 0.012 & $0.258\pm0.003$ \\
NGC\,6205  & 0.248 & 0.21 & 0.020 & $0.264\pm0.003$ \\
NGC\,6218  & 0.249 & 0.38 & 0.009 & $0.255\pm0.004$ \\
\hline
\end{tabular}
\]
\end{table}

\section{Data sets}
\label{datasets}

\begin{figure*}
\includegraphics{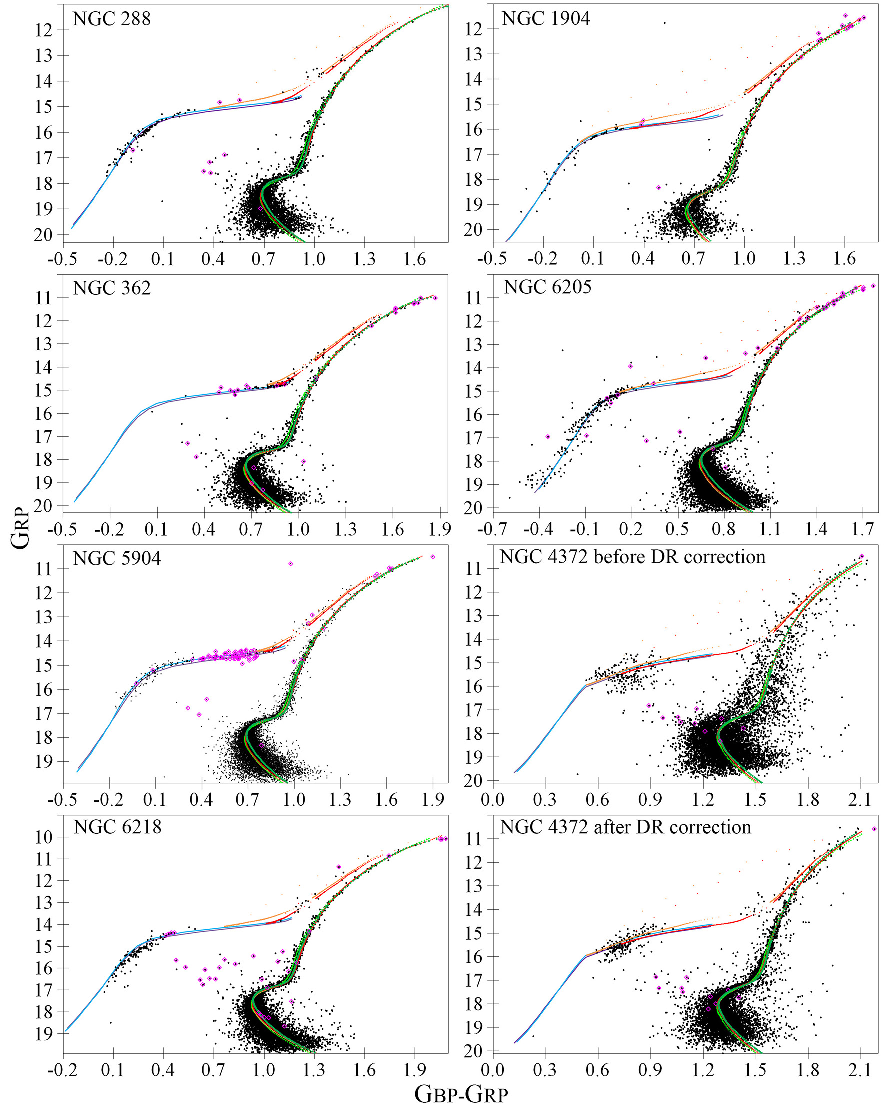}
\caption{$G_\mathrm{BP}-G_\mathrm{RP}$ versus $G_\mathrm{RP}$ CMDs for cluster members from the {\it Gaia} DR3.
The clusters are ordered by their [Fe/H]: those with [Fe/H]$\approx-1.3$ are in the left column, while NGC\,1904 and NGC\,6205 with [Fe/H]$\approx-1.6$ are in the top of the right column.
The NGC\,4372 CMDs before and after our DR correction are shown in the bottom of the right column. 
The isochrones for a primordial $Y\approx0.25$ from BaSTI (red), BaSTI ZAHB (purple), and DSED (green), isochrones for $Y=0.275$ from BaSTI (orange) and BaSTI ZAHB (blue), 
as well as isochrones for $Y=0.33$ from DSED (luminous green) are calculated with the best-fitting parameters from Table~\ref{cmds}.
Variable stars are shown by the magenta diamonds.
}
\label{gaia}
\end{figure*}

\begin{figure*}
\includegraphics{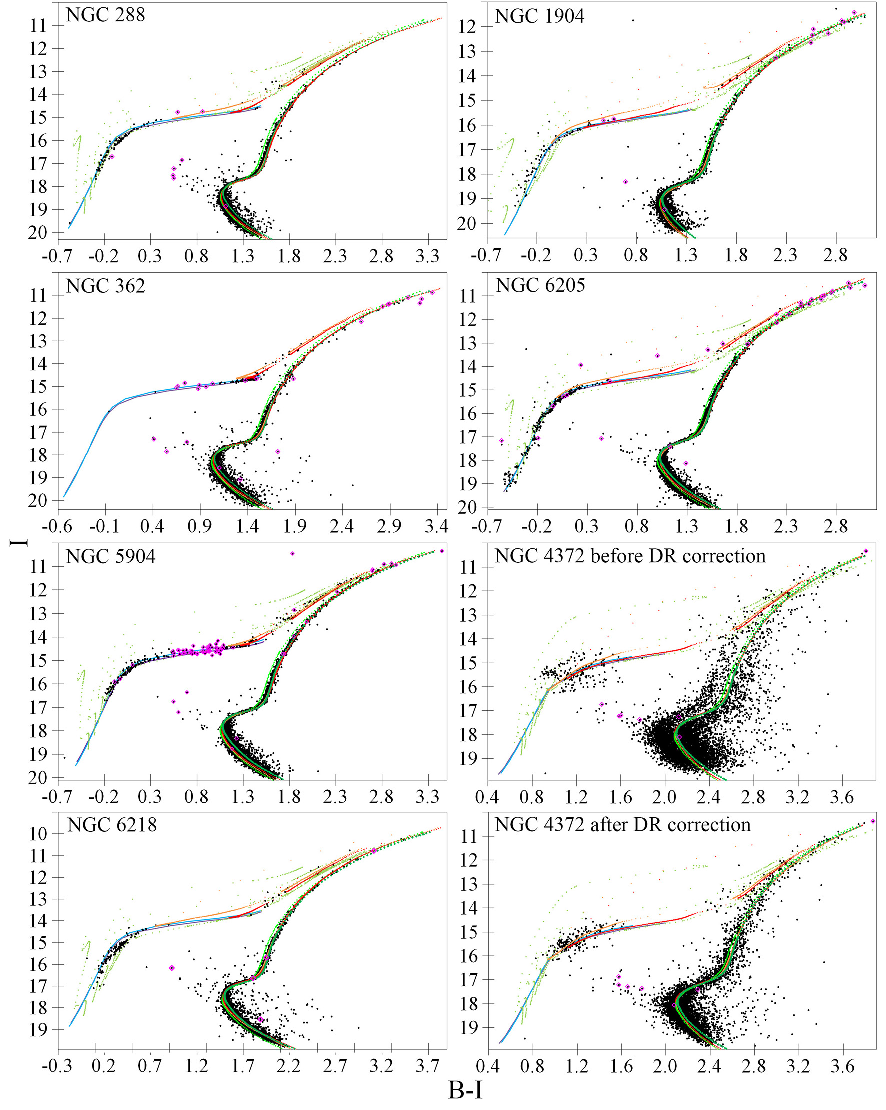}
\caption{$B-I$ versus $I$ CMDs for cluster members from the cross-identification of the {\it Gaia} DR3 and \citetalias{stetson2019} data sets.
The clusters are ordered by their [Fe/H] as in Fig.~\ref{gaia}.
The isochrones for a primordial $Y\approx0.25$ from BaSTI (red), BaSTI ZAHB (purple), and DSED (green), DSED HB/AGB tracks (light green), 
isochrones for $Y=0.275$ from BaSTI (orange) and BaSTI ZAHB (blue),  
as well as isochrones for $Y=0.33$ from DSED (luminous green) are calculated with the best-fitting parameters from Table~\ref{cmds}.
Variable stars are shown by the magenta diamonds.
}
\label{stetson}
\end{figure*}

\begin{figure*}
\includegraphics{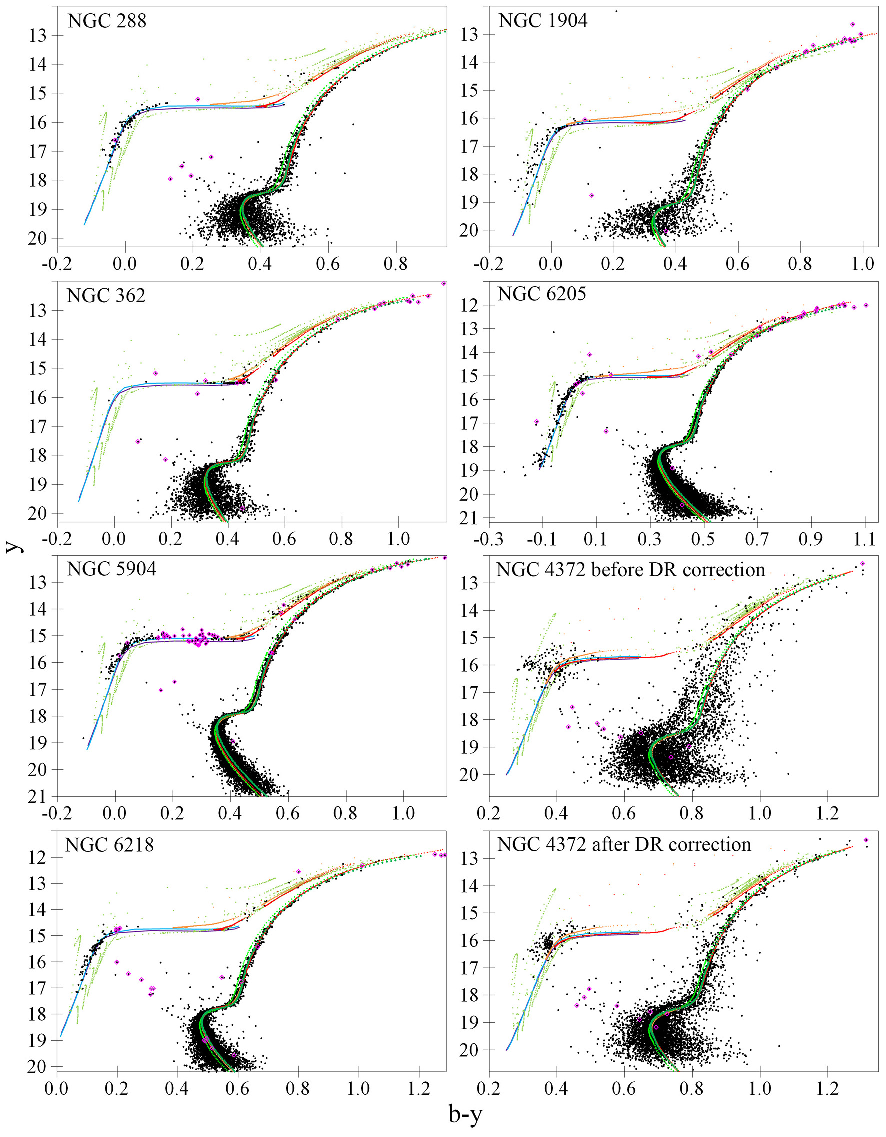}
\caption{The same as Fig.~\ref{stetson} but for the $b-y$ versus $y$ CMDs for cluster members from the cross-identification of the {\it Gaia} DR3 and Lee data sets.
For NGC\,5904 the ordinate is the $V$ magnitude.
}
\label{lee}
\end{figure*}

For all the clusters, we use cognate data sets acquired using the same telescope and/or processed through a consistent pipeline:
\begin{enumerate}
\item \label{filtergaia} {\it Gaia} DR3 photometry in the $G$, $G_\mathrm{BP}$ and $G_\mathrm{RP}$ filters \citep{riello2021}: 
4067, 4908, 1817, 6933, 9746, 10\,448, and 6918 cluster members in 
NGC\,288, NGC\,362, NGC\,1904, NGC\,4372, NGC\,5904, NGC\,6205, and NGC\,6218, respectively.\footnote{The DSED {\it Gaia} DR2 isochrones are equally suitable for DR3.}
The {\it Gaia} CMDs with our best isochrone fitting for all the clusters are shown in Fig.\ref{gaia}.\footnote{For the {\it Gaia} CMDs we select cluster members with both precise {\it Gaia} astrometry and photometry, while for the data sets cross-identified with {\it Gaia} we select cluster members with precise {\it Gaia} astrometry irrespective of their {\it Gaia} photometry. Therefore, sometimes we consider less cluster members in the former case than in the latter.}
\item \label{filterstetson} $UBVI$ photometry from various ground-based telescopes processed by \citetalias{stetson2019}:
4402, 4981, 2646, 6399, 10\,704, 9421, and 6805
cluster members, common in \citetalias{stetson2019} and {\it Gaia} DR3, in NGC\,288, NGC\,362, NGC\,1904, NGC\,4372, NGC\,5904, NGC\,6205, and NGC\,6218, respectively.
The \citetalias{stetson2019} CMDs with our best isochrone fitting for all the clusters are shown in Fig.\ref{stetson}.
\item \label{filterlee} Str\"omgren $by$\footnote{The filters used in these observations slightly differ from those adopted by BaSTI and DSED. We compensate this by adding $+0.005$, $-0.005$, and $+0.01$\,mag to the observed $b$, $y$, and $b-y$. $V$ instead of $y$ photometry is used and `$b-y$ versus $V$' CMD is considered for NGC\,5904.} photometry obtained with the 1-m telescope at CTIO and the 0.9-m telescope at KPNO for 3074, 5511, 1598, 4722, 16\,715, 16\,527, and 5541 cluster members, common with {\it Gaia} DR3, in NGC\,288, NGC\,362, NGC\,1904, NGC\,4372, NGC\,5904, NGC\,6205, and NGC\,6218, respectively.
These data sets are partially presented by \citet{lee2009} and \citet{lee2017,lee2021}, but entirely represented in this study.
The Lee CMDs with our best isochrone fitting for all the clusters are shown in Fig.\ref{lee}.
\item \label{filterwise} {\it Wide-field Infrared Survey Explorer (WISE}; \citealt{wise}) photometry in the $W1$ filter from the unWISE catalogue 
\citep{unwise}\footnote{\url{https://cdsarc.cds.unistra.fr/viz-bin/cat/II/363}}
for 470, 623, 333, 1101, 1513, 990, and 813 cluster members, common with {\it Gaia} DR3, in NGC\,288, NGC\,362, NGC\,1904, NGC\,4372, NGC\,5904, NGC\,6205, and NGC\,6218, respectively.
For the distant NGC\,1904, the unWISE photometry poorly covers the important TO domain of CMDs and, hence, is not used for the final estimates of extinction.
\end{enumerate}

The following data sets exist for some but not all the clusters:
\begin{enumerate}
\setcounter{enumi}{4}
\item \label{filterhst} The {\it HST} WFC3 UV Legacy Survey of Galactic Globular Clusters (the $F275W$, $F336W$, and $F438W$ filters) and 
the ACS (the $F606W$ and $F814W$ filters) survey of Galactic globular clusters \citep{piotto2015}, 
\citepalias{nardiello2018}:
15\,743, 45\,479, 61\,724, 70\,408, 22\,921 cluster members in NGC\,288, NGC\,362, NGC\,5904, NGC\,6205, and NGC\,6218, respectively.
\item \label{filterps1} The Panoramic Survey Telescope and Rapid Response System Data Release I (Pan-STARRS, PS1) photometry in the 
$g_\mathrm{PS1}$, $r_\mathrm{PS1}$, $i_\mathrm{PS1}$, $z_\mathrm{PS1}$, and $y_\mathrm{PS1}$ filters \citep{chambers2016}: 
4485, 1909, 9723, 10\,699, and 6918 cluster members, common with {\it Gaia} DR3, in NGC\,288, NGC\,1904, NGC\,5904, NGC\,6205, NGC\,6218, respectively.
\item \label{filtersdss} The Sloan Digital Sky Survey (SDSS) photometry in the $u_\mathrm{SDSS}$, $g_\mathrm{SDSS}$, $r_\mathrm{SDSS}$, $i_\mathrm{SDSS}$, and $z_\mathrm{SDSS}$ filters 
\citep{an2008}:\footnote{\url{http://classic.sdss.org/dr6/products/value_added/anjohnson08_clusterphotometry.htm.} We correct the SDSS magnitudes following \citet{an2009} and 
\citet{eisenstein2006}.}
9963 and 9632 cluster members, common with {\it Gaia} DR3, in NGC\,5904 and NGC\,6205 respectively.
\item \label{filtersmss} The SMSS DR4 photometry in the $g_\mathrm{SMSS}$, $r_\mathrm{SMSS}$, $i_\mathrm{SMSS}$, and $z_\mathrm{SMSS}$ filters 
for 2571, 2003, 1065, 2734, 4701, 3071 cluster members, common with {\it Gaia} DR3, in 
NGC\,288, NGC\,362, NGC\,1904, NGC\,4372, NGC\,5904, and NGC\,6218, respectively.
\item \label{filtergrundahl} The Str\"omgren $uvby$ photometry of 9316, 2559, 1889, 6549, 11\,413 cluster members, common with the {\it Gaia} DR3 or \citetalias{nardiello2018} {\it HST} data sets,\footnote{{\it HST} covers a few central arcminutes of the cluster fields, while {\it Gaia} DR3 does their periphery. Therefore, we select cluster members among the \citetalias{grundahl1999} stars by their cross-identification with {\it HST} or {\it Gaia} DR3 depending on what part of the field they cover.} 
in NGC\,288, NGC\,362, NGC\,1904, NGC\,5904, NGC\,6205, respectively, obtained by \citet[][hereafter GCL99]{grundahl1999} with the Nordic Optical Telescope (NOT), La Palma.
\item \label{filtervista} The $Y_\mathrm{VISTA}$, $J_\mathrm{VISTA}$ and $K_\mathrm{VISTA}$ photometry with the VISTA Hemisphere Survey with the VIRCAM instrument on the Visible and Infrared Survey Telescope for Astronomy (VISTA, VHS DR5) \citep{vista}\footnote{\url{https://www.vista-vhs.org}, \url{https://cdsarc.u-strasbg.fr/cgi-bin/Cat?II/367}}
for 3202, 1863, 7527, and 8267 cluster members, common with {\it Gaia} DR3, in NGC\,362, NGC\,1904, NGC\,4372, and NGC\,6218, respectively.
\item \label{filterukidss} The $Y_\mathrm{UKIDSS}$, $J_\mathrm{UKIDSS}$ and $K_\mathrm{UKIDSS}$ photometry with the United Kingdom Infrared Telescope Infrared Deep Sky Survey (UKIDSS)
\citep{ukidss}\footnote{\url{http://www.ukidss.org}. There are much more cluster members with photometry in $K_\mathrm{UKIDSS}$ than in $Y_\mathrm{UKIDSS}$ or $J_\mathrm{UKIDSS}$.
DSED and BaSTI do not provide isochrones for the $VISTA$ and $UKIDSS$ filters, respectively, hence, 
we substitute these similar filters by each other.}
for 4127 cluster members, common with {\it Gaia} DR3, in NGC\,5904.
\item \label{filterbellazzini} The $VI$ photometry with the 2.2 m ESO/MPI telescope, La Silla, equipped with the EFOSC2 camera, as well as the fiducial sequence \citep{bellazzini2001} for 2002 and 3243 cluster members, common with {\it Gaia} DR3, in NGC\,288 and NGC\,362, respectively.
\item \label{filternarloch} The $BV$ photometry with the 1-m Swope telescope of the Las Campanas Observatory \citep{narloch2017} for 
4135, 7728, and 5973 cluster members, common with {\it Gaia} DR3, in NGC\,362, NGC\,5904, and NGC\,6218, respectively.
\item \label{filtermcps} The $UBVI$ photometry of 4118 cluster members, common with {\it Gaia} DR3, in NGC\,362 obtained within the Magellanic Clouds Photometric Survey (MCPS) with the 1-m Swope telescope of the Las Campanas Observatory \citep{zaritsky2002}.
\item \label{filterhargis} The $BVI$ photometry with the Kitt Peak National Observatory 0.9-m telescope \citep{hargis2004} for 5276 cluster members, common with {\it Gaia} DR3, in NGC\,6218.
\item \label{filterzloczewski} The $BV$ photometry with the 2.5-m du Pont telescope of the Las Campanas Observatory \citep{zloczewski2012,kaluzny2015} for 3620 cluster members, common with {\it Gaia} DR3, in NGC\,6218.
\item \label{filterdes} The DES photometry in the $g_\mathrm{DECam}$, $r_\mathrm{DECam}$, $i_\mathrm{DECam}$, and $z_\mathrm{DECam}$ filters obtained by \citet{des} with the Dark Energy Camera (DECam) mounted on the 4-m Blanco telescope at CTIO for 898 cluster members, common with {\it Gaia} DR3, in NGC\,1904.\footnote{Similar DES DR2 data set for NGC\,288 appears saturated for its HB, AGB, and RGB stars making its useless.}
\item \label{filterkravtsov} The $UBV$ photometry with the 3.5-m New Technology Telescope (NTT) telescope at the European Southern Observatory, La Silla, with the EMMI camera for 1078 cluster members, common with {\it Gaia} DR3, in NGC\,1904 \citep{kravtsov1997}.\footnote{\url{https://cdsarc.cds.unistra.fr/viz-bin/cat/J/A+AS/125/1}}
\item \label{filtercoppola} The $K_\mathrm{2MASS}$ ($K_s$)\footnote{We do not use the instrumental photometry in the $J$ filter from the same data set.} photometry in the Two Micron All-Sky Survey (2MASS, \citealt{2mass}) system obtained with SOFI at the NTT\footnote{\url{http://www.eso.org/sci/facilities/lasilla/instruments/sofi/overview.html}} \citep{coppola2011} 
for 2710 cluster members, common with {\it Gaia} DR3, in NGC\,5904.
\item \label{filterrey} The $BV$ photometry of 2075 stars in NGC\,6205 with the 2.4-m telescope at Michigan-Dartmouth-MIT Observatory 
\citep{rey2001}.\footnote{\url{https://cdsarc.cds.unistra.fr/viz-bin/cat/J/AJ/122/3219}}
\end{enumerate}

We do not use some data sets for these clusters (including some data sets considered in \citetalias{ngc5904}--\citetalias{ngc288}) due to the reasons explained in \ref{rejected}.

All the data sets with the same filters are independent, e.g. \citetalias{grundahl1999} versus \citet{lee2017,lee2021}.
The \citetalias{stetson2019} data sets include photometry from various initial sources but not from the other data sets under consideration.

As in our previous papers, we perform comprehensive cross-identification of the data sets.

In total, 26, 24, 28, 17, 35, 27, and 27 filters are used for NGC\,288, NGC\,362, NGC\,1904, NGC\,4372, NGC\,5904, NGC\,6205, and NGC\,6218, respectively.
Each star has photometry in some but not all filters.
The filters used and cleaning of the data sets are described in \ref{lambda}.
After the cleaning, the median photometric uncertainty, derived from the data set authors' uncertainty statements, is a few hundredths of a magnitude for all the filters.
The uncertainty statements are used to assess the statistical uncertainty of our results, though the systematic uncertainty is higher, as demonstrated in Sect.~\ref{systematics} and \ref{results}.

\subsection{Cluster members}
\label{members}

\begin{table*}
\def\baselinestretch{1}\normalsize\normalsize
\caption[]{The cluster systemic PMs (mas\,yr$^{-1}$). The provided total uncertainties are dominated by the same systematic uncertainties as indicated by \citetalias{vasiliev2021}.
}
\label{systemic}
\[
\begin{tabular}{llcc}
\hline
\noalign{\smallskip}
Cluster & Source & $\mu_{\alpha}\cos(\delta)$ & $\mu_{\delta}$ \\
\hline
\noalign{\smallskip}
          & This study                & $4.152\pm0.024$ & $-5.707\pm0.025$ \\
NGC\,288  & \citetalias{vasiliev2021} & $4.164\pm0.024$ & $-5.705\pm0.025$ \\
          & \citet{vitral2021}        & $4.154\pm0.024$ & $-5.700\pm0.025$ \\
\noalign{\smallskip}
          & This study                & $6.694\pm0.024$ & $-2.549\pm0.024$ \\
NGC\,362  & \citetalias{vasiliev2021} & $6.694\pm0.024$ & $-2.536\pm0.024$ \\
          & \citet{vitral2021}        & $6.680\pm0.024$ & $-2.545\pm0.024$ \\
\noalign{\smallskip}
          & This study                & $2.457\pm0.025$ & $-1.597\pm0.026$ \\
NGC\,1904 & \citetalias{vasiliev2021} & $2.469\pm0.025$ & $-1.593\pm0.026$ \\
          & \citet{vitral2021}        & $2.475\pm0.025$ & $-1.599\pm0.026$ \\
\noalign{\smallskip}
          & This study                & $-6.411\pm0.024$ & $3.287\pm0.024$ \\
NGC\,4372 & \citetalias{vasiliev2021} & $-6.408\pm0.024$ & $3.297\pm0.024$ \\
          & \citet{vitral2021}        & $-6.409\pm0.024$ & $3.307\pm0.024$ \\
\noalign{\smallskip}
          & This study                & $4.072\pm0.023$ & $-9.865\pm0.023$ \\
NGC\,5904 & \citetalias{vasiliev2021} & $4.085\pm0.023$ & $-9.868\pm0.023$ \\
          & \citet{vitral2021}        & $4.070\pm0.023$ & $-9.866\pm0.023$ \\
\noalign{\smallskip}
          & This study                & $-3.137\pm0.023$ & $-2.573\pm0.023$ \\
NGC\,6205 & \citetalias{vasiliev2021} & $-3.149\pm0.023$ & $-2.573\pm0.023$ \\
          & \citet{vitral2021}        & $-3.139\pm0.023$ & $-2.570\pm0.023$ \\
\noalign{\smallskip}
          & This study                & $-0.204\pm0.024$ & $-6.806\pm0.024$ \\
NGC\,6218 & \citetalias{vasiliev2021} & $-0.191\pm0.024$ & $-6.801\pm0.024$ \\
          & \citet{vitral2021}        & $-0.213\pm0.024$ & $-6.811\pm0.024$ \\
\hline
\end{tabular}
\]
\end{table*}
\begin{table*}
\def\baselinestretch{1}\normalsize\normalsize
\caption[]{Parallax estimates (mas) with their total (statistic and systematic) uncertainties for clusters under consideration.
}
\label{parallax}
\[
\begin{tabular}{lcccc}
\hline
\noalign{\smallskip}
   & \citetalias{vasiliev2021} & \citetalias{baumgardt2021} & \multicolumn{2}{c}{This study} \\
Cluster   & {\it Gaia} DR3 astrometry & Various methods & {\it Gaia} DR3 astrometry & Isochrone fitting \\
\hline
\noalign{\smallskip}
NGC\,288  & $0.137\pm0.011$ & $0.111\pm0.001$ & $0.108\pm0.011$ & $0.113\pm0.004$ \\ 
NGC\,362  & $0.110\pm0.011$ & $0.113\pm0.002$ & $0.120\pm0.011$ & $0.111\pm0.004$ \\ 
NGC\,1904 & $0.084\pm0.011$ & $0.076\pm0.002$ & $0.080\pm0.011$ & $0.079\pm0.003$ \\ 
NGC\,4372 & $0.183\pm0.010$ & $0.175\pm0.007$ & $0.186\pm0.010$ & $0.193\pm0.006$ \\ 
NGC\,5904 & $0.138\pm0.010$ & $0.134\pm0.001$ & $0.138\pm0.010$ & $0.138\pm0.004$ \\ 
NGC\,6205 & $0.123\pm0.010$ & $0.135\pm0.001$ & $0.128\pm0.010$ & $0.135\pm0.004$ \\ 
NGC\,6218 & $0.204\pm0.010$ & $0.196\pm0.002$ & $0.194\pm0.010$ & $0.203\pm0.007$ \\ 
\hline
\end{tabular}
\]
\end{table*}

Table~\ref{properties} presents rather different tidal radius estimates for each cluster.
Accordingly, we consider the initial {\it Gaia} DR3 samples within some initial radii that exceed any previous radius estimates.
We determine truncation radii, listed in Table 1, as the radii at which the cluster member count surface density decreases to the Galactic background (see section 4.2 in \citetalias{ngc6362}).
We truncate all the data sets at these radii to reduce contamination from non-members.
This truncation allows us to create very clean samples.
While we may lose a small number of cluster members beyond the truncation radii, their absence is negligible and does not significantly influence our results.

As in our previous papers, accurate {\it Gaia} DR3 parallaxes and proper motions (PMs) are used to select cluster members and derive systemic parallaxes and PMs. We now briefly describe this procedure.

We leave only stars with parallaxes and PMs.\footnote{This means that we also eliminate all the stars without the {\it Gaia} parallaxes and PMs from all the data sets cross-identified with {\it Gaia}, i.e. all but the \citetalias{nardiello2018}, \citet{rey2001}, and some \citetalias{grundahl1999} data sets.} This selects almost all stars brighter than a magnitude in the middle of the MS, as seen in Figs~\ref{gaia}--\ref{lee}.
Foreground and background stars are rejected as those with measured parallax $\varpi>1/R+3\sigma_{\varpi}$ or $\varpi<1/R-3\sigma_{\varpi}$, where $\sigma_{\varpi}$ is the stated parallax uncertainty and $R$ is the distance from the Sun. Initially, we adopt the $R$ estimates from \citet[][hereafter BV21]{baumgardt2021} presented in Table~\ref{properties} and then replace them by the $R$ estimates derived from our isochrone fitting repeating the rejection of foreground and background stars iteratively.
To select cluster members as stars with appropriate PMs, we start with the initial systemic PM components $\overline{\mu_{\alpha}\cos(\delta)}$ and $\overline{\mu_{\delta}}$ from \citet[][hereafter VB21]{vasiliev2021},
calculate the standard deviations $\sigma_{\mu_{\alpha}\cos(\delta)}$ and $\sigma_{\mu_{\delta}}$, and
select cluster members as stars within $3\sigma$, i.e. with \\
$\sqrt{(\mu_{\alpha}\cos(\delta)-\overline{\mu_{\alpha}\cos(\delta)})^2+(\mu_{\delta}-\overline{\mu_{\delta}})^2}<3\sqrt{\sigma_{\mu_{\alpha}\cos(\delta)}^2+\sigma_{\mu_{\delta}}^2}$.
With the refined list of the members, we recalculate the median systemic PM components, truncation radii, and coordinates of the cluster centres as medians of the member coordinates (the latter is initially taken from Table~\ref{properties} and then change negligibly).
This selection of members is repeated iteratively until we stop losing stars in the $3\sigma$ cut.
The importance and efficiency of the selection of cluster members using the {\it Gaia} parallaxes and proper motions is illustrated in \ref{details}.

Similar to clusters in our previous papers, the final empirical standard deviations $\sigma_{\mu_{\alpha}\cos(\delta)}$ and $\sigma_{\mu_{\delta}}$ are reasonable, but slightly higher than the mean stated PM uncertainties, which may mean an underestimation of the latter.

Our final systemic parallaxes and PMs are determined as the median values for the cluster members.
Hence, our current PM estimates differ from those presented in \citetalias{ngc288} due to the adoption of larger truncation radii for these clusters and the use of median values instead of weighted means. 
Our systemic PMs are presented in Table~\ref{systemic} in comparison with those from \citetalias{vasiliev2021} and \citet{vitral2021}, which are also derived from {\it Gaia} DR3 but using different approaches. 
The total (statistic plus systematic) uncertainty of all these PM estimates is vastly dominated by the same systematics, as shown by \citetalias{vasiliev2021}, and, hence, is the same regardless of approach.
Table~\ref{systemic} shows that all the systemic PM estimates agree within the total uncertainties.

We correct the parallaxes of cluster members for the parallax zero-point following \citet{lindegren2021} and present the median corrected parallaxes in Table~\ref{parallax} for comparison with other estimates in Sect.~\ref{results}.
We adopt the total uncertainty of {\it Gaia} DR3 parallaxes, as determined by \citetalias{vasiliev2021}, to be 0.01 mas.

The \citet{rey2001}, \citetalias{nardiello2018}, and some \citetalias{grundahl1999} data sets are not cross-identified with {\it Gaia}. The former is cleared from non-members by its authors.
The cluster members are effectively selected from the \citetalias{nardiello2018} data sets (and some \citetalias{grundahl1999} data sets cross-identified with them) by their authors using dedicated {\it HST} PMs (see discussion in \ref{details}).

\section{Analysis}
\label{analysis}

\subsection{Differential reddening}
\label{difred}

\begin{figure*}
\includegraphics{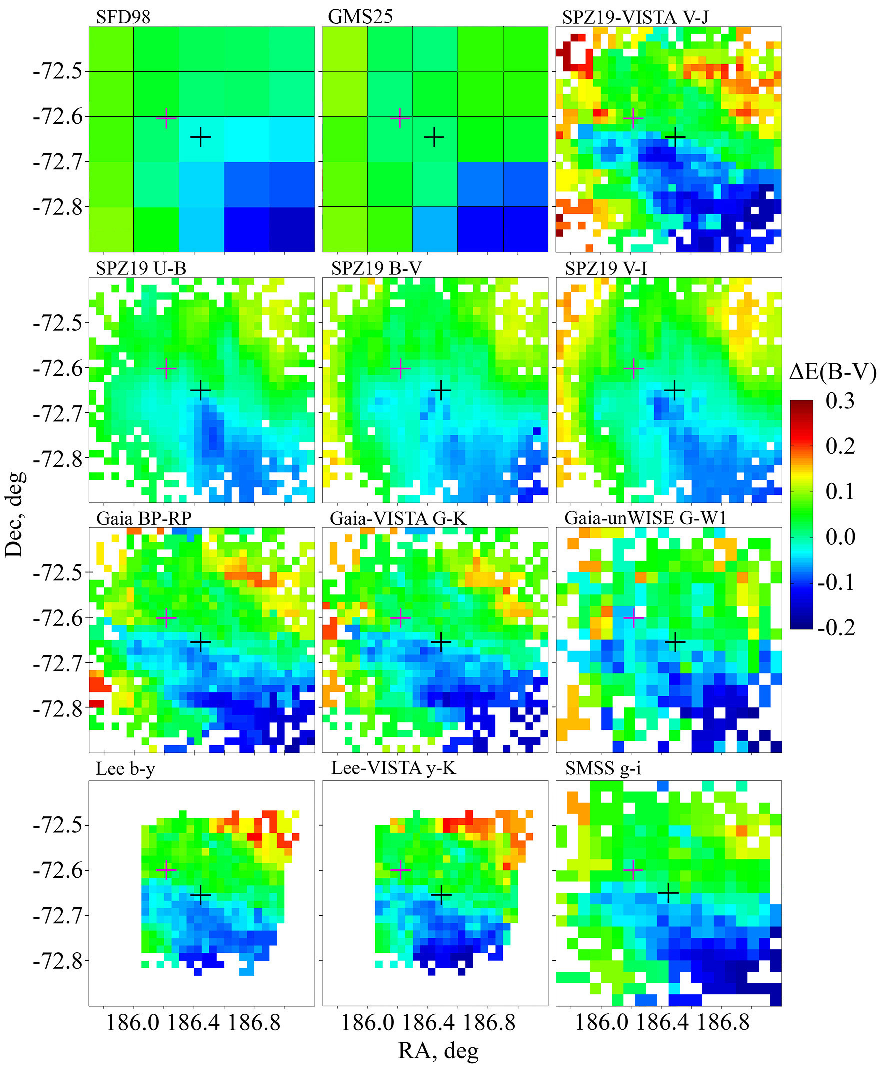}
\caption{DR maps derived from \citetalias{sfd98}, \citetalias{gms2025}, and various CMDs for the same NGC\,4372 field.
The DR maps derived from CMDs are converted from the initial adaptive angular resolution to the constant resolution of 1.5 arcmin for {\it Gaia}--unWISE and SMSS maps and 1 arcmin for the remaining CMD-based maps. \citetalias{sfd98} and \citetalias{gms2025} have the resolution of 6.1 arcmin.
All the maps are converted into $\Delta E(B-V)$ using the \citetalias{ccm89} extinction law with $R_\mathrm{V}=3.1$.
The white areas have no estimates.
The cluster centre is the black cross, the position of bright star HD\,107947 is marked by the magenta cross.}
\label{ngc4372_drmap}
\end{figure*}

Following the method of \citet[][hereafter BCK13]{bonatto2013}, we calculate variations of foreground reddening across the cluster field, i.e. differential reddening (DR) for all CMDs, which are based on a data set or a pair of cross-identified data sets with enough ($>3000$) stars.
Briefly, the cluster field is divided into a grid of cells, with higher angular resolution in regions containing more stars.
Then, the stellar density Hess diagram (including photometric errors) of each cell is matched to the diagram, averaged over the entire cluster field, by its shift along the reddening vector. This shift is then converted into DR in the cell. The same DR correction is applied to all stars in one cell. For illustrative purposes, to present the derived DR maps in Figs~\ref{ngc4372_drmap} and \ref{ngc6218_drmap}, we recalculate them for cells of a constant angular resolution averaging the individual DR corrections for stars in each cell.

We do not calculate DR for NGC\,1904, since none of its CMDs contains more than 3000 stars.
However, NGC\,1904 DR estimated by \citet{pancino2024} appears within $-0.005<dE(B-V)<0.005$\,mag, i.e. with a negligible impact on its CMDs.

Generally, NGC\,4372 and NGC\,6218 demonstrate a considerable DR ($\Delta E(B-V)>0.03$\,mag) versus a negligible DR for the remaining clusters. This is in line with the DR estimates in Table~\ref{properties} obtained by \citet{jang2022} and \citet{pancino2024}, but not with those of \citetalias{bonatto2013}, since the latter use the \citetalias{nardiello2018} data sets covering only a few central arcminutes of the cluster fields, where DR is relatively low and rather uncertain for all these clusters.

CMDs without enough stars are not corrected for DR, except for NGC\,4372 and NGC\,6218 whose such CMDs are corrected using our DR maps derived from the {\it Gaia} CMDs (since all the data sets are cross-identified with Gaia).

Examples of CMDs for NGC\,4372 before and after applying our DR correction are displayed in Figs.~\ref{gaia}--\ref{lee}, with additional examples provided in \ref{details}.  
The effectiveness of our corrections is clearly demonstrated in these diagrams by the reduced scatter of points around the isochrones.

Fig.~\ref{ngc4372_drmap} presents several DR maps for NGC\,4372,\footnote{The DR maps derived for the {\it Gaia}--unWISE and SMSS CMDs are shown for illustrative purposes only, since the corresponding data sets contain $<3000$ stars. Yet, they show the same pattern as in other DR maps.} including those based on global extinction/reddening maps, which are the most suitable for this cluster field:
reddening map of \citet[][hereafter SFD98]{sfd98} and our recent 3D extinction map based on the {\it Gaia} DR3 astrometry and multi-band photometry \citep[][hereafter GMS25]{gms2025}.
Note that other global reddening maps are less appropriate or informative in this field: 
the \citet[][hereafter GSZ19]{green2019} reddening map does not cover the NGC\,4372 field, 
the \citet{schlaflyfinkbeiner2011} map differs from \citetalias{sfd98} by only a constant term,
while the \citet{lallement2019} map predicts an unrealistically small reddening (see  Table~\ref{properties}) and DR of only $\Delta E(B-V)\approx0.03$\,mag.
Fig.~\ref{ngc4372_drmap} shows nearly the same pattern in all the DR maps: reddening increases from the lower right (South-East) to the upper left (North-West) corner by $\Delta E(B-V)\approx0.15-0.25$\,mag.
This agrees with the maps presented by \citet{kacharov2014} in their figure 4, \citet{jang2022} in their figures 3 and 4, and by \citet{pancino2024} in their figure A1.

%

\begin{figure}
\includegraphics{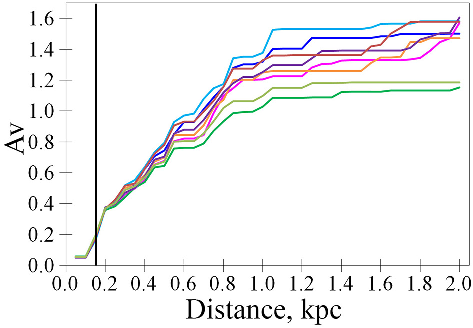}
\caption{The dependence of extinction $A_\mathrm{V}$ on distance $R$ for eight lines-of-sight within the NGC\,4372 field from the 3D extinction map of \citetalias{gms2025} -- colour curves. Black line denotes the steepest increase of $A_\mathrm{V}$ in the Musca dark nebula at about 150 pc from the Sun.
}
\label{ngc4372_gms25}
\end{figure}

The high interstellar extinction and high DR in the NGC\,4372 field is partially due to the obscuration by the Musca dark nebula (also known as [DB2002b]~G300.68-9.40, Sandquist~149 or the Dark Doodad Nebula; \citealt{sandquist1977,dutra2002,dobashi2005,hacar2016}).
This nebula is a part of the Musca--Chamaeleonis molecular cloud complex.
The distance to this nebula can be estimated as $109\pm11$ pc from the {\it Gaia} parallaxes of several young OBA stars, probable members of the Lower Centaurus--Crux association, likely embedded into the nebula.
They include, among others, gamma~Musca, HD~107947, HD~106676, HD~108735.

However, \citetalias{gms2025} show dust clouds in the NGC\,4372 field beyond the Musca dark nebula as presented in Fig.~\ref{ngc4372_gms25}. The steepest increases of the extinction profiles in Fig.~\ref{ngc4372_gms25} correspond to the Musca dark nebula at $150\pm50$ pc in a good agreement with the parallaxes of the embedded stars, as well as significant clouds at about 500, 800 and some others up to 1700 pc (i.e. down to $Z=-300$ pc below the Galactic mid-plane). It is seen that these clouds provide a difference between the extinction profiles up to $\Delta A_\mathrm{V}\approx0.5$\,mag at 2 kpc ($Z=-343$ pc), i.e. at the edge of the Galactic dust layer. This difference may well explain the DR observed in the NGC\,4372 field and presented in Fig.~\ref{ngc4372_drmap}.

The high DR in the NGC\,4372 field prevails over other systematics.
This is not the case for other clusters under consideration.
For example, DR maps for NGC\,6218 are presented in \ref{details}.

Since some data sets do not cover the cluster centres, we do not refer our final reddening/extinction estimates to the centres. 
Instead, our isochrone fitting provides us with an average reddening of all cluster members under consideration, the same before and after our DR correction.
The DR correction is applied in addition to the average reddening. 
Consequently, for each CMD, the DR correction is positive for some stars and negative for others, while the DR correction averaged for all stars in the field is exactly zero.

\subsection{Isochrone-to-data fitting}
\label{fitting}

To fit CMDs, we use the $\alpha$--enhanced [$\alpha$/Fe]$=+0.4$ theoretical models of stellar evolution and corresponding isochrones from 
BaSTI with initial solar $Z=0.0172$ and $Y=0.2695$, overshooting, diffusion, mass loss efficiency $\eta=0.3$, where $\eta$ is the free parameter in Reimers law \citep{reimers}, 
as well as from DSED with solar $Z=0.0189$ and no mass loss.
Also, as in \citetalias{ngc288}--\citetalias{ngc5024}, we use the BaSTI extended set of zero-age horizontal branch (ZAHB) models with a stochastic mass loss between the MS and HB,
whereas the DSED HB and AGB evolution tracks for a primordial $Y\approx0.25$, which exist for some filters, are used for comparison purposes.
We use the isochrones from both the models for the primordial $Y\approx0.25$, the BaSTI isochrone for $Y=0.275$, and the DSED one for $Y=0.33$ in order to interpolate the isochrones for the adopted $Y_\mathrm{mix}$,
Such interpolation produces a negligible uncertainty of $<0.01$\,mag in any CMD, as the initial isochrones are closely spaced and represented by the same evolutionary points.

In addition to Figs.~\ref{gaia}--\ref{lee}, \ref{smss_init}--\ref{ngc4372_dr}, and \ref{ngc6205_stetson}--\ref{ngc6218_adjustment}, some CMDs with our fitting are presented as a supplemental material or are available upon request.

We fit isochrones to hundreds CMDs for our globular clusters. 
Namely, we fit 6, 6, 3, 1, 4, 6, and 4 reliable independent UV (i.e. with filters within $\lambda_\mathrm{eff}<440$ nm, see \ref{lambda}) and UV--optical CMDs, such as \citetalias{grundahl1999} $u-v$;
15, 12, 18, 8, 19, 16, and 17 (105 in total) reliable independent optical CMDs of 8, 9, 8, 4, 9, 8, and 9 data sets, such as \citetalias{stetson2019} $B-I$;
7, 26, 9, 14, 18, 5, and 24 reliable independent optical--IR (i.e. with filters within $\lambda_\mathrm{eff}>1000$ nm) CMDs, such as \citetalias{stetson2019}-unWISE $V-W1$, 
for NGC\,288, NGC\,362, NGC\,1904, NGC\,4372, NGC\,5904, NGC\,6205, and NGC\,6218, respectively,
as well as a hundred CMDs obtained in our cross-identification, such as {\it Gaia}-\citetalias{stetson2019} $G_\mathrm{BP}-V$.
The optical--IR CMDs involve IR filters: 
unWISE $W1$ or VISTA $Y_\mathrm{VISTA}$, $J_\mathrm{VISTA}$, $K_\mathrm{VISTA}$ or UKIDSS $J_\mathrm{UKIDSS}$, $K_\mathrm{UKIDSS}$ or 2MASS $J_\mathrm{2MASS}$, $Ks_\mathrm{2MASS}$.
As in our previous studies, the UV, UV--optical, and optical--IR CMDs provide less reliable and less precise estimates of the derived cluster parameters than the optical ones due to higher random and systematic uncertainties of the UV and IR photometry, lower accuracy of models/isochrones for the UV and IR CMDs, and significant segregation of cluster generations in the UV CMDs.
Therefore, we use only the aforementioned 105 optical CMDs to derive our parameter estimates, average them into the final ones, and to estimate their uncertainties. The remaining CMDs are used to draw empirical extinction laws based on the derived reddenings and covering a range from the UV to IR.

The distribution of stars in our CMDs is well defined due to the accurate selection of cluster members.
Therefore, as in \citetalias{ngc6397} and \citetalias{ngc5024}, we fit isochrones directly to the bulk of cluster members, without calculating a fiducial sequence.
To balance the contributions of different CMD domains, we assign a weight to each data point.
The weight is inversely proportional to the number of stars of a given magnitude, i.e. it represents the luminosity function of a given data set.
Modern computers enable us to evaluate numerous parameter sets ([Fe/H], distance from the Sun $R$, reddening, and age) in their 4-dimensional space with their steps of 0.1 dex, 0.01 kpc, 0.001 mag, and 0.5 Gyr, respectively, for each CMD-model pair.
For each set of parameters, we calculate the sum of the squares of the residuals between the isochrones and the data points.
Table~\ref{cmds} in \ref{details} presents the best solutions, i.e. those with the minimal sum of the squares of the residuals.

As in \citetalias{ngc5024} (see its figure 6), we exclude four CMD domains from the fitting: 
the extremely blue HB (i.e. the area bluer than the turn of the observed HB downward or, in other words, stars with $T_\mathrm{eff}>9000$ K), 
RR~Lyrae stars, some other variable stars, and blue stragglers.
As shown in Figs.~\ref{gaia}--\ref{lee}, some of these clusters contain a significant number of RR~Lyrae and other variable stars \citep{arellano2022,arellano2024}.
They are identified using the \citetalias{stetson2019} \verb"Vary" (Welch-Stetson variability index) and \verb"Weight" (weight of the variability index) parameters 
or {\it Gaia} \verb"VarFlag=VARIABLE" parameter. The latter allows us to detect variable stars in all the data sets cross-identified with {\it Gaia} DR3.
In addition, we detect variable stars in all the data sets using the variable star database of
\citet{clement2017}\footnote{\url{https://www.astro.utoronto.ca/~cclement/cat/listngc.html}}.
The detection and subsequent removal of variable stars is highly beneficial for accurately determining the HB magnitude and, consequently, the cluster distance.
This is particularly important for NGC\,362 and NGC\,5904, as shown in Figs.~\ref{gaia}--\ref{lee}, where RR~Lyrae variables appear both above and below the BaSTI ZAHB for a primordial $Y\approx0.25$ (purple curve), which is very close to the ZAHB for $Y_\mathrm{mix}$ fitted as the lower bound of the non-variable HB stars.

\subsection{Systematics}
\label{systematics}

As in our previous studies, the results from all CMDs of the same data set (e.g. \citetalias{stetson2019} $U-B$, $B-V$ and $V-I$) appear consistent. Two examples of such a CMD-to-CMD consistency are provided in \ref{details}. As a result, the reddening estimates from different CMDs of the same data set always draw a rather smooth and meaningful empirical extinction law as discussed in Sect.~\ref{results}.

Our comparison of photometry in the same or similar filters of cross-identified data sets allows us to find systematic differences in colours and magnitudes, i.e. set-to-set systematics.
Typically, these differences are within 0.05 mag, while they tend to be slightly higher for distant NGC\,1904 or highly reddened NGC\,4372. Such differences have been found in our previous studies (see, for example, figure~1 in \citetalias{ngc288}).
They are expected due to photometry zero-point variations, point-spread function variations, telescope focus change, distortion, stellar content variations, and other systematics \citep{anderson2008}.

Generally, we have no information to determine which of the compared data sets is closer to the truth. Therefore, we do not correct any colour or magnitude of any data set (however, we adjust data sets to draw extinction laws as described in Sect.~\ref{results} and \ref{adjustment}). Instead, we derive parameter estimates from fitting of the uncorrected data sets and average the estimates for our final results.
The systematic differences in magnitudes and colours are translated into set-to-set systematic differences in the derived parameters. These systematic differences are used to determine the total uncertainties of the parameters. Also, we take into account model-to-model systematics, i.e. between the BaSTI and DSED results. We believe that half the difference between maximal and minimal estimates (i.e. half the range) can represent the uncertainty of averaged systematically different estimates. Finally, the total uncertainty of a parameter for a cluster is calculated as the quadrature sum of half the range of estimates for all data sets used (representing the set-to-set systematics), half the difference between the BaSTI and DSED estimates averaged over all data sets used (representing the model-to-model systematics), and the random errors.
The latter are much lower than the systematics, as discussed, for example, in appendix A of \citetalias{ngc6205} and section 3.1 of \citetalias{ngc6362}.

\section{Results}
\label{results}

\begin{table*}
\def\baselinestretch{1}\normalsize\scriptsize
\caption[]{Our [Fe/H] (dex), age (Gyr), distance (kpc), distance modulus (mag), $E(B-V)$ (mag), and apparent $V$-band distance modulus (mag) estimates.
The previous estimates from \citetalias{ngc5904}, \citetalias{ngc6205}, and \citetalias{ngc288} (except those adopted from the published results) are presented in the right column.
The $E(B-V)$ estimates are calculated from the derived reddenings by use of extinction coefficients from \citet{casagrande2014,casagrande2018a,casagrande2018b} or \citetalias{ccm89} with $R_\mathrm{V}=3.1$.
The uncertainties after the values are standard deviations of the mean.
`Model $\Delta$' and `Total' are half the difference between the models and total uncertainty of the average value, respectively.
}
\label{estimates}
\[
\begin{tabular}{lccccccc}
\hline
\noalign{\smallskip}
Cluster     & BaSTI            & DSED             &    Average value   &   Model $\Delta$   & Total & \citetalias{ngc5904}--\citetalias{ngc288}  \\    
\hline
\noalign{\smallskip}
        \multicolumn{7}{c}{Age}    \\
\noalign{\smallskip}
NGC\,288    &  $13.06\pm0.07$  &  $12.81\pm0.10$  &   $12.94\pm0.06$   &        0.13        &   0.76   & $13.5\pm1.1$  \\  
NGC\,362    &  $10.33\pm0.09$  &  $10.33\pm0.13$  &   $10.33\pm0.07$   &        0.00        &   0.75   & $11.0\pm0.6$  \\ 
NGC\,1904   &  $13.25\pm0.17$  &  $13.06\pm0.07$  &   $13.16\pm0.09$   &        0.09        &   0.76   & \\
NGC\,4372   &  $12.63\pm0.36$  &  $13.00\pm0.33$  &   $12.81\pm0.22$   &        0.19        &   0.81   &  \\
NGC\,5904   &  $11.44\pm0.14$  &  $11.61\pm0.12$  &   $11.53\pm0.09$   &        0.08        &   0.76   & $12.15\pm1.0$  \\ 
NGC\,6205   &  $12.81\pm0.17$  &  $12.69\pm0.14$  &   $12.75\pm0.11$   &        0.06        &   0.76   & $13.6\pm1.3$ \\
NGC\,6218   &  $13.33\pm0.15$  &  $12.72\pm0.09$  &   $13.03\pm0.11$   &        0.31        &   0.81   & $13.8\pm1.1$  \\
\noalign{\smallskip}
        \multicolumn{7}{c}{Distance} \\
\noalign{\smallskip}
NGC\,288    &  $8.83\pm0.05$  &  $8.82\pm0.05$  &   $8.83\pm0.03$   &           0.01        &  0.21   & $8.96\pm0.07$  \\  
NGC\,362    &  $9.06\pm0.05$  &  $8.94\pm0.04$  &   $9.00\pm0.03$   &           0.06        &  0.21   & $8.98\pm0.06$  \\ 
NGC\,1904   &  $12.70\pm0.10$ &  $12.63\pm0.07$ &   $12.66\pm0.06$  &           0.04        &  0.36   &  \\
NGC\,4372   &  $5.16\pm0.03$  &  $5.18\pm0.10$  &   $5.17\pm0.04$   &           0.01        &  0.15   &  \\
NGC\,5904   &  $7.34\pm0.03$  &  $7.14\pm0.04$  &   $7.24\pm0.03$   &           0.10        &  0.16   & $7.40\pm0.30$  \\ 
NGC\,6205   &  $7.43\pm0.02$  &  $7.35\pm0.03$  &   $7.39\pm0.02$   &           0.04        &  0.08   & $7.40\pm0.20$ \\ 
NGC\,6218   &  $4.91\pm0.03$  &  $4.94\pm0.04$  &   $4.92\pm0.02$   &           0.03        &  0.13   & $5.04\pm0.05$  \\
\noalign{\smallskip}
        \multicolumn{7}{c}{Distance modulus} \\
\noalign{\smallskip}
NGC\,288    &  $14.73\pm0.01$  &  $14.73\pm0.01$  &   $14.73\pm0.01$   &        0.00        &  0.05   & $14.76\pm0.02$  \\  
NGC\,362    &  $14.78\pm0.01$  &  $14.76\pm0.01$  &   $14.77\pm0.01$   &        0.01        &  0.05   & $14.77\pm0.02$  \\ 
NGC\,1904   &  $15.52\pm0.02$  &  $15.51\pm0.02$  &   $15.51\pm0.01$   &        0.01        &  0.06   &  \\
NGC\,4372   &  $13.56\pm0.01$  &  $13.57\pm0.04$  &   $13.57\pm0.02$   &        0.00        &  0.06   &  \\
NGC\,5904   &  $14.33\pm0.01$  &  $14.27\pm0.01$  &   $14.30\pm0.01$   &        0.03        &  0.05   & $14.35\pm0.09$  \\ 
NGC\,6205   &  $14.35\pm0.01$  &  $14.33\pm0.01$  &   $14.34\pm0.01$   &        0.01        &  0.02   & $14.35\pm0.06$ \\
NGC\,6218   &  $13.45\pm0.01$  &  $13.47\pm0.02$  &   $13.46\pm0.01$   &        0.01        &  0.06   & $13.51\pm0.02$ \\
\noalign{\smallskip}
          \multicolumn{7}{c}{[Fe/H]}   \\
\noalign{\smallskip}
NGC\,288    &  $-1.23\pm0.02$  &  $-1.34\pm0.02$  &   $-1.28\pm0.02$   &        0.06        &   0.08   &   \\  
NGC\,362    &  $-1.20\pm0.00$  &  $-1.32\pm0.02$  &   $-1.26\pm0.02$   &        0.06        &   0.07   &  \\ 
NGC\,1904   &  $-1.61\pm0.02$  &  $-1.68\pm0.02$  &   $-1.64\pm0.02$   &        0.03        &   0.09   & \\
NGC\,4372   &  $-2.25\pm0.03$  &  $-2.30\pm0.05$  &   $-2.28\pm0.03$   &        0.02        &   0.09   &  \\
NGC\,5904   &  $-1.28\pm0.02$  &  $-1.38\pm0.02$  &   $-1.33\pm0.02$   &        0.05        &   0.10   &   \\ 
NGC\,6205   &  $-1.53\pm0.02$  &  $-1.59\pm0.02$  &   $-1.56\pm0.02$   &        0.03        &   0.09   & \\
NGC\,6218   &  $-1.21\pm0.01$  &  $-1.32\pm0.02$  &   $-1.27\pm0.02$   &        0.06        &   0.10   & \\
\noalign{\smallskip}
        \multicolumn{7}{c}{$E(B-V)$} \\
\noalign{\smallskip}
NGC\,288    &  $0.009\pm0.004$ &  $0.036\pm0.006$ &   $0.022\pm0.005$  &      0.014       &  0.024  & $0.014\pm0.010$  \\  
NGC\,362    &  $0.013\pm0.004$ &  $0.044\pm0.005$ &   $0.029\pm0.005$  &      0.016       &  0.025  & $0.028\pm0.011$  \\  
NGC\,1904   &  $0.022\pm0.003$ &  $0.040\pm0.004$ &   $0.031\pm0.004$  &      0.009       &  0.018  &                  \\  
NGC\,4372   &  $0.542\pm0.013$ &  $0.549\pm0.017$ &   $0.545\pm0.009$  &      0.004       &  0.032  &                  \\  
NGC\,5904   &  $0.031\pm0.005$ &  $0.060\pm0.006$ &   $0.045\pm0.005$  &      0.015       &  0.027  & $0.056\pm0.006$  \\  
NGC\,6205   &  $0.013\pm0.004$ &  $0.035\pm0.005$ &   $0.024\pm0.004$  &      0.011       &  0.021  & $0.040\pm0.010$  \\  
NGC\,6218   &  $0.196\pm0.005$ &  $0.225\pm0.006$ &   $0.210\pm0.005$  &      0.015       &  0.028  & $0.189\pm0.010$  \\  
\noalign{\smallskip}
        \multicolumn{7}{c}{Apparent $V$-band distance modulus} \\
\noalign{\smallskip}
NGC\,288    &  $14.76\pm0.01$  &  $14.85\pm0.02$  &   $14.80\pm0.02$   &    0.05        &  0.08   & $14.81\pm0.04$  \\  
NGC\,362    &  $14.83\pm0.01$  &  $14.91\pm0.02$  &   $14.87\pm0.01$   &    0.04        &  0.07   & $14.86\pm0.04$  \\ 
NGC\,1904   &  $15.59\pm0.01$  &  $15.64\pm0.02$  &   $15.62\pm0.01$   &    0.02        &  0.07   &  \\
NGC\,4372   &  $15.41\pm0.03$  &  $15.44\pm0.04$  &   $15.42\pm0.02$   &    0.01        &  0.08   &  \\
NGC\,5904   &  $14.43\pm0.01$  &  $14.47\pm0.02$  &   $14.45\pm0.01$   &    0.02        &  0.08   & $14.54\pm0.09$  \\ 
NGC\,6205   &  $14.40\pm0.01$  &  $14.45\pm0.01$  &   $14.43\pm0.01$   &    0.03        &  0.06   & $14.49\pm0.06$ \\
NGC\,6218   &  $14.12\pm0.01$  &  $14.23\pm0.02$  &   $14.18\pm0.02$   &    0.06        &  0.09   & $14.17\pm0.04$ \\
\hline
\end{tabular}
\]
\end{table*}

Table~\ref{estimates} presents our final estimates of [Fe/H], age, distance, distance modulus $(m-M)_0$, $E(B-V)$, and apparent $V$-band distance modulus $(m-M)_\mathrm{V}$ in comparison with our estimates from \citetalias{ngc5904}--\citetalias{ngc288} by use of the same or similar models when possible.
The uncertainties of our previous estimates are adopted taking into account the significant differences between these estimates for the BaSTI and DSED models, especially for age. 
Table~\ref{estimates} shows that the current estimates of reddening and apparent $V$-band distance modulus are similar to the previous ones, while those of age, distance, and distance modulus differ systematically.

Namely, the age estimates, averaged for BaSTI and DSED, become approximately 0.6--0.8 Gyr younger for all the clusters, as expected after the revision of BaSTI in 2021. Our age estimates for the oldest clusters, NGC\,288 and NGC\,6218, align more closely with those of \citet{dotter2010} and \citet{vandenberg2013}, as shown in Table~\ref{properties}, than with those of \citet{valcin2020}.
The relative ages from \citetalias{ceccarelli2025} agree with ours in the sense that NGC\,288 and NGC\,6205 are approximately the same age and much older than NGC\,362.

Our current distance estimates are lower than not only our previous ones, but also those from a comprehensive compilation of \citetalias{baumgardt2021}\footnote{\citetalias{baumgardt2021} compile distance estimates obtained by various methods.} for all the clusters but NGC\,362. 
This discrepancy is in line with the finding of \citetalias{baumgardt2021} that isochrone based distance moduli are systematically larger than RR~Lyrae based ones by about 0.02 mag.
This may be explained by the ignoring of RR~Lyrae variable stars in some of isochrone-fitting studies compiled by \citetalias{baumgardt2021}.
Observed at random phase, RR~Lyrae in globular clusters may appear much above (brighter) or below (fainter) the observed HB. When one derives distance by fitting a theoretical ZAHB to the lower bound of the observed HB and ignores the RR~Lyrae, which appear below the non-variable HB stars, this causes a bias to a higher derived distance or distance modulus. Accordingly, this can explain NGC\,362 as the exception: Figs~\ref{gaia}--\ref{lee} show that its RR~Lyrae differ in colour (systematically bluer) from the remaining HB stars and, hence, do not contaminate them to derive distance.
The detection and removal of the RR~Lyrae in this study must eliminate this bias, in contrast to \citetalias{ngc5904}--\citetalias{ngc288}, where we ignored the RR~Lyrae.

Anyway, our distance estimates agree with those from \citetalias{baumgardt2021} (see Table~\ref{properties}) within 
$0.5\sigma$, $0.5\sigma$, $0.6\sigma$, $1.5\sigma$, $1.1\sigma$, $0.3\sigma$, and $0.9\sigma$ of their total uncertainties 
for NGC\,288, NGC\,362, NGC\,1904, NGC\,4372, NGC\,5904, NGC\,6205, and NGC\,6218, respectively.
Also, our distance estimates agree with those from \citetalias{ceccarelli2025}.
There is significantly less agreement between our distance estimates and those of \citet{valcin2020} in Table~\ref{properties}: the latter are systematically higher probably due to ignoring of RR~Lyrae.

We convert our distances and those of \citetalias{baumgardt2021}, along with their uncertainties, into parallaxes and their corresponding uncertainties to facilitate comparison in Table~\ref{parallax} with parallaxes obtained from our analysis and \citetalias{vasiliev2021}, both derived from Gaia DR3 astrometry. 

The total uncertainty of any astrometric estimate of {\it Gaia} DR3 parallax cannot be better than 0.01~mas \citepalias{vasiliev2021}, 
while the total uncertainty of parallax, derived in isochrone fitting, increases with parallax 
(i.e. the relative parallax uncertainty is nearly constant as evident from Table~\ref{parallax}).
Accordingly, Table~\ref{parallax} indicates that the parallax estimates from the {\it Gaia} DR3 astrometry are less precise than those from our isochrone fitting for such distant clusters.\footnote{This agrees with the note of \citetalias{baumgardt2021}: `The accuracy of parallax distances also quickly deteriorates with increasing distance and drops below the accuracy of CMD fitting distances beyond distances of about 5 kpc.'}
\citetalias{baumgardt2021} estimates also exhibit higher precision than those derived from {\it Gaia} DR3 astrometry.
Anyway, Table~\ref{parallax} shows that all the parallax estimates are consistent, with only two exceptions:
the \citetalias{vasiliev2021} estimate is an outlier for NGC\,288
and the \citetalias{baumgardt2021} estimate contradicts that from our isochrone-fitting for NGC\,4372.
The latter discrepancy can be attributed to biases and errors in previous estimates, compiled by \citetalias{baumgardt2021}, for such a reddened cluster.

The total uncertainties in Table~\ref{estimates} appear comparable with those we adopt in \citetalias{ngc6397} and \citetalias{ngc5024} for [Fe/H], age, and distance modulus: $0.1$ dex, 0.8 Gyr, and 0.07 mag, respectively.
Comparing half the differences between the models with total uncertainties presented in Table~\ref{estimates}, we can conclude that the models agree each other in their estimates of age, distance, and distance modulus, but differ in those of [Fe/H] and reddening.
This is expected after our previous studies, since DSED provides a systematically lower [Fe/H] and, hence, a higher reddening than BaSTI.
It is known that the lower the metallicity of an isochrone, the higher the reddening derived from its fitting to a CMD and, consequently, the higher the $(m-M)_\mathrm{V}$.
Accordingly, the systematic uncertainty about $0.1$ dex of [Fe/H] is a significant contributor to the systematic uncertainties of our reddening and extinction estimates, resulting in values equivalent to $\sigma_{E(B-V)}\approx0.015$ and $\sigma_{A_\mathrm{V}}\approx0.05$\,mag, respectively.

Comparing our [Fe/H] estimates with values from the published results (see Table~\ref{properties}), we find that, unlike the comparison for Galactic globular clusters of a lower metallicity in \citetalias{ngc5024}, our estimates align more closely with those of \citet{carretta2009} than with those of \citet{meszaros2020}. 
Additionally, they fall between the spectroscopic and photometric estimates provided by \citet{jurcsik2023}.
Thus, the arguments of \citet{mucciarelli2020} in favour of photometrically and against spectroscopically derived [Fe/H] of low-metallicity globular clusters may not be valid for higher metallicity ones.
Note that our estimate [Fe/H]$=-2.28\pm0.09$ for NGC\,4372 agrees well with $-2.19\pm0.03$ from high-resolution spectroscopy presented by \citet{sanroman2015}.

\subsection{Reddening, extinction, and extinction law}
\label{extinction}

\begin{figure*}
\includegraphics{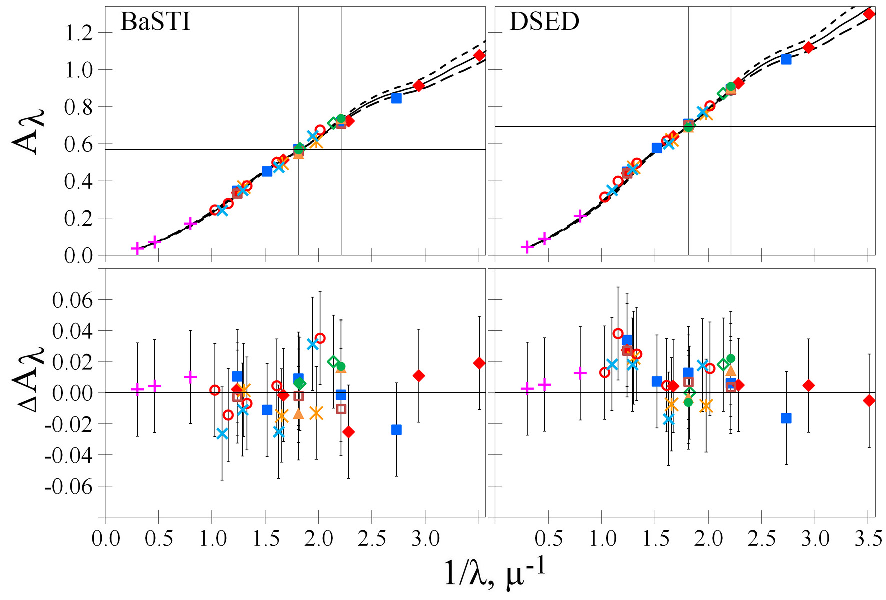}
\caption{Top: the empirical extinction law for NGC\,6218 from the isochrone fitting by the different models.
The $B$ and $V$ filters are denoted by the vertical lines.
The black dotted, solid, and dashed curves show the extinction law of \citetalias{ccm89} with $R_\mathrm{V}=2.9$, $3.1$ and $3.3$, respectively, with 
the derived $A_\mathrm{V}$, which is shown by the horizontal line.
The error bars are not shown in the top diagram, since they are about the height of the symbols used.
Bottom: the data set residuals around the extinction law of \citetalias{ccm89} with $R_\mathrm{V}=3.2$.
The data sets are:
\citetalias{nardiello2018} -- red diamonds;
Lee data set -- open green diamonds;
{\it Gaia} -- yellow snowflakes;
\citetalias{stetson2019} -- blue squares;
\citet{narloch2017} -- green circles;
\citet{zloczewski2012} -- yellow triangles;
\citet{hargis2004} -- open brown squares;
PS1 -- open red circles;
SMSS -- blue inclined crosses;
VISTA and unWISE -- purple upright crosses.
}
\label{ngc6218law}
\end{figure*}

As in our previous studies, we use the optical--IR CMDs to convert the derived reddenings into extinction for each filter under consideration. 
For example, the extinction $A_\mathrm{V}$ in the $V$ filter can be calculated as
\begin{equation}
\label{avaw1}
A_\mathrm{V}=(A_\mathrm{V}-A_\mathrm{W1})+A_\mathrm{W1}=E(V-W1)+A_\mathrm{W1},
\end{equation}
where reddening $E(V-W1)$ is obtained from a CMD, while very low extinction $A_\mathrm{W1}$ in the $W1$ filter is calculated using the 
\citet[][hereafter CCM89]{ccm89} extinction law with a certain $R_\mathrm{V}$,\footnote{Extinction-to-reddening ratio 
$R_\mathrm{V}\equiv A_\mathrm{V}/E(B-V)=3.1$ is defined for early type MS stars, while the observed ratio $A_\mathrm{V}/E(B-V)$ depends on the intrinsic spectral energy distribution 
of the stars under consideration \citep{casagrande2014}. 
For rather cool (the median effective temperature 5700~K) and metal-poor stars of Galactic globular clusters the observed extinction in the $V$ filter is calculated as $A_\mathrm{V}\approx3.4\,E(B-V)$.} 
which, in turn, is derived from the fitting of extinctions in all the filters by the \citetalias{ccm89} extinction law.
$R_\mathrm{V}$ and the extinctions in all the filters are updated iteratively, starting with an initial value of $R_\mathrm{V}=3.1$.
For all the clusters under consideration, any variation of $R_\mathrm{V}$ between 2.2 and 5.0 leads to a variation of $A_\mathrm{W1}$ within $\pm0.01$\,mag 
($\pm0.025$\,mag for NGC\,4372 due to its higher extinction). 
Equation~(\ref{avaw1}) shows that this variation becomes an additional uncertainty of any derived extinction, albeit negligible w.r.t. its total uncertainty.

For each combination of cluster, data set, and model we derive a set of $R_\mathrm{V}$ and extinctions, which draws an independent empirical extinction law.
To draw it more accurately, we reduce the scatter of extinction estimates by adjustment of data sets as described in our previous papers and in \ref{adjustment}.
An example of the empirical extinction law for NGC\,6218 by use of two models and various data sets is presented in Fig.~\ref{ngc6218law}.
A rather low scatter is seen not only for different CMDs of the same data set (the series of the same symbols in Fig.~\ref{ngc6218law}) but also for different data sets (the different series of the symbols in Fig.~\ref{ngc6218law}).
Finally, Fig.~\ref{ngc6218law} displays for NGC\,6218 an excellent agreement of the extinctions, derived for all the filters, following the \citetalias{ccm89} extinction law with $R_\mathrm{V}=3.2\pm0.1$.
Note that the VISTA and UKIDSS IR photometry appears especially useful in this study to obtain more precise $R_\mathrm{V}$ and extinction estimates than those in \citetalias{ngc5904}--\citetalias{ngc288}.

Our final estimates with their total uncertainties are
$A_\mathrm{V}=0.09\pm0.06$, $0.09\pm0.06$, $0.11\pm0.06$, $1.58\pm0.06$, $0.13\pm0.06$, $0.09\pm0.06$, and $0.67\pm0.06$,
as well as
$R_\mathrm{V}=3.9\pm0.7$,   $3.0\pm0.5$,   $3.8\pm0.5$,   $2.9\pm0.4$,   $2.9\pm0.2$,   $3.6\pm0.7$, and $3.2\pm0.1$
for NGC\,288, NGC\,362, NGC\,1904, NGC\,4372, NGC\,5904, NGC\,6205, and NGC\,6218, respectively.
Thus, NGC\,288 and NGC\,1904 exhibit an extinction law with $R_\mathrm{V}>3.1$, while the extinction laws for the remaining clusters are consistent with the common value of $R_\mathrm{V}=3.1$.

Comparing our reddening estimates with those in Tables~\ref{properties}, we find that our values align closely with those of \citetalias{sfd98}, \citet{schlaflyfinkbeiner2011}, and \citet{planck} for all the clusters.
The estimates from \citet{harris} and \citet{lallement2019} for NGC\,4372, as well as the \citet{lallement2019} estimate for NGC\,6218, are outliers and seem to be erroneous.

Six of the clusters are located at middle or high Galactic latitudes. The lowest $A_\mathrm{V}$ they exhibit is 0.09 mag, though this value is somewhat uncertain.
Hence, these clusters confirm our conclusion from \citetalias{ngc5024}, based on our estimates for five other high-latitude clusters, 
that the typical total Galactic extinction from the Sun to extragalactic objects at high latitudes is $A_\mathrm{V}>0.08$\,mag.

\subsection{Relative estimates}
\label{relative}

\begin{table*}
\centering
\def\baselinestretch{1}\normalsize\scriptsize
\caption{The relative estimates presented as cluster sequences along ascending or descending parameter. 
}
\label{sequences}
\[
\begin{tabular}{lc}
\hline
\noalign{\smallskip}
Parameter                                           & Sequence \\
\hline
\noalign{\smallskip}
$\Delta$Age (Gyr) from old to young      & NGC\,1904 - \textcolor{red}{0.33} -  NGC\,6218 - \textcolor{red}{0.00} - NGC\,288 - \textcolor{red}{0.17} - NGC\,6205 - \textcolor{red}{0.16} - NGC\,4372 - \textcolor{red}{1.17} - NGC\,5904 - \textcolor{red}{1.00} - NGC\,362 \\
$\Delta R$ (kpc) from distant to nearby     & NGC\,1904 - \textcolor{red}{3.67} - NGC\,362 - \textcolor{red}{0.08} - NGC\,288 - \textcolor{red}{1.50} - NGC\,6205 - \textcolor{red}{0.11} - NGC\,5904 - \textcolor{red}{2.14} - NGC\,4372 - \textcolor{red}{0.21} - NGC\,6218 \\ 
$\Delta$[Fe/H] (dex) from metal-poor to metal-rich   & NGC\,4372 - \textcolor{red}{0.63} - NGC\,1904 - \textcolor{red}{0.04} - NGC\,6205 - \textcolor{red}{0.25} - NGC\,5904 - \textcolor{red}{0.07} - NGC\,288 - \textcolor{red}{0.00} - NGC\,6218 - \textcolor{red}{0.00} - NGC\,362 \\
$\Delta E(B-V)$ (mag) from less to more reddened   & NGC\,288 - \textcolor{red}{0.008} - NGC\,362 - \textcolor{red}{0.002} - NGC\,6205 - \textcolor{red}{0.000} - NGC\,1904 - \textcolor{red}{0.020} - NGC\,5904 - \textcolor{red}{0.160} - NGC\,6218 - \textcolor{red}{0.339} - NGC\,4372 \\
\hline
\end{tabular}
\]
\end{table*}

The systematic differences between the models are relatively large. 
However, these differences are expected to cancel out in relative estimates of the cluster parameters, especially for [Fe/H] and $E(B-V)$ demonstrating a high model-to-model systematics. Their relative estimates can be obtained with the uncertainties 0.05 dex and 0.005 mag, respectively.
The relative estimates of age and distance can possibly be obtained with the uncertainties 0.4 Gyr and 0.1 kpc, respectively.
To derive the relative estimates, we use three optical CMDs available for all the clusters: (i) {\it Gaia} DR3 $G_\mathrm{BP}-G_\mathrm{RP}$, (ii) \citetalias{stetson2019} $B-I$, and (iii) Lee's $b-y$.

Table~\ref{sequences} presents the relative estimates of the parameters by red-coloured values between the cluster names ranked by ascending or descending parameter. A small value between a pair of clusters means their similarity in this parameter, while a large value means a significant difference between them.
Table~\ref{sequences} confirms nearly identical [Fe/H] values for the quartet NGC\,5904, NGC\,288, NGC\,6218, and NGC\,362, as well as for NGC\,1904 and NGC\,6205.
For NGC\,1904, NGC\,6218, NGC\,288, NGC\,6205, and NGC\,4372, the age decreases slightly but, taking into account its uncertainty, it can still be regarded as nearly the same, around 13~Gyr. 
In contrast, NGC\,5904 is significantly younger, and NGC\,362 is even younger, potentially making it one of the youngest globular clusters in the Galaxy.
The relative age estimates for the quartet ensure that NGC\,288 and NGC\,6218 have nearly the same age, significantly older ($1.5\pm0.4$ Gyr) than NGC\,5904, which is older ($1.0\pm0.4$ Gyr) than NGC\,362. In total, NGC\,288 and NGC\,6218 are $2.5\pm0.4$ Gyr older than NGC\,362. 
Consequently, this confirms that age can be the second parameter influencing HB morphology of this quartet.
Also, Table~\ref{sequences} confirms that NGC\,288, NGC\,362, NGC\,6205, and NGC\,1904 have nearly the same reddening.

\subsection{The HB morphology}
\label{hb}

Tables~\ref{estimates} and \ref{sequences} show a negligible difference between the derived [Fe/H] for the quartet of clusters with [Fe/H]$\approx-1.3$.
Therefore, their HB morphology differences cannot be attributed to metallicity. 
However, variations in their age estimates can mainly explain the observed differences in their HB morphology.

Similarly, the HB morphology difference between NGC\,1904, NGC\,6205, and NGC\,5272 (the latter from \citetalias{ngc5024}) with [Fe/H]$\approx-1.6$ can be attributed to their age difference, although age alone does not provide a complete explanation, as discussed later.

Comparing NGC\,4372 with the metal-poor clusters NGC\,5024, NGC\,5053, NGC\,5466, and NGC\,7099 from \citetalias{ngc5024}, we note that 
they have very similar ages of approximately 12.8 Gyr, except for the slightly younger NGC\,5466. 
[Fe/H] of NGC\,4372 is slightly lower than that those of the remaining clusters, while its HB closely resembles those of NGC\,5053 and NGC\,7099 in the sparse population of their blue HB.

\begin{table*}
\def\baselinestretch{1}\normalsize\normalsize
\caption{Our count of the blue HB, RR~Lyrae, and red HB stars and calculated HB type of the clusters. 
The clusters are divided into 3 groups with similar [Fe/H] and sorted by their derived age (Gyr) within each group. The [Fe/H] and age estimates are taken from Table~\ref{estimates}
}
\label{hbtype}
\[
\begin{tabular}{lcccccc}
\hline
\noalign{\smallskip}
Cluster   & Blue HB & RR Lyrae & Red HB & HB type       & [Fe/H] & Age \\
\hline
\noalign{\smallskip}
NGC\,4372 &   191   &     0    &   0    & $+1.00$        &  $-2.28$ & 12.8 \\
\hline
\noalign{\smallskip}
NGC\,1904 &    73   &     8    &   0    & $+0.90^{+0.06}_{-0.06}$ & $-1.64$ & 13.2 \\
\noalign{\smallskip}
NGC\,6205 &   555   &     8    &   3    & $+0.98^{+0.01}_{-0.01}$  & $-1.56$ & 12.8 \\
\hline
\noalign{\smallskip}
NGC\,6218 &   169   &     0    &   2    & $+0.98^{+0.01}_{-0.03}$  & $-1.27$ & 13.0 \\
\noalign{\smallskip}
NGC\,288  &   119   &     2    &   3    & $+0.94^{+0.02}_{-0.02}$  & $-1.28$ & 12.9 \\
\noalign{\smallskip}
NGC\,5904 &   316   &   127    &  70    & $+0.48^{+0.04}_{-0.04}$  & $-1.33$ & 11.5 \\
\noalign{\smallskip}
NGC\,362  &    18   &    32    & 335    & $-0.82^{+0.05}_{-0.05}$  & $-1.26$ & 10.3 \\
\hline
\end{tabular}
\]
\end{table*}

We calculate HB types of the clusters by counting the HB stars in the {\it Gaia}, \citetalias{stetson2019}, and \citetalias{nardiello2018} data sets, except for NGC\,1904 and NGC\,4372, for which only the {\it Gaia} and \citetalias{stetson2019} data sets are used. Our choice of the data sets is due to \citetalias{nardiello2018} covers only few central arcminutes of the cluster fields, while 
{\it Gaia} and \citetalias{stetson2019}, cross-identified with {\it Gaia}, cover the remaining outer parts of the fields. In addition, {\it Gaia} and \citetalias{stetson2019} allow us to detect some variable stars omitted in the \citet{clement2017} data base using their variability parameters mentioned in Sect.~\ref{fitting}.\footnote{In case of discrepancy, we prefer the \citet{clement2017} data base. For example, {\it Gaia} DR3 declares four HB variables in NGC\,6218 at its RR~Lyrae magnitudes and colours, which are absent in the \citet{clement2017} data base. We do not consider them as RR~Lyrae yet, but we increase the HB type uncertainty accordingly.} All the three data sets are cross-identified each other in order to count each HB star once. This provides a complete count of the HB stars of the clusters presented in Table~\ref{hbtype} together with the calculated HB types.
Table~\ref{properties} shows a good agreement of our HB types with published results.

The uncertainties of our HB type estimates are calculated by a Monte-Carlo simulation (except for NGC\,4372 due to its high DR, although it seems that NGC\,4372 has no red HB stars and RR~Lyrae and, hence, its HB type is exactly $+1$).
Briefly, we take into account the probability for a blue HB, RR~Lyra or red HB star to be out of the correspondent CMD domain or the probability for an extraneous star to be in these domains based on the star's photometric and DR uncertainty and variability amplitude. The star position in the CMD and the residual contamination of the clean samples by field stars are simulated with the TRILEGAL code \citep{trilegal}.
It is worth noting that the accurate count of the red HB stars and RR~Lyrae affects the HB types much more than that of the blue HB stars, as indicated in \citetalias{ngc5024}.

The clusters in Table~\ref{hbtype} are divided into 3 groups with similar [Fe/H] and sorted by their derived age within each group.
It is seen that this sorting almost perfectly follows the decrease of their HB type. This confirms that metallicity is the first and age is the second parameter of the
HB morphology.
The only exception is the pair of NGC\,1904 and NGC\,6205.
Table~\ref{sequences} indicates that the former is $\Delta$[Fe/H]$=0.04$ metal poorer and 0.5 Gyr older, i.e. the both factors are in contradiction with its shorter tail of the blue HB stars in Figs.~\ref{gaia}--\ref{lee} and corresponding lower HB type in Table~\ref{hbtype}.
%
%

To understand the HB morphology, it is important to note that HB stars demonstrate a strong correlation between their mass, effective temperature, and colour.
The mass increases (from about 0.48 to 0.80 solar masses) as the temperature decreases (from about 32\,000 to 5300~K) and the dereddened colour index increases
[from about $(B-I)_0=-0.57$ to $+1.5$], moving from left to right across the CMDs shown in Figs.~\ref{gaia}--\ref{lee}.

The best-fitting BaSTI isochrones for the MS, TO, SGB, RGB, and AGB, i.e. the red and orange curves in Figs.~\ref{gaia}--\ref{lee} for a primordial $Y\approx0.25$ and $0.275$, respectively, 
describe regular stellar evolution including a moderate mass loss with its efficiency $\eta=0.3$.\footnote{We do not consider the DSED isochrones in this subsection, since they do not take into account mass loss and different $Y$ at the HB and AGB.}

In contrast, any HB or AGB star far from a red or orange curve in Figs.~\ref{gaia}--\ref{lee} needs an additional explanation beyond the regular stellar evolution.
The BaSTI extended set of ZAHB models with a stochastic mass loss between the MS and HB, presented in Figs.~\ref{gaia}--\ref{lee} by the purple and blue curves, gives us
such an additional explanation.
Namely, a higher mass loss efficiency produces less massive (hotter and bluer) HB stars and, hence, shifts them leftward along purple or blue curves in Figs.~\ref{gaia}--\ref{lee}.
Conversely, a lower mass-loss efficiency would shift HB stars to the right on the CMD.

Figs.~\ref{gaia}--\ref{lee} show that all the clusters, except NGC\,362 and NGC\,4372, demonstrate many HB stars located to the left of the red or orange regular evolution isochrone curves.\footnote{For comparison, there was only one such cluster NGC\,5272 among five in \citetalias{ngc5024}.}
Moreover, the distribution of these HB stars along their mass (as well as effective temperature and colour), i.e. along the purple or blue ZAHB curves in Figs.~\ref{gaia}--\ref{lee}, 
is nearly Gaussian, as expected in the case of stochastic mass loss \citep{catelan2009}.
This means that the HB morphology of these five clusters can be explained by a higher mass loss efficiency $\eta>0.3$ as the third parameter after metallicity and age.
Moreover, the longer the cluster's blue HB tail along the purple/blue ZAHB curves, the higher its mass loss efficiency.
In particular, the NGC\,6205 mass loss efficiency is much higher than that in NGC\,1904, while their [Fe/H] and age are nearly the same.
Also, a comparison of Figs.~\ref{gaia}--\ref{lee} with figures 1 and 3 from \citetalias{ngc5024} for NGC\,5272 indicates that the latter is significantly younger than NGC\,1904 and NGC\,6205. 
However, NGC\,5272 also demonstrates a higher mass loss efficiency $\eta>0.3$, as evident from the presence of numerous HB stars to the left of all regular evolution isochrones, as discussed in \citetalias{ngc5024}.

A higher helium mass fraction, such as $Y>0.3$, as alternative to a higher mass loss efficiency, cannot account for the long blue HB tails of these clusters, since both theory (e.g. the best-fitting BaSTI isochrones in Figs~\ref{gaia}--\ref{lee}) and
observations ($Y$ estimates for different generations of these clusters) deny very high $Y$ for these clusters.
Moreover, it is difficult to disentangle the effects of increased helium mass fraction and mass loss on the observed HBs, as both lead to a decrease in the mass of HB stars, making them bluer.
However, \citet{tailo2020} appear to resolve this degeneracy with their approach. 
They derive estimates of $\eta$, albeit for only a subset of clusters, which align with and confirm our suggestions:
NGC\,288, NGC\,5904, NGC\,6205 and NGC\,6218 demonstrate mass loss much higher than the common $\eta=0.3$:
$\eta=0.52\pm0.02$, $0.40\pm0.03$, $0.53\pm0.03$, and $0.54\pm0.03$, respectively.
Moreover, \citet{tailo2020} estimates divide other clusters from our studies into NGC\,5272, NGC\,6362, NGC\,6397, NGC\,6723, and NGC\,6809 with rather high $\eta=0.46\pm0.02$, $0.48\pm0.03$, $0.37\pm0.02$, $0.40\pm0.03$, and $0.37\pm0.02$, respectively, versus 
NGC\,5024, NGC\,5053, NGC\,5466, and NGC\,7099 with lower $\eta=0.26\pm0.02$, $0.32\pm0.02$, $0.26\pm0.02$, and $0.19\pm0.02$, respectively, in a good agreement with the existence and length of their blue HB tails.

Thus, the majority (at least, 9 among 16) of clusters studied by us indicate a high mass-loss efficiency with $\eta>0.3$, which suggests a substantial amount of expelled intracluster gas and dust. Recent research by \citet{pancino2024} provides insights into the characteristics and distribution of this intracluster medium.


No \citet{tailo2020} estimates for NGC\,362, NGC\,1904, and NGC\,4372.
The long blue HB tail suggesting a high mass-loss efficiency is evident for NGC\,1904 in Figs~\ref{gaia}--\ref{lee}, while the case of NGC\,362 and NGC\,4372 is uncertain.
NGC\,362 has a nearly Gaussian distribution of the HB stars in colour and, accordingly, in mass and effective temperature,\footnote{The hottest NGC\,362 HB member {\it Gaia} DR3 4690839418131927680 with a mass as low as $0.60\pm0.01$ solar masses is very remarkable.} while NGC\,4372 demonstrates a non-Gaussian distribution with an abrupt cutoff on its HB blue side.
This cutoff corresponds to the elimination of all HB stars with masses $<0.642\pm0.017$ solar masses \citep{gratton2010} or $<0.66\pm0.01$ solar masses by our estimate using BaSTI. Note that these estimates differ due to an older model and different DR correction used by \citet{gratton2010}.

In \citetalias{ngc5024}, we discussed such an abrupt cutoff in the distribution of the HB stars or, in other words, the absence of the lowest mass HB stars, for the 
core collapse cluster NGC\,7099 and the loose clusters NGC\,5053 and NGC\,5466, attributed to their dynamical evolution and mass segregation.
In particular, loose cluster must lose stars due to two-body encounters and a tidal shock in a rapidly changing Galactic potential during its crossing of the Galactic disk \citep{odenkirchen2004,sollima2017}.
Thus, probably, both core-collapse and loose clusters lose low-mass stars more efficiently than clusters with intermediate star concentrations \citep{meylan1997}.
NGC\,5053, NGC\,5466, and NGC\,7099 demonstrate nearly the same cutoff level as NGC\,4372: $0.660\pm0.005$, $0.670\pm0.005$, and $0.650\pm0.005$, respectively 
(the higher value for NGC\,5466 is due to its younger age).
For comparison, NGC\,5024, with similar metallicity and age but without such a cutoff, demonstrates a lower minimum HB mass of $0.640\pm0.005$ solar masses.
Thus, NGC\,4372 can be considered an analog of NGC\,5053 and NGC\,7099.
However, NGC\,4372 seems to be neither a core-collapse nor a loose cluster, although \citet{kacharov2014} suggest it as a re-bounced, post-core-collapse cluster.
On the other hand, NGC\,362 and NGC\,1904 are core-collapse clusters (as confirmed by their high core density and the ratio of tidal to core radii presented in Table~\ref{properties}), but they do not show an evident loss of low-mass HB stars.

It seems that a detailed analysis of cluster evolution and orbit may resolve this issue, as it has been proposed by \citet{meylan1997} for the trio of core-collapse clusters with a similar [Fe/H], age, but different blue HB tails -- NGC\,6397, NGC\,7078, and NGC\,7099. To lose low-mass stars, it may be more important to have many slow disk crossings than to be a core-collapse or loose cluster.
Therefore, it is probably important that the Galactic orbit of NGC\,4372 involves many slow disk crossings \citep{kacharov2014,baumgardt2021}. 
This follows from the NGC\,4372's
rather short orbital period of $98^{+2}_{-4}$~Myr, very low orbital inclination of $27\pm1^{\circ}$, and recent time $45\pm1$~Myr since the last disk crossing compared to such values 
$220^{+6}_{-4}$ Myr, $68^{+24}_{-9}$$^{\circ}$, and $92\pm1$ Myr for the unlike NGC\,1904
and 
$344^{+30}_{-14}$ Myr, $75\pm1^{\circ}$, and $188\pm2$ Myr for the unlike NGC\,5024 \citep{bajkova2022}.
The impact of evolution history of Galactic globular clusters into their HB morphology should be investigated further.

Thus, the HB morphology difference between metal-poor clusters from \citetalias{ngc5024} and this study can be explained by their age as the second parameter. Additionally, their regular stellar evolution with moderate mass loss ($\eta\approx0.3$) generates few low-mass blue HB stars making mass loss efficiency the third parameter. The different evolutionary histories of the clusters seem to be the fourth parameter. These histories eliminate low-mass HB stars in the loose clusters NGC\,5053 and NGC\,5466, the core-collapse cluster NGC\,7099, and NGC\,4372, which is probably a re-bounced, post-core-collapse cluster or has been disrupted by its frequent slow crossings of the Galactic disk.

\section{Conclusions}
\label{conclusions}

Following our approach developed in \citetalias{ngc5904}--\citetalias{ngc5024}, 
we have estimated metallicity [Fe/H], age, distance $R$ from the Sun, reddenings, extinctions in various filters, and extinction-to-reddening ratio $R_\mathrm{V}$ for NGC\,1904 and NGC\,4372. 
We have also re-estimated these parameters for NGC\,288, NGC\,362, NGC\,5904, NGC\,6205, and NGC\,6218 from \citetalias{ngc5904}--\citetalias{ngc288} to account for the significant upgrade of the data sets and isochrones in recent years.
We fitted BaSTI and DSED theoretical isochrones for [$\alpha$/Fe]$=+0.4$ to CMDs based on multi-band photometry from the {\it HST}, {\it Gaia} DR3, PS1, SDSS, SMSS DR4, UKIDSS, VISTA VHS DR5, unWISE, a large compilation of the $UBVRI$ ground-based observations by \citetalias{stetson2019}, and other data sets.
The filters under consideration span a wide wavelength range from the UV to mid-IR. 
{\it Gaia} DR3 and {\it HST} proper motions and parallaxes are used to select cluster members in almost all the data sets and to calculate the median parallax and systemic proper motion of the clusters.
Besides the selection of members, cross-identification of the data sets to each other allowed us to consider systematic differences between them and calculate the most probable empirical extinction law for each cluster.
These laws are generally close to the \citetalias{ccm89} extinction law with $R_\mathrm{V}=3.1$, except for NGC\,288 and NGC\,1904, which have $R_\mathrm{V}=3.9\pm0.7$ and $3.8\pm0.5$, respectively.

We present the obtained estimates of [Fe/H], age, $R$, distance modulus, reddening $E(B-V)$, and apparent $V$-band distance modulus for all the clusters in Table~\ref{estimates}.
We estimated the severe differential reddening up to $\Delta E(B-V)=0.5$ in the NGC\,4372 field due to foreground Musca dark nebula and other dust clouds detectable in our 3D extinction map.
Six of the clusters are located at middle or high Galactic latitudes. With the lowest extinction estimate $A_\mathrm{V}=0.09$\,mag for them, these clusters confirm our conclusion for five other high-latitude clusters in \citetalias{ngc5024} that 
the typical total Galactic extinction from the Sun to extragalactic objects at middle and high latitudes is $A_\mathrm{V}>0.08$\,mag.

We calculated the HB types of the clusters under consideration and analysed the distribution of their HB stars in colour, effective temperature, and mass using the BaSTI isochrones. 
This allowed us to confirm the suggestion of \citetalias{ngc5024} that the HB morphology difference of Galactic globular clusters can be explained by their different metallicity, age, mass loss efficiency, and loss of low-mass members, including the bluest HB stars, 
during cluster evolution. The latter process seems to be particularly important for NGC\,4372.
The majority of 16 clusters from our studies indicate a relatively high mass-loss efficiency, consistent with the Reimers mass-loss law with $\eta > 0.3$.

\section*{Acknowledgements}

J.-W.\ Lee performed Str\"omgren observations with the 1-m telescope at CTIO and the 0.9-m telescope at KPNO for all the clusters (see Sect.~\ref{datasets}), 
their processing and analysis with the financial support from the Basic Science Research Program (grant Nos. 2019R1A2C2086290 and RS-2025-20252972) through the National Research Foundation of Korea (NRF).
All the remaining investigations were performed with the financial support from the Russian Science Foundation (grant no. 20--72--10052).

We thank the anonymous reviewer for useful comments.
We thank
Armando Arellano Ferro for very fruitful discussion of the cluster RR~Lyrae stars,
Anisa Bajkova for discussion of cluster orbital parameters,
Vadim Bobylev for discussion of the NGC\,4372 area,
Santi Cassisi for providing the valuable BaSTI isochrones with his exceptionally useful comments,
Aaron Dotter for his comments on DSED,
Gregory Green for discussion of extinction/reddening estimates in the fields of globular clusters,
Frank Grundahl for providing his valuable Str\"omgren data sets with very useful comments,
Weronika Narloch for her comments on some data sets and variable stars,
Christopher Onken for his help in using the SkyMapper Southern Sky Survey,
Peter Stetson for providing and having discussion of his valuable $UBVRI$ photometry,
Don VandenBerg for discussion of many aspects of globular clusters,
Eugene Vasiliev for his very useful comments on the cluster properties.

This work has made use of BaSTI and DSED web tools;
Filtergraph \citep{filtergraph}, an online data visualization tool developed at Vanderbilt University through the Vanderbilt Initiative in 
Data-intensive Astrophysics (VIDA) and the Frist Center for Autism and Innovation (FCAI, \url{https://filtergraph.com});
the resources of the Centre de Donn\'ees astronomiques de Strasbourg, Strasbourg, France (\url{http://cds.u-strasbg.fr}), including the SIMBAD database, 
the VizieR catalogue access tool \citep{vizier} and the X-Match service;
observations made with the NASA/ESA {\it Hubble Space Telescope};
data products from the {\it Wide-field Infrared Survey Explorer}, which is a joint project of the University of California, Los Angeles, and the Jet Propulsion 
Laboratory/California Institute of Technology;
data products from the Pan-STARRS Surveys (PS1);
data products from the Sloan Digital Sky Survey;
data products from the SkyMapper Southern Sky Survey, SkyMapper is owned and operated by The Australian National University's Research School of Astronomy and Astrophysics,
the SkyMapper survey data were processed and provided by the SkyMapper Team at ANU, the SkyMapper node of the All-Sky Virtual Observatory (ASVO) is hosted at the National 
Computational Infrastructure (NCI);
data products from the Two Micron All Sky Survey, which is a joint project of the University of Massachusetts and the Infrared Processing and Analysis Center/California 
Institute of Technology, funded by the National Aeronautics and Space Administration and the National Science Foundation;
data from the European Space Agency (ESA) mission {\it Gaia} (\url{https://www.cosmos.esa.int/gaia}), processed by the {\it Gaia} Data Processing and Analysis Consortium 
(DPAC, \url{https://www.cosmos.esa.int/web/gaia/dpac/consortium}), and {\it Gaia} archive website (\url{https://archives.esac.esa.int/gaia});
data products from the VISTA Hemisphere Survey catalog DR5 based on observations made with ESO Telescopes at the La Silla or Paranal Observatories under
programme ID(s) 179.A-2010(A), 179.A-2010(B), 179.A-2010(C), 179.A-2010(D), 179.A-2010(E), 179.A-2010(F), 179.A-2010(G),
179.A-2010(H), 179.A-2010(I), 179.A-2010(J), 179.A-2010(K), 179.A-2010(L), 179.A-2010(M), 179.A-2010(N), 179.A-2010(O).

\section*{Competing interests}

None.

\section*{Data availability statement}

The data underlying this article will be shared on reasonable request to the corresponding author.


\printbibliography

@ARTICLE{des,
       author = {{Abbott}, T.~M.~C. and {Adam{\'o}w}, M. and {Aguena}, M. and {Allam}, S. and {Amon}, A. and {Annis}, J. and {Avila}, S. and {Bacon}, D. and {Banerji}, M. and {Bechtol}, K. and {Becker}, M.~R. and {Bernstein}, G.~M. and {Bertin}, E. and {Bhargava}, S. and {Bridle}, S.~L. and {Brooks}, D. and {Burke}, D.~L. and {Carnero Rosell}, A. and {Carrasco Kind}, M. and {Carretero}, J. and {Castander}, F.~J. and {Cawthon}, R. and {Chang}, C. and {Choi}, A. and {Conselice}, C. and {Costanzi}, M. and {Crocce}, M. and {da Costa}, L.~N. and {Davis}, T.~M. and {De Vicente}, J. and {DeRose}, J. and {Desai}, S. and {Diehl}, H.~T. and {Dietrich}, J.~P. and {Drlica-Wagner}, A. and {Eckert}, K. and {Elvin-Poole}, J. and {Everett}, S. and {Evrard}, A.~E. and {Ferrero}, I. and {Fert{\'e}}, A. and {Flaugher}, B. and {Fosalba}, P. and {Friedel}, D. and {Frieman}, J. and {Garc{\'\i}a-Bellido}, J. and {Gaztanaga}, E. and {Gelman}, L. and {Gerdes}, D.~W. and {Giannantonio}, T. and {Gill}, M.~S.~S. and {Gruen}, D. and {Gruendl}, R.~A. and {Gschwend}, J. and {Gutierrez}, G. and {Hartley}, W.~G. and {Hinton}, S.~R. and {Hollowood}, D.~L. and {Honscheid}, K. and {Huterer}, D. and {James}, D.~J. and {Jeltema}, T. and {Johnson}, M.~D. and {Kent}, S. and {Kron}, R. and {Kuehn}, K. and {Kuropatkin}, N. and {Lahav}, O. and {Li}, T.~S. and {Lidman}, C. and {Lin}, H. and {MacCrann}, N. and {Maia}, M.~A.~G. and {Manning}, T.~A. and {Maloney}, J.~D. and {March}, M. and {Marshall}, J.~L. and {Martini}, P. and {Melchior}, P. and {Menanteau}, F. and {Miquel}, R. and {Morgan}, R. and {Myles}, J. and {Neilsen}, E. and {Ogando}, R.~L.~C. and {Palmese}, A. and {Paz-Chinch{\'o}n}, F. and {Petravick}, D. and {Pieres}, A. and {Plazas}, A.~A. and {Pond}, C. and {Rodriguez-Monroy}, M. and {Romer}, A.~K. and {Roodman}, A. and {Rykoff}, E.~S. and {Sako}, M. and {Sanchez}, E. and {Santiago}, B. and {Scarpine}, V. and {Serrano}, S. and {Sevilla-Noarbe}, I. and {Smith}, J. Allyn and {Smith}, M. and {Soares-Santos}, M. and {Suchyta}, E. and {Swanson}, M.~E.~C. and {Tarle}, G. and {Thomas}, D. and {To}, C. and {Tremblay}, P.~E. and {Troxel}, M.~A. and {Tucker}, D.~L. and {Turner}, D.~J. and {Varga}, T.~N. and {Walker}, A.~R. and {Wechsler}, R.~H. and {Weller}, J. and {Wester}, W. and {Wilkinson}, R.~D. and {Yanny}, B. and {Zhang}, Y. and {Nikutta}, R. and {Fitzpatrick}, M. and {Jacques}, A. and {Scott}, A. and {Olsen}, K. and {Huang}, L. and {Herrera}, D. and {Juneau}, S. and {Nidever}, D. and {Weaver}, B.~A. and {Adean}, C. and {Correia}, V. and {de Freitas}, M. and {Freitas}, F.~N. and {Singulani}, C. and {Vila-Verde}, G. and {Linea Science Server}},
        title = "{The Dark Energy Survey Data Release 2}",
      journal = {\apjs},
         year = 2021,
        month = aug,
       volume = {255},
       number = {2},
          eid = {20},
        pages = {20},
          doi = {10.3847/1538-4365/ac00b3},
archivePrefix = {arXiv},
       eprint = {2101.05765},
 primaryClass = {astro-ph.IM},
       adsurl = {https://ui.adsabs.harvard.edu/abs/2021ApJS..255...20A}
}

@ARTICLE{aguado2025,
       author = {{Aguado-Agelet}, Fernando and {Massari}, Davide and {Monelli}, Matteo and {Cassisi}, Santi and {Gallart}, Carme and {Ceccarelli}, Edoardo and {Gonz{\'a}lez-Koda}, Yllari Kay and {Ruiz-Lara}, Tom{\'a}s and {Pancino}, Elena and {Saracino}, Sara and {Salaris}, Maurizio},
        title = "{Cluster Ages to Reconstruct the Milky Way Assembly (CARMA). II. The age-metallicity relation of Gaia-Sausage-Enceladus globular clusters}",
      journal = {arXiv e-prints},
         year = 2025,
        month = feb,
          eid = {arXiv:2502.20436},
        pages = {arXiv:2502.20436},
          doi = {10.48550/arXiv.2502.20436},
archivePrefix = {arXiv},
       eprint = {2502.20436},
 primaryClass = {astro-ph.GA},
       adsurl = {https://ui.adsabs.harvard.edu/abs/2025arXiv250220436A}
}

@ARTICLE{an2008,
       author = {{An}, Deokkeun and {Johnson}, Jennifer A. and {Clem}, James L. and {Yanny}, Brian and {Rockosi}, Constance M. and {Morrison}, Heather L. and {Harding}, Paul and {Gunn}, James E. and {Allende Prieto}, Carlos and {Beers}, Timothy C. and {Cudworth}, Kyle M. and {Ivans}, Inese I. and {Ivezi{\'c}}, {\v{Z}}eljko and {Lee}, Young Sun and {Lupton}, Robert H. and {Bizyaev}, Dmitry and {Brewington}, Howard and {Malanushenko}, Elena and {Malanushenko}, Viktor and {Oravetz}, Dan and {Pan}, Kaike and {Simmons}, Audrey and {Snedden}, Stephanie and {Watters}, Shannon and {York}, Donald G.},
        title = "{Galactic Globular and Open Clusters in the Sloan Digital Sky Survey. I. Crowded-Field Photometry and Cluster Fiducial Sequences in ugriz}",
      journal = {\apjs},
         year = 2008,
        month = dec,
       volume = {179},
       number = {2},
        pages = {326-354},
          doi = {10.1086/592090},
archivePrefix = {arXiv},
       eprint = {0808.0001},
 primaryClass = {astro-ph},
       adsurl = {https://ui.adsabs.harvard.edu/abs/2008ApJS..179..326A}
}

@ARTICLE{an2009,
       author = {{An}, Deokkeun and {Pinsonneault}, Marc H. and {Masseron}, Thomas and {Delahaye}, Franck and {Johnson}, Jennifer A. and {Terndrup}, Donald M. and {Beers}, Timothy C. and {Ivans}, Inese I. and {Ivezi{\'c}}, {\v{Z}}eljko},
        title = "{Galactic Globular and Open Clusters in the Sloan Digital Sky Survey. II. Test of Theoretical Stellar Isochrones}",
      journal = {\apj},
         year = 2009,
        month = jul,
       volume = {700},
       number = {1},
        pages = {523-544},
          doi = {10.1088/0004-637X/700/1/523},
archivePrefix = {arXiv},
       eprint = {0905.3743},
 primaryClass = {astro-ph.SR},
       adsurl = {https://ui.adsabs.harvard.edu/abs/2009ApJ...700..523A}
}

@ARTICLE{anderson2008,
       author = {{Anderson}, Jay and {Sarajedini}, Ata and {Bedin}, Luigi R. and {King}, Ivan R. and {Piotto}, Giampaolo and {Reid}, I. Neill and {Siegel}, Michael and {Majewski}, Steven R. and {Paust}, Nathaniel E.~Q. and {Aparicio}, Antonio and {Milone}, Antonino P. and {Chaboyer}, Brian and {Rosenberg}, Alfred},
        title = "{The Acs Survey of Globular Clusters. V. Generating a Comprehensive Star Catalog for each Cluster}",
      journal = {\aj},
         year = 2008,
        month = jun,
       volume = {135},
       number = {6},
        pages = {2055-2073},
          doi = {10.1088/0004-6256/135/6/2055},
archivePrefix = {arXiv},
       eprint = {0804.2025},
 primaryClass = {astro-ph},
       adsurl = {https://ui.adsabs.harvard.edu/abs/2008AJ....135.2055A}
}

@ARTICLE{arellano2022,
       author = {{Arellano Ferro}, A.},
        title = "{A Vindication of the RR Lyrae Fourier Light Curve Decomposition for the Calculation of Metallicity and Distance in Globular Clusters}",
      journal = {Revista Mexicana de Astronom\'ia y Astrof\'isica},
         year = 2022,
        month = oct,
       volume = {58},
        pages = {257-271},
          doi = {10.22201/ia.01851101p.2022.58.02.08},
archivePrefix = {arXiv},
       eprint = {2204.01953},
 primaryClass = {astro-ph.SR},
       adsurl = {https://ui.adsabs.harvard.edu/abs/2022RMxAA..58..257A}
}

@ARTICLE{arellano2024,
       author = {{Arellano Ferro}, A.},
        title = "{Globular cluster metallicities and distances from disentangling their RR Lyrae light curves}",
      journal = {IAU Symposium},
         year = 2024,
        month = jan,
       volume = {376},
        pages = {222-238},
          doi = {10.1017/S1743921323002880},
archivePrefix = {arXiv},
       eprint = {2306.01175},
 primaryClass = {astro-ph.SR},
       adsurl = {https://ui.adsabs.harvard.edu/abs/2024IAUS..376..222A}
}

@ARTICLE{bajkova2022,
       author = {{Bajkova}, A.~T. and {Bobylev}, V.~V.},
        title = "{A New Catalog of orbits of 152 Globular Clusters from Gaia EDR3}",
      journal = {arXiv e-prints},
         year = 2022,
        month = nov,
          eid = {arXiv:2212.00739},
        pages = {arXiv:2212.00739},
          doi = {10.48550/arXiv.2212.00739},
archivePrefix = {arXiv},
       eprint = {2212.00739},
 primaryClass = {astro-ph.GA},
       adsurl = {https://ui.adsabs.harvard.edu/abs/2022arXiv221200739B}
}

@ARTICLE{baumgardt2021,
       author = {{Baumgardt}, H. and {Vasiliev}, E.},
        title = "{Accurate distances to Galactic globular clusters through a combination of Gaia EDR3, HST, and literature data}",
      journal = {\mnras},
         year = 2021,
        month = aug,
       volume = {505},
       number = {4},
        pages = {5957-5977},
          doi = {10.1093/mnras/stab1474},
archivePrefix = {arXiv},
       eprint = {2105.09526},
 primaryClass = {astro-ph.GA},
       adsurl = {https://ui.adsabs.harvard.edu/abs/2021MNRAS.505.5957B}
}

@ARTICLE{bellazzini2001,
       author = {{Bellazzini}, Michele and {Pecci}, Flavio Fusi and {Ferraro}, Francesco R. and {Galleti}, Silvia and {Catelan}, M{\'a}rcio and {Landsman}, Wayne B.},
        title = "{Age as the Second Parameter in NGC 288/NGC 362? I. Turnoff Ages: A Purely Differential Comparison}",
      journal = {\aj},
         year = 2001,
        month = nov,
       volume = {122},
       number = {5},
        pages = {2569-2586},
          doi = {10.1086/323711},
archivePrefix = {arXiv},
       eprint = {astro-ph/0109028},
 primaryClass = {astro-ph},
       adsurl = {https://ui.adsabs.harvard.edu/abs/2001AJ....122.2569B}
}

@ARTICLE{bica2019,
       author = {{Bica}, Eduardo and {Pavani}, Daniela B. and {Bonatto}, Charles J. and {Lima}, Eliade F.},
        title = "{A Multi-band Catalog of 10978 Star Clusters, Associations, and Candidates in the Milky Way}",
      journal = {\aj},
         year = 2019,
        month = jan,
       volume = {157},
       number = {1},
          eid = {12},
        pages = {12},
          doi = {10.3847/1538-3881/aaef8d},
archivePrefix = {arXiv},
       eprint = {1812.10292},
 primaryClass = {astro-ph.GA},
       adsurl = {https://ui.adsabs.harvard.edu/abs/2019AJ....157...12B}
}

@ARTICLE{bolte1992,
       author = {{Bolte}, Michael},
        title = "{CCD Photometry in the Globular Cluster NGC 288. I. Blue Stragglers and Main-Sequence Binary Stars}",
      journal = {\apjs},
         year = 1992,
        month = sep,
       volume = {82},
        pages = {145},
          doi = {10.1086/191712},
       adsurl = {https://ui.adsabs.harvard.edu/abs/1992ApJS...82..145B}
}

@ARTICLE{bonatto2013,
       author = {{Bonatto}, C. and {Campos}, Fab{\'\i}ola and {Kepler}, S.~O.},
        title = "{Mapping the differential reddening in globular clusters}",
      journal = {\mnras},
         year = 2013,
        month = oct,
       volume = {435},
       number = {1},
        pages = {263-272},
          doi = {10.1093/mnras/stt1304},
archivePrefix = {arXiv},
       eprint = {1307.3935},
 primaryClass = {astro-ph.GA},
       adsurl = {https://ui.adsabs.harvard.edu/abs/2013MNRAS.435..263B}
}

@ARTICLE{brasseur2010,
       author = {{Brasseur}, Crystal M. and {Stetson}, Peter B. and {VandenBerg}, Don A. and {Casagrande}, Luca and {Bono}, Giuseppe and {Dall'Ora}, Massimo},
        title = "{Fiducial Stellar Population Sequences for the VJK$_{S}$ Photometric System}",
      journal = {\aj},
         year = 2010,
        month = dec,
       volume = {140},
       number = {6},
        pages = {1672-1686},
          doi = {10.1088/0004-6256/140/6/1672},
archivePrefix = {arXiv},
       eprint = {1010.0247},
 primaryClass = {astro-ph.SR},
       adsurl = {https://ui.adsabs.harvard.edu/abs/2010AJ....140.1672B}
}

@ARTICLE{filtergraph,
       author = {{Burger}, Dan and {Stassun}, Keivan G. and {Pepper}, Joshua and {Siverd}, Robert J. and {Paegert}, Martin and {De Lee}, Nathan M. and {Robinson}, William H.},
        title = "{Filtergraph: An interactive web application for visualization of astronomy datasets}",
      journal = {Astronomy and Computing},
         year = 2013,
        month = aug,
       volume = {2},
        pages = {40-45},
          doi = {10.1016/j.ascom.2013.06.002},
archivePrefix = {arXiv},
       eprint = {1307.4000},
 primaryClass = {astro-ph.IM},
       adsurl = {https://ui.adsabs.harvard.edu/abs/2013A&C.....2...40B}
}

@ARTICLE{ccm89,
       author = {{Cardelli}, Jason A. and {Clayton}, Geoffrey C. and {Mathis}, John S.},
        title = "{The Relationship between Infrared, Optical, and Ultraviolet Extinction}",
      journal = {\apj},
         year = 1989,
        month = oct,
       volume = {345},
        pages = {245},
          doi = {10.1086/167900},
       adsurl = {https://ui.adsabs.harvard.edu/abs/1989ApJ...345..245C}
}

@ARTICLE{carretta2009,
       author = {{Carretta}, E. and {Bragaglia}, A. and {Gratton}, R. and {D'Orazi}, V. and {Lucatello}, S.},
        title = "{Intrinsic iron spread and a new metallicity scale for globular clusters}",
      journal = {\aap},
         year = 2009,
        month = dec,
       volume = {508},
       number = {2},
        pages = {695-706},
          doi = {10.1051/0004-6361/200913003},
archivePrefix = {arXiv},
       eprint = {0910.0675},
 primaryClass = {astro-ph.GA},
       adsurl = {https://ui.adsabs.harvard.edu/abs/2009A&A...508..695C}
}

@ARTICLE{carretta2010,
       author = {{Carretta}, E. and {Bragaglia}, A. and {Gratton}, R.~G. and {Recio-Blanco}, A. and {Lucatello}, S. and {D'Orazi}, V. and {Cassisi}, S.},
        title = "{Properties of stellar generations in globular clusters and relations with global parameters}",
      journal = {\aap},
         year = 2010,
        month = jun,
       volume = {516},
          eid = {A55},
        pages = {A55},
          doi = {10.1051/0004-6361/200913451},
archivePrefix = {arXiv},
       eprint = {1003.1723},
 primaryClass = {astro-ph.GA},
       adsurl = {https://ui.adsabs.harvard.edu/abs/2010A&A...516A..55C}
}

@ARTICLE{casagrande2014,
       author = {{Casagrande}, L. and {VandenBerg}, Don A.},
        title = "{Synthetic stellar photometry - I. General considerations and new transformations for broad-band systems}",
      journal = {\mnras},
         year = 2014,
        month = oct,
       volume = {444},
       number = {1},
        pages = {392-419},
          doi = {10.1093/mnras/stu1476},
archivePrefix = {arXiv},
       eprint = {1407.6095},
 primaryClass = {astro-ph.SR},
       adsurl = {https://ui.adsabs.harvard.edu/abs/2014MNRAS.444..392C}
}

@ARTICLE{casagrande2018a,
       author = {{Casagrande}, L. and {VandenBerg}, Don A.},
        title = "{Synthetic Stellar Photometry - II. Testing the bolometric flux scale and tables of bolometric corrections for the Hipparcos/Tycho, Pan-STARRS1, SkyMapper, and JWST systems}",
      journal = {\mnras},
         year = 2018,
        month = apr,
       volume = {475},
       number = {4},
        pages = {5023-5040},
          doi = {10.1093/mnras/sty149},
archivePrefix = {arXiv},
       eprint = {1801.05508},
 primaryClass = {astro-ph.SR},
       adsurl = {https://ui.adsabs.harvard.edu/abs/2018MNRAS.475.5023C}
}

@ARTICLE{casagrande2018b,
       author = {{Casagrande}, L. and {VandenBerg}, Don A.},
        title = "{On the use of Gaia magnitudes and new tables of bolometric corrections}",
      journal = {\mnras},
         year = 2018,
        month = sep,
       volume = {479},
       number = {1},
        pages = {L102-L107},
          doi = {10.1093/mnrasl/sly104},
archivePrefix = {arXiv},
       eprint = {1806.01953},
 primaryClass = {astro-ph.SR},
       adsurl = {https://ui.adsabs.harvard.edu/abs/2018MNRAS.479L.102C}
}

@ARTICLE{catelan2009,
       author = {{Catelan}, M.},
        title = "{Horizontal branch stars: the interplay between observations and theory, and insights into the formation of the Galaxy}",
      journal = {\apss},
         year = 2009,
        month = apr,
       volume = {320},
        pages = {261-309},
          doi = {10.1007/s10509-009-9987-8},
archivePrefix = {arXiv},
       eprint = {astro-ph/0507464},
 primaryClass = {astro-ph},
       adsurl = {https://ui.adsabs.harvard.edu/abs/2009Ap&SS.320..261C}
}

@ARTICLE{ceccarelli2025,
       author = {{Ceccarelli}, E. and {Massari}, D. and {Aguado-Agelet}, F. and {Mucciarelli}, A. and {Cassisi}, S. and {Monelli}, M. and {Pancino}, E. and {Salaris}, M. and {Saracino}, S.},
        title = "{Cluster Ages to Reconstruct the Milky Way Assembly (CARMA). III. NGC 288 as the first Splashed globular cluster}",
      journal = {arXiv e-prints},
         year = 2025,
        month = mar,
          eid = {arXiv:2503.02939},
        pages = {arXiv:2503.02939},
          doi = {10.48550/arXiv.2503.02939},
archivePrefix = {arXiv},
       eprint = {2503.02939},
 primaryClass = {astro-ph.GA},
       adsurl = {https://ui.adsabs.harvard.edu/abs/2025arXiv250302939C}
}

@ARTICLE{chambers2016,
       author = {{Chambers}, K.~C. and {Magnier}, E.~A. and {Metcalfe}, N. and {Flewelling}, H.~A. and {Huber}, M.~E. and {Waters}, C.~Z. and {Denneau}, L. and {Draper}, P.~W. and {Farrow}, D. and {Finkbeiner}, D.~P. and {Holmberg}, C. and {Koppenhoefer}, J. and {Price}, P.~A. and {Rest}, A. and {Saglia}, R.~P. and {Schlafly}, E.~F. and {Smartt}, S.~J. and {Sweeney}, W. and {Wainscoat}, R.~J. and {Burgett}, W.~S. and {Chastel}, S. and {Grav}, T. and {Heasley}, J.~N. and {Hodapp}, K.~W. and {Jedicke}, R. and {Kaiser}, N. and {Kudritzki}, R. -P. and {Luppino}, G.~A. and {Lupton}, R.~H. and {Monet}, D.~G. and {Morgan}, J.~S. and {Onaka}, P.~M. and {Shiao}, B. and {Stubbs}, C.~W. and {Tonry}, J.~L. and {White}, R. and {Ba{\~n}ados}, E. and {Bell}, E.~F. and {Bender}, R. and {Bernard}, E.~J. and {Boegner}, M. and {Boffi}, F. and {Botticella}, M.~T. and {Calamida}, A. and {Casertano}, S. and {Chen}, W. -P. and {Chen}, X. and {Cole}, S. and {Deacon}, N. and {Frenk}, C. and {Fitzsimmons}, A. and {Gezari}, S. and {Gibbs}, V. and {Goessl}, C. and {Goggia}, T. and {Gourgue}, R. and {Goldman}, B. and {Grant}, P. and {Grebel}, E.~K. and {Hambly}, N.~C. and {Hasinger}, G. and {Heavens}, A.~F. and {Heckman}, T.~M. and {Henderson}, R. and {Henning}, T. and {Holman}, M. and {Hopp}, U. and {Ip}, W. -H. and {Isani}, S. and {Jackson}, M. and {Keyes}, C.~D. and {Koekemoer}, A.~M. and {Kotak}, R. and {Le}, D. and {Liska}, D. and {Long}, K.~S. and {Lucey}, J.~R. and {Liu}, M. and {Martin}, N.~F. and {Masci}, G. and {McLean}, B. and {Mindel}, E. and {Misra}, P. and {Morganson}, E. and {Murphy}, D.~N.~A. and {Obaika}, A. and {Narayan}, G. and {Nieto-Santisteban}, M.~A. and {Norberg}, P. and {Peacock}, J.~A. and {Pier}, E.~A. and {Postman}, M. and {Primak}, N. and {Rae}, C. and {Rai}, A. and {Riess}, A. and {Riffeser}, A. and {Rix}, H.~W. and {R{\"o}ser}, S. and {Russel}, R. and {Rutz}, L. and {Schilbach}, E. and {Schultz}, A.~S.~B. and {Scolnic}, D. and {Strolger}, L. and {Szalay}, A. and {Seitz}, S. and {Small}, E. and {Smith}, K.~W. and {Soderblom}, D.~R. and {Taylor}, P. and {Thomson}, R. and {Taylor}, A.~N. and {Thakar}, A.~R. and {Thiel}, J. and {Thilker}, D. and {Unger}, D. and {Urata}, Y. and {Valenti}, J. and {Wagner}, J. and {Walder}, T. and {Walter}, F. and {Watters}, S.~P. and {Werner}, S. and {Wood-Vasey}, W.~M. and {Wyse}, R.},
        title = "{The Pan-STARRS1 Surveys}",
      journal = {arXiv e-prints},
         year = 2016,
        month = dec,
          eid = {arXiv:1612.05560},
        pages = {arXiv:1612.05560},
          doi = {10.48550/arXiv.1612.05560},
archivePrefix = {arXiv},
       eprint = {1612.05560},
 primaryClass = {astro-ph.IM},
       adsurl = {https://ui.adsabs.harvard.edu/abs/2016arXiv161205560C}
}

@ARTICLE{clem2008,
       author = {{Clem}, James L. and {Vanden Berg}, Don A. and {Stetson}, Peter B.},
        title = "{Fiducial Stellar Population Sequences for the u'g'r'i'z' System}",
      journal = {\aj},
         year = 2008,
        month = feb,
       volume = {135},
       number = {2},
        pages = {682-692},
          doi = {10.1088/0004-6256/135/2/682},
archivePrefix = {arXiv},
       eprint = {0711.4045},
 primaryClass = {astro-ph},
       adsurl = {https://ui.adsabs.harvard.edu/abs/2008AJ....135..682C}
}

@INPROCEEDINGS{clement2017,
       author = {{Clement}, Christine},
        title = "{Catalogue of variable stars in Milky Way globular clusters}",
    booktitle = {European Physical Journal Web of Conferences},
         year = 2017,
       series = {European Physical Journal Web of Conferences},
       volume = {152},
        month = sep,
          eid = {01021},
        pages = {01021},
          doi = {10.1051/epjconf/201715201021},
       adsurl = {https://ui.adsabs.harvard.edu/abs/2017EPJWC.15201021C}
}

@ARTICLE{cohen1997,
       author = {{Cohen}, Randi L. and {Guhathakurta}, Puragra and {Yanny}, Brian and {Schneider}, Donald P. and {Bahcall}, John N.},
        title = "{Globular Cluster Photometry with the Hubble Space Telescope.VI.WF/PC-I Observations of the Stellar Populations in the Core of M13 (NGC 6205)}",
      journal = {\aj},
         year = 1997,
        month = feb,
       volume = {113},
        pages = {669-681},
          doi = {10.1086/118285},
archivePrefix = {arXiv},
       eprint = {astro-ph/9611151},
 primaryClass = {astro-ph},
       adsurl = {https://ui.adsabs.harvard.edu/abs/1997AJ....113..669C}
}

@ARTICLE{cohen2015,
       author = {{Cohen}, Roger E. and {Hempel}, Maren and {Mauro}, Francesco and {Geisler}, Douglas and {Alonso-Garcia}, Javier and {Kinemuchi}, Karen},
        title = "{Wide Field Near-infrared Photometry of 12 Galactic Globular Clusters: Observations Versus Models on the Red Giant Branch}",
      journal = {\aj},
         year = 2015,
        month = dec,
       volume = {150},
       number = {6},
          eid = {176},
        pages = {176},
          doi = {10.1088/0004-6256/150/6/176},
archivePrefix = {arXiv},
       eprint = {1509.01470},
 primaryClass = {astro-ph.SR},
       adsurl = {https://ui.adsabs.harvard.edu/abs/2015AJ....150..176C}
}

@ARTICLE{coppola2011,
       author = {{Coppola}, G. and {Dall'Ora}, M. and {Ripepi}, V. and {Marconi}, M. and {Musella}, I. and {Bono}, G. and {Piersimoni}, A.~M. and {Stetson}, P.~B. and {Storm}, J.},
        title = "{Distance to Galactic globulars using the near-infrared magnitudes of RR Lyrae stars - IV. The case of M5 (NGC 5904)}",
      journal = {\mnras},
         year = 2011,
        month = sep,
       volume = {416},
       number = {2},
        pages = {1056-1066},
          doi = {10.1111/j.1365-2966.2011.19102.x},
archivePrefix = {arXiv},
       eprint = {1105.4031},
 primaryClass = {astro-ph.GA},
       adsurl = {https://ui.adsabs.harvard.edu/abs/2011MNRAS.416.1056C}
}

@ARTICLE{dalessandro2013,
       author = {{Dalessandro}, E. and {Salaris}, M. and {Ferraro}, F.~R. and {Mucciarelli}, A. and {Cassisi}, S.},
        title = "{The horizontal branch in the UV colour-magnitude diagrams - II. The case of M3, M13 and M79}",
      journal = {\mnras},
         year = 2013,
        month = mar,
       volume = {430},
       number = {1},
        pages = {459-471},
          doi = {10.1093/mnras/sts644},
archivePrefix = {arXiv},
       eprint = {1212.4419},
 primaryClass = {astro-ph.SR},
       adsurl = {https://ui.adsabs.harvard.edu/abs/2013MNRAS.430..459D}
}

@ARTICLE{davidge1995,
       author = {{Davidge}, T.~J. and {Harris}, W.~E.},
        title = "{Deep Infrared Array Photometry of Globular Clusters. III. M13}",
      journal = {\apj},
         year = 1995,
        month = may,
       volume = {445},
        pages = {211},
          doi = {10.1086/175687},
       adsurl = {https://ui.adsabs.harvard.edu/abs/1995ApJ...445..211D}
}

@ARTICLE{davidge1997,
       author = {{Davidge}, T.~J. and {Harris}, W.~E.},
        title = "{Deep Near-Infrared Array Photometry of Globular Clusters. VI. NGC 288}",
      journal = {\apj},
         year = 1997,
        month = feb,
       volume = {475},
       number = {2},
        pages = {584-593},
          doi = {10.1086/303566},
       adsurl = {https://ui.adsabs.harvard.edu/abs/1997ApJ...475..584D}
}

@ARTICLE{dobashi2005,
       author = {{Dobashi}, Kazuhito and {Uehara}, Hayato and {Kandori}, Ryo and {Sakurai}, Tohko and {Kaiden}, Masahiro and {Umemoto}, Tomofumi and {Sato}, Fumio},
        title = "{Atlas and Catalog of Dark Clouds Based on Digitized Sky Survey I}",
      journal = {\pasj},
         year = 2005,
        month = feb,
       volume = {57},
        pages = {S1-S386},
          doi = {10.1093/pasj/57.sp1.S1},
       adsurl = {https://ui.adsabs.harvard.edu/abs/2005PASJ...57S...1D}
}

@ARTICLE{dondoglio2021,
       author = {{Dondoglio}, E. and {Milone}, A.~P. and {Lagioia}, E.~P. and {Marino}, A.~F. and {Tailo}, M. and {Cordoni}, G. and {Jang}, S. and {Carlos}, M.},
        title = "{Multiple Stellar Populations along the Red Horizontal Branch and Red Clump of Globular Clusters}",
      journal = {\apj},
         year = 2021,
        month = jan,
       volume = {906},
       number = {2},
          eid = {76},
        pages = {76},
          doi = {10.3847/1538-4357/abc882},
archivePrefix = {arXiv},
       eprint = {2011.03283},
 primaryClass = {astro-ph.GA},
       adsurl = {https://ui.adsabs.harvard.edu/abs/2021ApJ...906...76D}
}

@ARTICLE{dotter2007,
       author = {{Dotter}, Aaron and {Chaboyer}, Brian and {Jevremovi{\'c}}, Darko and {Baron}, E. and {Ferguson}, Jason W. and {Sarajedini}, Ata and {Anderson}, Jay},
        title = "{The ACS Survey of Galactic Globular Clusters. II. Stellar Evolution Tracks, Isochrones, Luminosity Functions, and Synthetic Horizontal-Branch Models}",
      journal = {\aj},
         year = 2007,
        month = jul,
       volume = {134},
       number = {1},
        pages = {376-390},
          doi = {10.1086/517915},
archivePrefix = {arXiv},
       eprint = {0706.0847},
 primaryClass = {astro-ph},
       adsurl = {https://ui.adsabs.harvard.edu/abs/2007AJ....134..376D}
}

@ARTICLE{dotter2008,
       author = {{Dotter}, Aaron and {Chaboyer}, Brian and {Jevremovi{\'c}}, Darko and {Kostov}, Veselin and {Baron}, E. and {Ferguson}, Jason W.},
        title = "{The Dartmouth Stellar Evolution Database}",
      journal = {\apjs},
         year = 2008,
        month = sep,
       volume = {178},
       number = {1},
        pages = {89-101},
          doi = {10.1086/589654},
archivePrefix = {arXiv},
       eprint = {0804.4473},
 primaryClass = {astro-ph},
       adsurl = {https://ui.adsabs.harvard.edu/abs/2008ApJS..178...89D}
}

@ARTICLE{dotter2010,
       author = {{Dotter}, Aaron and {Sarajedini}, Ata and {Anderson}, Jay and {Aparicio}, Antonio and {Bedin}, Luigi R. and {Chaboyer}, Brian and {Majewski}, Steven and {Mar{\'\i}n-Franch}, A. and {Milone}, Antonino and {Paust}, Nathaniel and {Piotto}, Giampaolo and {Reid}, I. Neill and {Rosenberg}, Alfred and {Siegel}, Michael},
        title = "{The ACS Survey of Galactic Globular Clusters. IX. Horizontal Branch Morphology and the Second Parameter Phenomenon}",
      journal = {\apj},
         year = 2010,
        month = jan,
       volume = {708},
       number = {1},
        pages = {698-716},
          doi = {10.1088/0004-637X/708/1/698},
archivePrefix = {arXiv},
       eprint = {0911.2469},
 primaryClass = {astro-ph.SR},
       adsurl = {https://ui.adsabs.harvard.edu/abs/2010ApJ...708..698D}
}

@ARTICLE{dotter2011,
       author = {{Dotter}, Aaron and {Sarajedini}, Ata and {Anderson}, Jay},
        title = "{Globular Clusters in the Outer Galactic Halo: New Hubble Space Telescope/Advanced Camera for Surveys Imaging of Six Globular Clusters and the Galactic Globular Cluster Age-metallicity Relation}",
      journal = {\apj},
         year = 2011,
        month = sep,
       volume = {738},
       number = {1},
          eid = {74},
        pages = {74},
          doi = {10.1088/0004-637X/738/1/74},
archivePrefix = {arXiv},
       eprint = {1106.4307},
 primaryClass = {astro-ph.GA},
       adsurl = {https://ui.adsabs.harvard.edu/abs/2011ApJ...738...74D}
}

@ARTICLE{dutra2002,
       author = {{Dutra}, C.~M. and {Bica}, E.},
        title = "{A catalogue of dust clouds in the Galaxy}",
      journal = {\aap},
         year = 2002,
        month = feb,
       volume = {383},
        pages = {631-635},
          doi = {10.1051/0004-6361:20011761},
archivePrefix = {arXiv},
       eprint = {astro-ph/0203256},
 primaryClass = {astro-ph},
       adsurl = {https://ui.adsabs.harvard.edu/abs/2002A&A...383..631D}
}

@ARTICLE{eisenstein2006,
       author = {{Eisenstein}, Daniel J. and {Liebert}, James and {Harris}, Hugh C. and {Kleinman}, S.~J. and {Nitta}, Atsuko and {Silvestri}, Nicole and {Anderson}, Scott A. and {Barentine}, J.~C. and {Brewington}, Howard J. and {Brinkmann}, J. and {Harvanek}, Michael and {Krzesi{\'n}ski}, Jurek and {Neilsen}, Eric H., Jr. and {Long}, Dan and {Schneider}, Donald P. and {Snedden}, Stephanie A.},
        title = "{A Catalog of Spectroscopically Confirmed White Dwarfs from the Sloan Digital Sky Survey Data Release 4}",
      journal = {\apjs},
         year = 2006,
        month = nov,
       volume = {167},
       number = {1},
        pages = {40-58},
          doi = {10.1086/507110},
archivePrefix = {arXiv},
       eprint = {astro-ph/0606700},
 primaryClass = {astro-ph},
       adsurl = {https://ui.adsabs.harvard.edu/abs/2006ApJS..167...40E}
}

@ARTICLE{trilegal,
       author = {{Girardi}, L. and {Groenewegen}, M.~A.~T. and {Hatziminaoglou}, E. and {da Costa}, L.},
        title = "{Star counts in the Galaxy. Simulating from very deep to very shallow photometric surveys with the TRILEGAL code}",
      journal = {\aap},
         year = 2005,
        month = jun,
       volume = {436},
       number = {3},
        pages = {895-915},
          doi = {10.1051/0004-6361:20042352},
archivePrefix = {arXiv},
       eprint = {astro-ph/0504047},
 primaryClass = {astro-ph},
       adsurl = {https://ui.adsabs.harvard.edu/abs/2005A&A...436..895G}
}

@ARTICLE{goldsbury2010,
       author = {{Goldsbury}, Ryan and {Richer}, Harvey B. and {Anderson}, Jay and {Dotter}, Aaron and {Sarajedini}, Ata and {Woodley}, Kristin},
        title = "{The ACS Survey of Galactic Globular Clusters. X. New Determinations of Centers for 65 Clusters}",
      journal = {\aj},
         year = 2010,
        month = dec,
       volume = {140},
       number = {6},
        pages = {1830-1837},
          doi = {10.1088/0004-6256/140/6/1830},
archivePrefix = {arXiv},
       eprint = {1008.2755},
 primaryClass = {astro-ph.GA},
       adsurl = {https://ui.adsabs.harvard.edu/abs/2010AJ....140.1830G}
}

@ARTICLE{sage,
       author = {{Gordon}, K.~D. and {Meixner}, M. and {Meade}, M.~R. and {Whitney}, B. and {Engelbracht}, C. and {Bot}, C. and {Boyer}, M.~L. and {Lawton}, B. and {Sewi{\l}o}, M. and {Babler}, B. and {Bernard}, J. -P. and {Bracker}, S. and {Block}, M. and {Blum}, R. and {Bolatto}, A. and {Bonanos}, A. and {Harris}, J. and {Hora}, J.~L. and {Indebetouw}, R. and {Misselt}, K. and {Reach}, W. and {Shiao}, B. and {Tielens}, X. and {Carlson}, L. and {Churchwell}, E. and {Clayton}, G.~C. and {Chen}, C. -H.~R. and {Cohen}, M. and {Fukui}, Y. and {Gorjian}, V. and {Hony}, S. and {Israel}, F.~P. and {Kawamura}, A. and {Kemper}, F. and {Leroy}, A. and {Li}, A. and {Madden}, S. and {Marble}, A.~R. and {McDonald}, I. and {Mizuno}, A. and {Mizuno}, N. and {Muller}, E. and {Oliveira}, J.~M. and {Olsen}, K. and {Onishi}, T. and {Paladini}, R. and {Paradis}, D. and {Points}, S. and {Robitaille}, T. and {Rubin}, D. and {Sandstrom}, K. and {Sato}, S. and {Shibai}, H. and {Simon}, J.~D. and {Smith}, L.~J. and {Srinivasan}, S. and {Vijh}, U. and {Van Dyk}, S. and {van Loon}, J. Th. and {Zaritsky}, D.},
        title = "{Surveying the Agents of Galaxy Evolution in the Tidally Stripped, Low Metallicity Small Magellanic Cloud (SAGE-SMC). I. Overview}",
      journal = {\aj},
         year = 2011,
        month = oct,
       volume = {142},
       number = {4},
          eid = {102},
        pages = {102},
          doi = {10.1088/0004-6256/142/4/102},
archivePrefix = {arXiv},
       eprint = {1107.4313},
 primaryClass = {astro-ph.CO},
       adsurl = {https://ui.adsabs.harvard.edu/abs/2011AJ....142..102G}
}

@ARTICLE{ngc5904,
       author = {{Gontcharov}, George A. and {Mosenkov}, Aleksandr V. and {Khovritchev}, Maxim Yu},
        title = "{Isochrone fitting of Galactic globular clusters - I. NGC 5904}",
      journal = {\mnras},
         year = 2019,
        month = mar,
       volume = {483},
       number = {4},
        pages = {4949-4967},
          doi = {10.1093/mnras/sty3439},
archivePrefix = {arXiv},
       eprint = {1812.06433},
 primaryClass = {astro-ph.GA},
       adsurl = {https://ui.adsabs.harvard.edu/abs/2019MNRAS.483.4949G}
}

@ARTICLE{ngc6205,
       author = {{Gontcharov}, George A. and {Khovritchev}, Maxim Yu and {Mosenkov}, Aleksandr V.},
        title = "{Isochrone fitting of Galactic globular clusters - II. NGC 6205 (M13)}",
      journal = {\mnras},
         year = 2020,
        month = sep,
       volume = {497},
       number = {3},
        pages = {3674-3693},
          doi = {10.1093/mnras/staa1694},
archivePrefix = {arXiv},
       eprint = {2008.10200},
 primaryClass = {astro-ph.GA},
       adsurl = {https://ui.adsabs.harvard.edu/abs/2020MNRAS.497.3674G}
}

@ARTICLE{ngc288,
       author = {{Gontcharov}, George A. and {Khovritchev}, Maxim Yu and {Mosenkov}, Aleksandr V. and {Il'in}, Vladimir B. and {Marchuk}, Alexander A. and {Savchenko}, Sergey S. and {Smirnov}, Anton A. and {Usachev}, Pavel A. and {Poliakov}, Denis M.},
        title = "{Isochrone fitting of Galactic globular clusters - III. NGC 288, NGC 362, and NGC 6218 (M12)}",
      journal = {\mnras},
         year = 2021,
        month = dec,
       volume = {508},
       number = {2},
        pages = {2688-2705},
          doi = {10.1093/mnras/stab2756},
archivePrefix = {arXiv},
       eprint = {2109.13115},
 primaryClass = {astro-ph.GA},
       adsurl = {https://ui.adsabs.harvard.edu/abs/2021MNRAS.508.2688G}
}

@ARTICLE{ngc6362,
       author = {{Gontcharov}, George A. and {Khovritchev}, Maxim Yu and {Mosenkov}, Aleksandr V. and {Il'in}, Vladimir B. and {Marchuk}, Alexander A. and {Poliakov}, Denis M. and {Ryutina}, Olga S. and {Savchenko}, Sergey S. and {Smirnov}, Anton A. and {Usachev}, Pavel A. and {Lee}, Jae-Woo and {Camacho}, Conner and {Hebdon}, Noah},
        title = "{Isochrone fitting of Galactic globular clusters - IV. NGC 6362 and NGC 6723}",
      journal = {\mnras},
         year = 2023,
        month = jan,
       volume = {518},
       number = {2},
        pages = {3036-3054},
          doi = {10.1093/mnras/stac3300},
archivePrefix = {arXiv},
       eprint = {2211.12684},
 primaryClass = {astro-ph.GA},
       adsurl = {https://ui.adsabs.harvard.edu/abs/2023MNRAS.518.3036G}
}

@ARTICLE{ngc6397,
       author = {{Gontcharov}, George A. and {Bonatto}, Charles J. and {Ryutina}, Olga S. and {Savchenko}, Sergey S. and {Mosenkov}, Aleksandr V. and {Il'in}, Vladimir B. and {Khovritchev}, Maxim Yu and {Marchuk}, Alexander A. and {Poliakov}, Denis M. and {Smirnov}, Anton A. and {Seguine}, Jonah},
        title = "{Isochrone fitting of Galactic globular clusters - V. NGC 6397 and NGC 6809 (M55)}",
      journal = {\mnras},
         year = 2023,
        month = dec,
       volume = {526},
       number = {4},
        pages = {5628-5647},
          doi = {10.1093/mnras/stad3134},
archivePrefix = {arXiv},
       eprint = {2402.06524},
 primaryClass = {astro-ph.GA},
       adsurl = {https://ui.adsabs.harvard.edu/abs/2023MNRAS.526.5628G}
}

@ARTICLE{ngc5024,
       author = {{Gontcharov}, G.~A. and {Savchenko}, S.~S. and {Marchuk}, A.~A. and {Bonatto}, C.~J. and {Ryutina}, O.~S. and {Khovritchev}, M. Yu. and {Il'in}, V.~B. and {Mosenkov}, A.~V. and {Poliakov}, D.~M. and {Smirnov}, A.~A.},
        title = "{Isochrone Fitting of Galactic Globular Clusters{\textemdash}VI. High-latitude Clusters NGC 5024 (M53), NGC 5053, NGC 5272 (M3), NGC 5466, and NGC 7099 (M30)}",
      journal = {Research in Astronomy and Astrophysics},
         year = 2024,
        month = jun,
       volume = {24},
       number = {6},
          eid = {065014},
        pages = {065014},
          doi = {10.1088/1674-4527/ad420f},
archivePrefix = {arXiv},
       eprint = {2404.14797},
 primaryClass = {astro-ph.GA},
       adsurl = {https://ui.adsabs.harvard.edu/abs/2024RAA....24f5014G}
}

@ARTICLE{gms2025,
       author = {{Gontcharov}, G.~A. and {Marchuk}, A.~A. and {Savchenko}, S.~S. and {Mosenkov}, A.~V. and {Il'in}, V.~B. and {Poliakov}, D.~M. and {Smirnov}, A.~A. and {Krayani}, H.},
        title = "{Foreground Extinction to Extended Celestial Objects. I. New Extinction Maps}",
      journal = {Research in Astronomy and Astrophysics},
         year = 2025,
        month = dec,
       volume = {25},
       number = {12},
          eid = {125016},
        pages = {125016},
          doi = {10.1088/1674-4527/ae12a6},
archivePrefix = {arXiv},
       eprint = {2510.02600},
 primaryClass = {astro-ph.GA},
       adsurl = {https://ui.adsabs.harvard.edu/abs/2025RAA....25l5016G}
}

@ARTICLE{gratton2010,
       author = {{Gratton}, R.~G. and {Carretta}, E. and {Bragaglia}, A. and {Lucatello}, S. and {D'Orazi}, V.},
        title = "{The second and third parameters of the horizontal branch in globular clusters}",
      journal = {\aap},
         year = 2010,
        month = jul,
       volume = {517},
          eid = {A81},
        pages = {A81},
          doi = {10.1051/0004-6361/200912572},
archivePrefix = {arXiv},
       eprint = {1004.3862},
 primaryClass = {astro-ph.SR},
       adsurl = {https://ui.adsabs.harvard.edu/abs/2010A&A...517A..81G}
}

@ARTICLE{green2019,
       author = {{Green}, Gregory M. and {Schlafly}, Edward and {Zucker}, Catherine and {Speagle}, Joshua S. and {Finkbeiner}, Douglas},
        title = "{A 3D Dust Map Based on Gaia, Pan-STARRS 1, and 2MASS}",
      journal = {\apj},
         year = 2019,
        month = dec,
       volume = {887},
       number = {1},
          eid = {93},
        pages = {93},
          doi = {10.3847/1538-4357/ab5362},
archivePrefix = {arXiv},
       eprint = {1905.02734},
 primaryClass = {astro-ph.GA},
       adsurl = {https://ui.adsabs.harvard.edu/abs/2019ApJ...887...93G}
}

@ARTICLE{grundahl1999,
       author = {{Grundahl}, F. and {Catelan}, M. and {Landsman}, W.~B. and {Stetson}, P.~B. and {Andersen}, M.~I.},
        title = "{Hot Horizontal-Branch Stars: The Ubiquitous Nature of the ``Jump'' in Str{\"o}mgren u, Low Gravities, and the Role of Radiative Levitation of Metals}",
      journal = {\apj},
         year = 1999,
        month = oct,
       volume = {524},
       number = {1},
        pages = {242-261},
          doi = {10.1086/307807},
archivePrefix = {arXiv},
       eprint = {astro-ph/9903120},
 primaryClass = {astro-ph},
       adsurl = {https://ui.adsabs.harvard.edu/abs/1999ApJ...524..242G}
}

@ARTICLE{hacar2016,
       author = {{Hacar}, A. and {Kainulainen}, J. and {Tafalla}, M. and {Beuther}, H. and {Alves}, J.},
        title = "{The Musca cloud: A 6 pc-long velocity-coherent, sonic filament}",
      journal = {\aap},
         year = 2016,
        month = mar,
       volume = {587},
          eid = {A97},
        pages = {A97},
          doi = {10.1051/0004-6361/201526015},
archivePrefix = {arXiv},
       eprint = {1511.06370},
 primaryClass = {astro-ph.GA},
       adsurl = {https://ui.adsabs.harvard.edu/abs/2016A&A...587A..97H}
}

@ARTICLE{hargis2004,
       author = {{Hargis}, Jonathan R. and {Sandquist}, Eric L. and {Bolte}, Michael},
        title = "{The Luminosity Function and Color-Magnitude Diagram of the Globular Cluster M12}",
      journal = {\apj},
         year = 2004,
        month = jun,
       volume = {608},
       number = {1},
        pages = {243-260},
          doi = {10.1086/386329},
archivePrefix = {arXiv},
       eprint = {astro-ph/0402460},
 primaryClass = {astro-ph},
       adsurl = {https://ui.adsabs.harvard.edu/abs/2004ApJ...608..243H}
}

@ARTICLE{harris,
       author = {{Harris}, William E.},
        title = "{A Catalog of Parameters for Globular Clusters in the Milky Way}",
      journal = {\aj},
         year = 1996,
        month = oct,
       volume = {112},
        pages = {1487},
          doi = {10.1086/118116},
       adsurl = {https://ui.adsabs.harvard.edu/abs/1996AJ....112.1487H}
}

@ARTICLE{ukidss,
       author = {{Hewett}, P.~C. and {Warren}, S.~J. and {Leggett}, S.~K. and {Hodgkin}, S.~T.},
        title = "{The UKIRT Infrared Deep Sky Survey ZY JHK photometric system: passbands and synthetic colours}",
      journal = {\mnras},
         year = 2006,
        month = apr,
       volume = {367},
       number = {2},
        pages = {454-468},
          doi = {10.1111/j.1365-2966.2005.09969.x},
archivePrefix = {arXiv},
       eprint = {astro-ph/0601592},
 primaryClass = {astro-ph},
       adsurl = {https://ui.adsabs.harvard.edu/abs/2006MNRAS.367..454H}
}

@ARTICLE{newbasti,
       author = {{Hidalgo}, Sebastian L. and {Pietrinferni}, Adriano and {Cassisi}, Santi and {Salaris}, Maurizio and {Mucciarelli}, Alessio and {Savino}, Alessandro and {Aparicio}, Antonio and {Silva Aguirre}, Victor and {Verma}, Kuldeep},
        title = "{The Updated BaSTI Stellar Evolution Models and Isochrones. I. Solar-scaled Calculations}",
      journal = {\apj},
         year = 2018,
        month = apr,
       volume = {856},
       number = {2},
          eid = {125},
        pages = {125},
          doi = {10.3847/1538-4357/aab158},
archivePrefix = {arXiv},
       eprint = {1802.07319},
 primaryClass = {astro-ph.GA},
       adsurl = {https://ui.adsabs.harvard.edu/abs/2018ApJ...856..125H}
}

@ARTICLE{jang2022,
       author = {{Jang}, S. and {Milone}, A.~P. and {Legnardi}, M.~V. and {Marino}, A.~F. and {Mastrobuono-Battisti}, A. and {Dondoglio}, E. and {Lagioia}, E.~P. and {Casagrande}, L. and {Carlos}, M. and {Mohandasan}, A. and {Cordoni}, G. and {Bortolan}, E. and {Lee}, Y. -W.},
        title = "{Chromosome maps of globular clusters from wide-field ground-based photometry}",
      journal = {\mnras},
         year = 2022,
        month = dec,
       volume = {517},
       number = {4},
        pages = {5687-5703},
          doi = {10.1093/mnras/stac3086},
archivePrefix = {arXiv},
       eprint = {2211.00650},
 primaryClass = {astro-ph.GA},
       adsurl = {https://ui.adsabs.harvard.edu/abs/2022MNRAS.517.5687J}
}

@ARTICLE{jang2025,
       author = {{Jang}, S. and {Milone}, A.~P. and {Marino}, A.~F. and {Tailo}, M. and {Dondoglio}, E. and {Legnardi}, M.~V. and {Cordoni}, G. and {Ziliotto}, T. and {Lagioia}, E.~P. and {Carlos}, M. and {Mohandasan}, A. and {Bortolan}, E. and {Lee}, Y.-W.},
        title = "{New Perspective on the Multiple-population Phenomenon in Galactic Globular Clusters from a Wide-field Photometric Survey}",
      journal = {\apj},
         year = 2025,
        month = mar,
       volume = {981},
       number = {1},
          eid = {57},
        pages = {57},
          doi = {10.3847/1538-4357/adafa1},
archivePrefix = {arXiv},
       eprint = {2502.02585},
 primaryClass = {astro-ph.GA},
       adsurl = {https://ui.adsabs.harvard.edu/abs/2025ApJ...981...57J}
}

@ARTICLE{johnson2006,
       author = {{Johnson}, Christian I. and {Pilachowski}, Catherine A.},
        title = "{A Moderate Sample Size, Multielement Analysis of the Globular Cluster M12 (NGC 6218)}",
      journal = {\aj},
         year = 2006,
        month = dec,
       volume = {132},
       number = {6},
        pages = {2346-2359},
          doi = {10.1086/508486},
       adsurl = {https://ui.adsabs.harvard.edu/abs/2006AJ....132.2346J}
}

@ARTICLE{jurcsik2023,
       author = {{Jurcsik}, Johanna and {Hajdu}, Gergely},
        title = "{Photometric metallicities of fundamental-mode RR Lyr stars from Gaia G band photometry of globular-cluster variables}",
      journal = {\mnras},
         year = 2023,
        month = nov,
       volume = {525},
       number = {3},
        pages = {3486-3498},
          doi = {10.1093/mnras/stad2510},
archivePrefix = {arXiv},
       eprint = {2308.08929},
 primaryClass = {astro-ph.SR},
       adsurl = {https://ui.adsabs.harvard.edu/abs/2023MNRAS.525.3486J}
}

@ARTICLE{kacharov2014,
       author = {{Kacharov}, N. and {Bianchini}, P. and {Koch}, A. and {Frank}, M.~J. and {Martin}, N.~F. and {van de Ven}, G. and {Puzia}, T.~H. and {McDonald}, I. and {Johnson}, C.~I. and {Zijlstra}, A.~A.},
        title = "{A study of rotating globular clusters. The case of the old, metal-poor globular cluster NGC 4372}",
      journal = {\aap},
         year = 2014,
        month = jul,
       volume = {567},
          eid = {A69},
        pages = {A69},
          doi = {10.1051/0004-6361/201423709},
archivePrefix = {arXiv},
       eprint = {1406.1552},
 primaryClass = {astro-ph.GA},
       adsurl = {https://ui.adsabs.harvard.edu/abs/2014A&A...567A..69K}
}

@ARTICLE{kaluzny2015,
       author = {{Kaluzny}, J. and {Thompson}, I.~B. and {Narloch}, W. and {Pych}, W. and {Rozyczka}, M.},
        title = "{The Clusters AgeS Experiment (CASE). Variable Stars in the Field of the Globular Cluster M12}",
      journal = {\actaa},
         year = 2015,
        month = sep,
       volume = {65},
       number = {3},
        pages = {267-282},
          doi = {10.48550/arXiv.1509.03094},
archivePrefix = {arXiv},
       eprint = {1509.03094},
 primaryClass = {astro-ph.SR},
       adsurl = {https://ui.adsabs.harvard.edu/abs/2015AcA....65..267K}
}

@ARTICLE{kravtsov1997,
       author = {{Kravtsov}, V. and {Ipatov}, A. and {Samus}, N. and {Smirnov}, O. and {Alcaino}, G. and {Liller}, W. and {Alvarado}, F.},
        title = "{NTT CCD photometry of the globular cluster M 79 = NGC 1904 in UBV}",
      journal = {\aaps},
         year = 1997,
        month = oct,
       volume = {125},
        pages = {1-9},
          doi = {10.1051/aas:1997210},
       adsurl = {https://ui.adsabs.harvard.edu/abs/1997A&AS..125....1K}
}

@ARTICLE{lallement2019,
       author = {{Lallement}, R. and {Babusiaux}, C. and {Vergely}, J.~L. and {Katz}, D. and {Arenou}, F. and {Valette}, B. and {Hottier}, C. and {Capitanio}, L.},
        title = "{Gaia-2MASS 3D maps of Galactic interstellar dust within 3 kpc}",
      journal = {\aap},
         year = 2019,
        month = may,
       volume = {625},
          eid = {A135},
        pages = {A135},
          doi = {10.1051/0004-6361/201834695},
archivePrefix = {arXiv},
       eprint = {1902.04116},
 primaryClass = {astro-ph.GA},
       adsurl = {https://ui.adsabs.harvard.edu/abs/2019A&A...625A.135L}
}

@ARTICLE{layden2005,
       author = {{Layden}, Andrew C. and {Sarajedini}, Ata and {von Hippel}, Ted and {Cool}, Adrienne M.},
        title = "{Deep Photometry of the Globular Cluster M5: Distance Estimates from White Dwarf and Main-Sequence Stars}",
      journal = {\apj},
         year = 2005,
        month = oct,
       volume = {632},
       number = {1},
        pages = {266-276},
          doi = {10.1086/444407},
archivePrefix = {arXiv},
       eprint = {astro-ph/0506727},
 primaryClass = {astro-ph},
       adsurl = {https://ui.adsabs.harvard.edu/abs/2005ApJ...632..266L}
}

@ARTICLE{lee1994,
       author = {{Lee}, Young-Wook and {Demarque}, Pierre and {Zinn}, Robert},
        title = "{The Horizontal-Branch Stars in Globular Clusters. II. The Second Parameter Phenomenon}",
      journal = {\apj},
         year = 1994,
        month = mar,
       volume = {423},
        pages = {248},
          doi = {10.1086/173803},
       adsurl = {https://ui.adsabs.harvard.edu/abs/1994ApJ...423..248L}
}

@ARTICLE{lee1999,
       author = {{Lee}, Jae-Woo and {Carney}, Bruce W.},
        title = "{RR Lyrae Luminosity Differences between Oosterhoff Group I and II Cluster Systems and the Origin of the Oosterhoff Dichotomy}",
      journal = {\aj},
         year = 1999,
        month = sep,
       volume = {118},
       number = {3},
        pages = {1373-1389},
          doi = {10.1086/301008},
       adsurl = {https://ui.adsabs.harvard.edu/abs/1999AJ....118.1373L}
}

@ARTICLE{lee2009,
       author = {{Lee}, Jae-Woo and {Kang}, Young-Woon and {Lee}, Jina and {Lee}, Young-Wook},
        title = "{Enrichment by supernovae in globular clusters with multiple populations}",
      journal = {\nat},
         year = 2009,
        month = nov,
       volume = {462},
       number = {7272},
        pages = {480-482},
          doi = {10.1038/nature08565},
archivePrefix = {arXiv},
       eprint = {0911.4798},
 primaryClass = {astro-ph.GA},
       adsurl = {https://ui.adsabs.harvard.edu/abs/2009Natur.462..480L}
}

@ARTICLE{lee2017,
       author = {{Lee}, Jae-Woo},
        title = "{Multiple Stellar Populations of Globular Clusters from Homogeneous Ca-CN Photometry. II. M5 (NGC 5904) and a New Filter System}",
      journal = {\apj},
         year = 2017,
        month = jul,
       volume = {844},
       number = {1},
          eid = {77},
        pages = {77},
          doi = {10.3847/1538-4357/aa7b8c},
archivePrefix = {arXiv},
       eprint = {1706.07969},
 primaryClass = {astro-ph.GA},
       adsurl = {https://ui.adsabs.harvard.edu/abs/2017ApJ...844...77L}
}

@ARTICLE{lee2021,
       author = {{Lee}, Jae-Woo},
        title = "{Formation of Multiple Populations of M5 (NGC 5904)}",
      journal = {\apjl},
         year = 2021,
        month = sep,
       volume = {918},
       number = {2},
          eid = {L24},
        pages = {L24},
          doi = {10.3847/2041-8213/ac1ffe},
archivePrefix = {arXiv},
       eprint = {2108.13756},
 primaryClass = {astro-ph.GA},
       adsurl = {https://ui.adsabs.harvard.edu/abs/2021ApJ...918L..24L}
}

@ARTICLE{lee2024,
       author = {{Lee}, Jae-Woo},
        title = "{A Comparative Study between M30 and M92: M92 is a Merger Remnant with a Large Helium Enhancement}",
      journal = {\apj},
         year = 2024,
        month = feb,
       volume = {961},
       number = {2},
          eid = {227},
        pages = {227},
          doi = {10.3847/1538-4357/ad12ca},
archivePrefix = {arXiv},
       eprint = {2312.02442},
 primaryClass = {astro-ph.GA},
       adsurl = {https://ui.adsabs.harvard.edu/abs/2024ApJ...961..227L}
}

@ARTICLE{libralato2022,
       author = {{Libralato}, Mattia and {Bellini}, Andrea and {Vesperini}, Enrico and {Piotto}, Giampaolo and {Milone}, Antonino P. and {van der Marel}, Roeland P. and {Anderson}, Jay and {Aparicio}, Antonio and {Barbuy}, Beatriz and {Bedin}, Luigi R. and {Borsato}, Luca and {Cassisi}, Santi and {Dalessandro}, Emanuele and {Ferraro}, Francesco R. and {King}, Ivan R. and {Lanzoni}, Barbara and {Nardiello}, Domenico and {Ortolani}, Sergio and {Sarajedini}, Ata and {Sohn}, Sangmo Tony},
        title = "{The Hubble Space Telescope UV Legacy Survey of Galactic Globular Clusters. XXIII. Proper-motion Catalogs and Internal Kinematics}",
      journal = {\apj},
         year = 2022,
        month = aug,
       volume = {934},
       number = {2},
          eid = {150},
        pages = {150},
          doi = {10.3847/1538-4357/ac7727},
archivePrefix = {arXiv},
       eprint = {2206.09924},
 primaryClass = {astro-ph.GA},
       adsurl = {https://ui.adsabs.harvard.edu/abs/2022ApJ...934..150L}
}

@ARTICLE{lindegren2021,
       author = {{Lindegren}, L. and {Bastian}, U. and {Biermann}, M. and {Bombrun}, A. and {de Torres}, A. and {Gerlach}, E. and {Geyer}, R. and {Hern{\'a}ndez}, J. and {Hilger}, T. and {Hobbs}, D. and {Klioner}, S.~A. and {Lammers}, U. and {McMillan}, P.~J. and {Ramos-Lerate}, M. and {Steidelm{\"u}ller}, H. and {Stephenson}, C.~A. and {van Leeuwen}, F.},
        title = "{Gaia Early Data Release 3. Parallax bias versus magnitude, colour, and position}",
      journal = {\aap},
         year = 2021,
        month = may,
       volume = {649},
          eid = {A4},
        pages = {A4},
          doi = {10.1051/0004-6361/202039653},
archivePrefix = {arXiv},
       eprint = {2012.01742},
 primaryClass = {astro-ph.IM},
       adsurl = {https://ui.adsabs.harvard.edu/abs/2021A&A...649A...4L}
}

@ARTICLE{mackey2019,
       author = {{Mackey}, Dougal and {Lewis}, Geraint F. and {Brewer}, Brendon J. and {Ferguson}, Annette M.~N. and {Veljanoski}, Jovan and {Huxor}, Avon P. and {Collins}, Michelle L.~M. and {C{\^o}t{\'e}}, Patrick and {Ibata}, Rodrigo A. and {Irwin}, Mike J. and {Martin}, Nicolas and {McConnachie}, Alan W. and {Pe{\~n}arrubia}, Jorge and {Tanvir}, Nial and {Wan}, Zhen},
        title = "{Two major accretion epochs in M31 from two distinct populations of globular clusters}",
      journal = {\nat},
         year = 2019,
        month = oct,
       volume = {574},
       number = {7776},
        pages = {69-71},
          doi = {10.1038/s41586-019-1597-1},
archivePrefix = {arXiv},
       eprint = {1910.00808},
 primaryClass = {astro-ph.GA},
       adsurl = {https://ui.adsabs.harvard.edu/abs/2019Natur.574...69M}
}

@ARTICLE{marin2009,
       author = {{Mar{\'\i}n-Franch}, Antonio and {Aparicio}, Antonio and {Piotto}, Giampaolo and {Rosenberg}, Alfred and {Chaboyer}, Brian and {Sarajedini}, Ata and {Siegel}, Michael and {Anderson}, Jay and {Bedin}, Luigi R. and {Dotter}, Aaron and {Hempel}, Maren and {King}, Ivan and {Majewski}, Steven and {Milone}, Antonino P. and {Paust}, Nathaniel and {Reid}, I. Neill},
        title = "{The ACS Survey of Galactic Globular Clusters. VII. Relative Ages}",
      journal = {\apj},
         year = 2009,
        month = apr,
       volume = {694},
       number = {2},
        pages = {1498-1516},
          doi = {10.1088/0004-637X/694/2/1498},
archivePrefix = {arXiv},
       eprint = {0812.4541},
 primaryClass = {astro-ph},
       adsurl = {https://ui.adsabs.harvard.edu/abs/2009ApJ...694.1498M}
}

@ARTICLE{massari2023,
       author = {{Massari}, Davide and {Aguado-Agelet}, Fernando and {Monelli}, Matteo and {Cassisi}, Santi and {Pancino}, Elena and {Saracino}, Sara and {Gallart}, Carme and {Ruiz-Lara}, Tom{\'a}s and {Fern{\'a}ndez-Alvar}, Emma and {Surot}, Francisco and {Stokholm}, Amalie and {Salaris}, Maurizio and {Miglio}, Andrea and {Ceccarelli}, Edoardo},
        title = "{Cluster Ages to Reconstruct the Milky Way Assembly (CARMA). I. The final word on the origin of NGC 6388 and NGC 6441}",
      journal = {\aap},
         year = 2023,
        month = dec,
       volume = {680},
          eid = {A20},
        pages = {A20},
          doi = {10.1051/0004-6361/202347289},
archivePrefix = {arXiv},
       eprint = {2310.01495},
 primaryClass = {astro-ph.GA},
       adsurl = {https://ui.adsabs.harvard.edu/abs/2023A&A...680A..20M}
}

@ARTICLE{masseron2019,
       author = {{Masseron}, T. and {Garc{\'\i}a-Hern{\'a}ndez}, D.~A. and {M{\'e}sz{\'a}ros}, Sz. and {Zamora}, O. and {Dell'Agli}, F. and {Allende Prieto}, C. and {Edvardsson}, B. and {Shetrone}, M. and {Plez}, B. and {Fern{\'a}ndez-Trincado}, J.~G. and {Cunha}, K. and {J{\"o}nsson}, H. and {Geisler}, D. and {Beers}, T.~C. and {Cohen}, R.~E.},
        title = "{Homogeneous analysis of globular clusters from the APOGEE survey with the BACCHUS code. I. The northern clusters}",
      journal = {\aap},
         year = 2019,
        month = feb,
       volume = {622},
          eid = {A191},
        pages = {A191},
          doi = {10.1051/0004-6361/201834550},
archivePrefix = {arXiv},
       eprint = {1812.08817},
 primaryClass = {astro-ph.SR},
       adsurl = {https://ui.adsabs.harvard.edu/abs/2019A&A...622A.191M}
}

@ARTICLE{vista,
       author = {{McMahon}, R.~G. and {Banerji}, M. and {Gonzalez}, E. and {Koposov}, S.~E. and {Bejar}, V.~J. and {Lodieu}, N. and {Rebolo}, R. and {VHS Collaboration}},
        title = "{First Scientific Results from the VISTA Hemisphere Survey (VHS)}",
      journal = {The Messenger},
         year = 2013,
        month = dec,
       volume = {154},
        pages = {35-37},
       adsurl = {https://ui.adsabs.harvard.edu/abs/2013Msngr.154...35M}
}

@ARTICLE{planck,
       author = {{Meisner}, Aaron M. and {Finkbeiner}, Douglas P.},
        title = "{Modeling Thermal Dust Emission with Two Components: Application to the Planck High Frequency Instrument Maps}",
      journal = {\apj},
         year = 2015,
        month = jan,
       volume = {798},
       number = {2},
          eid = {88},
        pages = {88},
          doi = {10.1088/0004-637X/798/2/88},
archivePrefix = {arXiv},
       eprint = {1410.7523},
 primaryClass = {astro-ph.GA},
       adsurl = {https://ui.adsabs.harvard.edu/abs/2015ApJ...798...88M}
}

@ARTICLE{meszaros2020,
       author = {{M{\'e}sz{\'a}ros}, Szabolcs and {Masseron}, Thomas and {Garc{\'\i}a-Hern{\'a}ndez}, D.~A. and {Allende Prieto}, Carlos and {Beers}, Timothy C. and {Bizyaev}, Dmitry and {Chojnowski}, Drew and {Cohen}, Roger E. and {Cunha}, Katia and {Dell'Agli}, Flavia and {Ebelke}, Garrett and {Fern{\'a}ndez-Trincado}, Jos{\'e} G. and {Frinchaboy}, Peter and {Geisler}, Doug and {Hasselquist}, Sten and {Hearty}, Fred and {Holtzman}, Jon and {Johnson}, Jennifer and {Lane}, Richard R. and {Lacerna}, Ivan and {Longa-Pe{\~n}a}, Penelop{\'e} and {Majewski}, Steven R. and {Martell}, Sarah L. and {Minniti}, Dante and {Nataf}, David and {Nidever}, David L. and {Pan}, Kaike and {Schiavon}, Ricardo P. and {Shetrone}, Matthew and {Smith}, Verne V. and {Sobeck}, Jennifer S. and {Stringfellow}, Guy S. and {Szigeti}, L{\'a}szl{\'o} and {Tang}, Baitian and {Wilson}, John C. and {Zamora}, Olga},
        title = "{Homogeneous analysis of globular clusters from the APOGEE survey with the BACCHUS code - II. The Southern clusters and overview}",
      journal = {\mnras},
         year = 2020,
        month = feb,
       volume = {492},
       number = {2},
        pages = {1641-1670},
          doi = {10.1093/mnras/stz3496},
archivePrefix = {arXiv},
       eprint = {1912.04839},
 primaryClass = {astro-ph.SR},
       adsurl = {https://ui.adsabs.harvard.edu/abs/2020MNRAS.492.1641M}
}

@ARTICLE{meylan1997,
       author = {{Meylan}, G. and {Heggie}, D.~C.},
        title = "{Internal dynamics of globular clusters}",
      journal = {\aapr},
         year = 1997,
        month = jan,
       volume = {8},
        pages = {1-143},
          doi = {10.1007/s001590050008},
archivePrefix = {arXiv},
       eprint = {astro-ph/9610076},
 primaryClass = {astro-ph},
       adsurl = {https://ui.adsabs.harvard.edu/abs/1997A&ARv...8....1M}
}

@ARTICLE{milone2017,
       author = {{Milone}, A.~P. and {Piotto}, G. and {Renzini}, A. and {Marino}, A.~F. and {Bedin}, L.~R. and {Vesperini}, E. and {D'Antona}, F. and {Nardiello}, D. and {Anderson}, J. and {King}, I.~R. and {Yong}, D. and {Bellini}, A. and {Aparicio}, A. and {Barbuy}, B. and {Brown}, T.~M. and {Cassisi}, S. and {Ortolani}, S. and {Salaris}, M. and {Sarajedini}, A. and {van der Marel}, R.~P.},
        title = "{The Hubble Space Telescope UV Legacy Survey of Galactic globular clusters - IX. The Atlas of multiple stellar populations}",
      journal = {\mnras},
         year = 2017,
        month = jan,
       volume = {464},
       number = {3},
        pages = {3636-3656},
          doi = {10.1093/mnras/stw2531},
archivePrefix = {arXiv},
       eprint = {1610.00451},
 primaryClass = {astro-ph.SR},
       adsurl = {https://ui.adsabs.harvard.edu/abs/2017MNRAS.464.3636M}
}

@ARTICLE{milone2018,
       author = {{Milone}, A.~P. and {Marino}, A.~F. and {Renzini}, A. and {D'Antona}, F. and {Anderson}, J. and {Barbuy}, B. and {Bedin}, L.~R. and {Bellini}, A. and {Brown}, T.~M. and {Cassisi}, S. and {Cordoni}, G. and {Lagioia}, E.~P. and {Nardiello}, D. and {Ortolani}, S. and {Piotto}, G. and {Sarajedini}, A. and {Tailo}, M. and {van der Marel}, R.~P. and {Vesperini}, E.},
        title = "{The Hubble Space Telescope UV legacy survey of galactic globular clusters - XVI. The helium abundance of multiple populations}",
      journal = {\mnras},
         year = 2018,
        month = dec,
       volume = {481},
       number = {4},
        pages = {5098-5122},
          doi = {10.1093/mnras/sty2573},
archivePrefix = {arXiv},
       eprint = {1809.05006},
 primaryClass = {astro-ph.SR},
       adsurl = {https://ui.adsabs.harvard.edu/abs/2018MNRAS.481.5098M}
}

@ARTICLE{monty2023,
       author = {{Monty}, Stephanie and {Yong}, David and {Massari}, Davide and {McKenzie}, Madeleine and {Myeong}, GyuChul and {Buder}, Sven and {Karakas}, Amanda I. and {Freeman}, Ken C. and {Marino}, Anna F. and {Belokurov}, Vasily and {Evans}, N. Wyn},
        title = "{Peeking beneath the precision floor - II. Probing the chemo-dynamical histories of the potential globular cluster siblings, NGC 288 and NGC 362}",
      journal = {\mnras},
         year = 2023,
        month = jul,
       volume = {522},
       number = {3},
        pages = {4404-4420},
          doi = {10.1093/mnras/stad1154},
archivePrefix = {arXiv},
       eprint = {2302.06644},
 primaryClass = {astro-ph.GA},
       adsurl = {https://ui.adsabs.harvard.edu/abs/2023MNRAS.522.4404M}
}

@ARTICLE{mucciarelli2020,
       author = {{Mucciarelli}, A. and {Bonifacio}, P.},
        title = "{Facing problems in the determination of stellar temperatures and gravities: Galactic globular clusters}",
      journal = {\aap},
         year = 2020,
        month = aug,
       volume = {640},
          eid = {A87},
        pages = {A87},
          doi = {10.1051/0004-6361/202037703},
archivePrefix = {arXiv},
       eprint = {2003.07390},
 primaryClass = {astro-ph.SR},
       adsurl = {https://ui.adsabs.harvard.edu/abs/2020A&A...640A..87M}
}

@ARTICLE{nardiello2018,
       author = {{Nardiello}, D. and {Libralato}, M. and {Piotto}, G. and {Anderson}, J. and {Bellini}, A. and {Aparicio}, A. and {Bedin}, L.~R. and {Cassisi}, S. and {Granata}, V. and {King}, I.~R. and {Lucertini}, F. and {Marino}, A.~F. and {Milone}, A.~P. and {Ortolani}, S. and {Platais}, I. and {van der Marel}, R.~P.},
        title = "{The Hubble Space Telescope UV Legacy Survey of Galactic Globular Clusters - XVII. Public Catalogue Release}",
      journal = {\mnras},
         year = 2018,
        month = dec,
       volume = {481},
       number = {3},
        pages = {3382-3393},
          doi = {10.1093/mnras/sty2515},
archivePrefix = {arXiv},
       eprint = {1809.04300},
 primaryClass = {astro-ph.SR},
       adsurl = {https://ui.adsabs.harvard.edu/abs/2018MNRAS.481.3382N}
}

@ARTICLE{narloch2017,
       author = {{Narloch}, W. and {Kaluzny}, J. and {Poleski}, R. and {Rozyczka}, M. and {Pych}, W. and {Thompson}, I.~B.},
        title = "{A ground-based proper motion study of 12 nearby globular clusters}",
      journal = {\mnras},
         year = 2017,
        month = oct,
       volume = {471},
       number = {2},
        pages = {1446-1467},
          doi = {10.1093/mnras/stx1637},
archivePrefix = {arXiv},
       eprint = {1706.09612},
 primaryClass = {astro-ph.SR},
       adsurl = {https://ui.adsabs.harvard.edu/abs/2017MNRAS.471.1446N}
}

@ARTICLE{vizier,
       author = {{Ochsenbein}, F. and {Bauer}, P. and {Marcout}, J.},
        title = "{The VizieR database of astronomical catalogues}",
      journal = {\aaps},
         year = 2000,
        month = apr,
       volume = {143},
        pages = {23-32},
          doi = {10.1051/aas:2000169},
archivePrefix = {arXiv},
       eprint = {astro-ph/0002122},
 primaryClass = {astro-ph},
       adsurl = {https://ui.adsabs.harvard.edu/abs/2000A&AS..143...23O}
}

@INPROCEEDINGS{odenkirchen2004,
       author = {{Odenkirchen}, M. and {Grebel}, E.~K.},
        title = "{The Tidal Perturbation of the Low Mass Globular Cluster NGC 5466}",
    booktitle = {Satellites and Tidal Streams},
         year = 2004,
       editor = {{Prada}, F. and {Martinez Delgado}, D. and {Mahoney}, T.~J.},
       series = {Astronomical Society of the Pacific Conference Series},
       volume = {327},
        month = dec,
        pages = {284},
          doi = {10.48550/arXiv.astro-ph/0307481},
archivePrefix = {arXiv},
       eprint = {astro-ph/0307481},
 primaryClass = {astro-ph},
       adsurl = {https://ui.adsabs.harvard.edu/abs/2004ASPC..327..284O}
}

@ARTICLE{onken2024,
       author = {{Onken}, Christopher A. and {Wolf}, Christian and {Bessell}, Michael S. and {Chang}, Seo-Won and {Luvaul}, Lance C. and {Tonry}, John L. and {White}, Marc C. and {Da Costa}, Gary S.},
        title = "{SkyMapper Southern Survey: Data release 4}",
      journal = {\pasa},
         year = 2024,
        month = oct,
       volume = {41},
          eid = {e061},
        pages = {e061},
          doi = {10.1017/pasa.2024.53},
archivePrefix = {arXiv},
       eprint = {2402.02015},
 primaryClass = {astro-ph.CO},
       adsurl = {https://ui.adsabs.harvard.edu/abs/2024PASA...41...61O}
}

@ARTICLE{paltrinieri1998,
       author = {{Paltrinieri}, Barbara and {Ferraro}, Francesco R. and {Carretta}, Eugenio and {Fusi Pecci}, Flavio},
        title = "{CCD photometry of the Galactic globular cluster M13}",
      journal = {\mnras},
         year = 1998,
        month = feb,
       volume = {293},
       number = {4},
        pages = {434-442},
          doi = {10.1046/j.1365-8711.1998.01175.x},
       adsurl = {https://ui.adsabs.harvard.edu/abs/1998MNRAS.293..434P}
}

@ARTICLE{pancino2024,
       author = {{Pancino}, E. and {Zocchi}, A. and {Rainer}, M. and {Monaci}, M. and {Massari}, D. and {Monelli}, M. and {Hunt}, L.~K. and {Monaco}, L. and {Mart{\'\i}nez-V{\'a}zquez}, C.~E. and {Sanna}, N. and {Bianchi}, S. and {Stetson}, P.~B.},
        title = "{Differential reddening in 48 globular clusters: An end to the quest for the intracluster medium}",
      journal = {\aap},
         year = 2024,
        month = jun,
       volume = {686},
          eid = {A283},
        pages = {A283},
          doi = {10.1051/0004-6361/202449462},
archivePrefix = {arXiv},
       eprint = {2404.05548},
 primaryClass = {astro-ph.GA},
       adsurl = {https://ui.adsabs.harvard.edu/abs/2024A&A...686A.283P}
}

@ARTICLE{pietrinferni2021,
       author = {{Pietrinferni}, Adriano and {Hidalgo}, Sebastian and {Cassisi}, Santi and {Salaris}, Maurizio and {Savino}, Alessandro and {Mucciarelli}, Alessio and {Verma}, Kuldeep and {Silva Aguirre}, Victor and {Aparicio}, Antonio and {Ferguson}, Jason W.},
        title = "{Updated BaSTI Stellar Evolution Models and Isochrones. II. {\ensuremath{\alpha}}-enhanced Calculations}",
      journal = {\apj},
         year = 2021,
        month = feb,
       volume = {908},
       number = {1},
          eid = {102},
        pages = {102},
          doi = {10.3847/1538-4357/abd4d5},
archivePrefix = {arXiv},
       eprint = {2012.10085},
 primaryClass = {astro-ph.SR},
       adsurl = {https://ui.adsabs.harvard.edu/abs/2021ApJ...908..102P}
}

@ARTICLE{piotto2002,
       author = {{Piotto}, G. and {King}, I.~R. and {Djorgovski}, S.~G. and {Sosin}, C. and {Zoccali}, M. and {Saviane}, I. and {De Angeli}, F. and {Riello}, M. and {Recio-Blanco}, A. and {Rich}, R.~M. and {Meylan}, G. and {Renzini}, A.},
        title = "{HST color-magnitude diagrams of 74 galactic globular clusters in the HST F439W and F555W bands}",
      journal = {\aap},
         year = 2002,
        month = sep,
       volume = {391},
        pages = {945-965},
          doi = {10.1051/0004-6361:20020820},
archivePrefix = {arXiv},
       eprint = {astro-ph/0207124},
 primaryClass = {astro-ph},
       adsurl = {https://ui.adsabs.harvard.edu/abs/2002A&A...391..945P}
}

@ARTICLE{piotto2015,
       author = {{Piotto}, G. and {Milone}, A.~P. and {Bedin}, L.~R. and {Anderson}, J. and {King}, I.~R. and {Marino}, A.~F. and {Nardiello}, D. and {Aparicio}, A. and {Barbuy}, B. and {Bellini}, A. and {Brown}, T.~M. and {Cassisi}, S. and {Cool}, A.~M. and {Cunial}, A. and {Dalessandro}, E. and {D'Antona}, F. and {Ferraro}, F.~R. and {Hidalgo}, S. and {Lanzoni}, B. and {Monelli}, M. and {Ortolani}, S. and {Renzini}, A. and {Salaris}, M. and {Sarajedini}, A. and {van der Marel}, R.~P. and {Vesperini}, E. and {Zoccali}, M.},
        title = "{The Hubble Space Telescope UV Legacy Survey of Galactic Globular Clusters. I. Overview of the Project and Detection of Multiple Stellar Populations}",
      journal = {\aj},
         year = 2015,
        month = mar,
       volume = {149},
       number = {3},
          eid = {91},
        pages = {91},
          doi = {10.1088/0004-6256/149/3/91},
archivePrefix = {arXiv},
       eprint = {1410.4564},
 primaryClass = {astro-ph.SR},
       adsurl = {https://ui.adsabs.harvard.edu/abs/2015AJ....149...91P}
}

@ARTICLE{reimers,
       author = {{Reimers}, D.},
        title = "{Circumstellar absorption lines and mass loss from red giants.}",
      journal = {Memoires of the Societe Royale des Sciences de Liege},
         year = 1975,
        month = jan,
       volume = {8},
        pages = {369-382},
       adsurl = {https://ui.adsabs.harvard.edu/abs/1975MSRSL...8..369R}
}

@ARTICLE{rey2001,
       author = {{Rey}, Soo-Chang and {Yoon}, Suk-Jin and {Lee}, Young-Wook and {Chaboyer}, Brian and {Sarajedini}, Ata},
        title = "{CCD Photometry of the Classic Second-Parameter Globular Clusters M3 and M13}",
      journal = {\aj},
         year = 2001,
        month = dec,
       volume = {122},
       number = {6},
        pages = {3219-3230},
          doi = {10.1086/324104},
archivePrefix = {arXiv},
       eprint = {astro-ph/0109203},
 primaryClass = {astro-ph},
       adsurl = {https://ui.adsabs.harvard.edu/abs/2001AJ....122.3219R}
}

@ARTICLE{riello2021,
       author = {{Riello}, M. and {De Angeli}, F. and {Evans}, D.~W. and {Montegriffo}, P. and {Carrasco}, J.~M. and {Busso}, G. and {Palaversa}, L. and {Burgess}, P.~W. and {Diener}, C. and {Davidson}, M. and {Rowell}, N. and {Fabricius}, C. and {Jordi}, C. and {Bellazzini}, M. and {Pancino}, E. and {Harrison}, D.~L. and {Cacciari}, C. and {van Leeuwen}, F. and {Hambly}, N.~C. and {Hodgkin}, S.~T. and {Osborne}, P.~J. and {Altavilla}, G. and {Barstow}, M.~A. and {Brown}, A.~G.~A. and {Castellani}, M. and {Cowell}, S. and {De Luise}, F. and {Gilmore}, G. and {Giuffrida}, G. and {Hidalgo}, S. and {Holland}, G. and {Marinoni}, S. and {Pagani}, C. and {Piersimoni}, A.~M. and {Pulone}, L. and {Ragaini}, S. and {Rainer}, M. and {Richards}, P.~J. and {Sanna}, N. and {Walton}, N.~A. and {Weiler}, M. and {Yoldas}, A.},
        title = "{Gaia Early Data Release 3. Photometric content and validation}",
      journal = {\aap},
         year = 2021,
        month = may,
       volume = {649},
          eid = {A3},
        pages = {A3},
          doi = {10.1051/0004-6361/202039587},
archivePrefix = {arXiv},
       eprint = {2012.01916},
 primaryClass = {astro-ph.IM},
       adsurl = {https://ui.adsabs.harvard.edu/abs/2021A&A...649A...3R}
}

@ARTICLE{rosenberg2000a,
       author = {{Rosenberg}, A. and {Piotto}, G. and {Saviane}, I. and {Aparicio}, A.},
        title = "{Photometric catalog of nearby globular clusters. I. A large homogeneous (V,I) color-magnitude diagram data-base}",
      journal = {\aaps},
         year = 2000,
        month = may,
       volume = {144},
        pages = {5-38},
          doi = {10.1051/aas:2000341},
archivePrefix = {arXiv},
       eprint = {astro-ph/0002205},
 primaryClass = {astro-ph},
       adsurl = {https://ui.adsabs.harvard.edu/abs/2000A&AS..144....5R}
}

@ARTICLE{rosenberg2000b,
       author = {{Rosenberg}, A. and {Aparicio}, A. and {Saviane}, I. and {Piotto}, G.},
        title = "{Photometric catalog of nearby globular clusters. II. A large homogeneous (V,I) color-magnitude diagram data-base}",
      journal = {\aaps},
         year = 2000,
        month = sep,
       volume = {145},
        pages = {451-465},
          doi = {10.1051/aas:2000356},
archivePrefix = {arXiv},
       eprint = {astro-ph/0006299},
 primaryClass = {astro-ph},
       adsurl = {https://ui.adsabs.harvard.edu/abs/2000A&AS..145..451R}
}

@ARTICLE{sanroman2015,
       author = {{San Roman}, I. and {Mu{\~n}oz}, C. and {Geisler}, D. and {Villanova}, S. and {Kacharov}, N. and {Koch}, A. and {Carraro}, G. and {Tautvai{\v{s}}iene}, G. and {Vallenari}, A. and {Alfaro}, E.~J. and {Bensby}, T. and {Flaccomio}, E. and {Francois}, P. and {Korn}, A.~J. and {Pancino}, E. and {Recio-Blanco}, A. and {Smiljanic}, R. and {Bergemann}, M. and {Costado}, M.~T. and {Damiani}, F. and {Heiter}, U. and {Hourihane}, A. and {Jofr{\'e}}, P. and {Lardo}, C. and {de Laverny}, P. and {Masseron}, T. and {Morbidelli}, L. and {Sbordone}, L. and {Sousa}, S.~G. and {Worley}, C.~C. and {Zaggia}, S.},
        title = "{The Gaia-ESO Survey: Detailed abundances in the metal-poor globular cluster NGC 4372}",
      journal = {\aap},
         year = 2015,
        month = jul,
       volume = {579},
          eid = {A6},
        pages = {A6},
          doi = {10.1051/0004-6361/201525722},
archivePrefix = {arXiv},
       eprint = {1504.03497},
 primaryClass = {astro-ph.SR},
       adsurl = {https://ui.adsabs.harvard.edu/abs/2015A&A...579A...6S}
}

@ARTICLE{sandquist1977,
       author = {{Sandqvist}, Aa.},
        title = "{More southern dark dust clouds.}",
      journal = {\aap},
         year = 1977,
        month = may,
       volume = {57},
        pages = {467-470},
       adsurl = {https://ui.adsabs.harvard.edu/abs/1977A&A....57..467S}
}

@ARTICLE{sarajedini2007,
       author = {{Sarajedini}, Ata and {Bedin}, Luigi R. and {Chaboyer}, Brian and {Dotter}, Aaron and {Siegel}, Michael and {Anderson}, Jay and {Aparicio}, Antonio and {King}, Ivan and {Majewski}, Steven and {Mar{\'\i}n-Franch}, A. and {Piotto}, Giampaolo and {Reid}, I. Neill and {Rosenberg}, Alfred},
        title = "{The ACS Survey of Galactic Globular Clusters. I. Overview and Clusters without Previous Hubble Space Telescope Photometry}",
      journal = {\aj},
         year = 2007,
        month = apr,
       volume = {133},
       number = {4},
        pages = {1658-1672},
          doi = {10.1086/511979},
archivePrefix = {arXiv},
       eprint = {astro-ph/0612598},
 primaryClass = {astro-ph},
       adsurl = {https://ui.adsabs.harvard.edu/abs/2007AJ....133.1658S}
}

@ARTICLE{savino2018,
       author = {{Savino}, A. and {Massari}, D. and {Bragaglia}, A. and {Dalessandro}, E. and {Tolstoy}, E.},
        title = "{M13 multiple stellar populations seen with the eyes of Str{\"o}mgren photometry}",
      journal = {\mnras},
         year = 2018,
        month = mar,
       volume = {474},
       number = {4},
        pages = {4438-4446},
          doi = {10.1093/mnras/stx3093},
archivePrefix = {arXiv},
       eprint = {1712.01284},
 primaryClass = {astro-ph.SR},
       adsurl = {https://ui.adsabs.harvard.edu/abs/2018MNRAS.474.4438S}
}

@ARTICLE{schlaflyfinkbeiner2011,
       author = {{Schlafly}, Edward F. and {Finkbeiner}, Douglas P.},
        title = "{Measuring Reddening with Sloan Digital Sky Survey Stellar Spectra and Recalibrating SFD}",
      journal = {\apj},
         year = 2011,
        month = aug,
       volume = {737},
       number = {2},
          eid = {103},
        pages = {103},
          doi = {10.1088/0004-637X/737/2/103},
archivePrefix = {arXiv},
       eprint = {1012.4804},
 primaryClass = {astro-ph.GA},
       adsurl = {https://ui.adsabs.harvard.edu/abs/2011ApJ...737..103S}
}

@ARTICLE{unwise,
       author = {{Schlafly}, Edward F. and {Meisner}, Aaron M. and {Green}, Gregory M.},
        title = "{The unWISE Catalog: Two Billion Infrared Sources from Five Years of WISE Imaging}",
      journal = {\apjs},
         year = 2019,
        month = feb,
       volume = {240},
       number = {2},
          eid = {30},
        pages = {30},
          doi = {10.3847/1538-4365/aafbea},
archivePrefix = {arXiv},
       eprint = {1901.03337},
 primaryClass = {astro-ph.IM},
       adsurl = {https://ui.adsabs.harvard.edu/abs/2019ApJS..240...30S}
}

@ARTICLE{sfd98,
       author = {{Schlegel}, David J. and {Finkbeiner}, Douglas P. and {Davis}, Marc},
        title = "{Maps of Dust Infrared Emission for Use in Estimation of Reddening and Cosmic Microwave Background Radiation Foregrounds}",
      journal = {\apj},
         year = 1998,
        month = jun,
       volume = {500},
       number = {2},
        pages = {525-553},
          doi = {10.1086/305772},
archivePrefix = {arXiv},
       eprint = {astro-ph/9710327},
 primaryClass = {astro-ph},
       adsurl = {https://ui.adsabs.harvard.edu/abs/1998ApJ...500..525S}
}

@ARTICLE{searle1978,
       author = {{Searle}, L. and {Zinn}, R.},
        title = "{Composition of halo clusters and the formation of the galactic halo.}",
      journal = {\apj},
         year = 1978,
        month = oct,
       volume = {225},
        pages = {357-379},
          doi = {10.1086/156499},
       adsurl = {https://ui.adsabs.harvard.edu/abs/1978ApJ...225..357S}
}

@ARTICLE{shanks2015,
       author = {{Shanks}, T. and {Metcalfe}, N. and {Chehade}, B. and {Findlay}, J.~R. and {Irwin}, M.~J. and {Gonzalez-Solares}, E. and {Lewis}, J.~R. and {Yoldas}, A. Kupcu and {Mann}, R.~G. and {Read}, M.~A. and {Sutorius}, E.~T.~W. and {Voutsinas}, S.},
        title = "{The VLT Survey Telescope ATLAS}",
      journal = {\mnras},
         year = 2015,
        month = aug,
       volume = {451},
       number = {4},
        pages = {4238-4252},
          doi = {10.1093/mnras/stv1130},
archivePrefix = {arXiv},
       eprint = {1502.05432},
 primaryClass = {astro-ph.GA},
       adsurl = {https://ui.adsabs.harvard.edu/abs/2015MNRAS.451.4238S}
}

@ARTICLE{simioni2018,
       author = {{Simioni}, M. and {Bedin}, L.~R. and {Aparicio}, A. and {Piotto}, G. and {Milone}, A.~P. and {Nardiello}, D. and {Anderson}, J. and {Bellini}, A. and {Brown}, T.~M. and {Cassisi}, S. and {Cunial}, A. and {Granata}, V. and {Ortolani}, S. and {van der Marel}, R.~P. and {Vesperini}, E.},
        title = "{The Hubble Space Telescope UV Legacy Survey of Galactic globular clusters - XIII. ACS/WFC parallel-field catalogues}",
      journal = {\mnras},
         year = 2018,
        month = may,
       volume = {476},
       number = {1},
        pages = {271-299},
          doi = {10.1093/mnras/sty177},
archivePrefix = {arXiv},
       eprint = {1801.07445},
 primaryClass = {astro-ph.GA},
       adsurl = {https://ui.adsabs.harvard.edu/abs/2018MNRAS.476..271S}
}

@ARTICLE{2mass,
       author = {{Skrutskie}, M.~F. and {Cutri}, R.~M. and {Stiening}, R. and {Weinberg}, M.~D. and {Schneider}, S. and {Carpenter}, J.~M. and {Beichman}, C. and {Capps}, R. and {Chester}, T. and {Elias}, J. and {Huchra}, J. and {Liebert}, J. and {Lonsdale}, C. and {Monet}, D.~G. and {Price}, S. and {Seitzer}, P. and {Jarrett}, T. and {Kirkpatrick}, J.~D. and {Gizis}, J.~E. and {Howard}, E. and {Evans}, T. and {Fowler}, J. and {Fullmer}, L. and {Hurt}, R. and {Light}, R. and {Kopan}, E.~L. and {Marsh}, K.~A. and {McCallon}, H.~L. and {Tam}, R. and {Van Dyk}, S. and {Wheelock}, S.},
        title = "{The Two Micron All Sky Survey (2MASS)}",
      journal = {\aj},
         year = 2006,
        month = feb,
       volume = {131},
       number = {2},
        pages = {1163-1183},
          doi = {10.1086/498708},
       adsurl = {https://ui.adsabs.harvard.edu/abs/2006AJ....131.1163S}
}

@ARTICLE{sollima2016,
       author = {{Sollima}, A. and {Ferraro}, F.~R. and {Lovisi}, L. and {Contenta}, F. and {Vesperini}, E. and {Origlia}, L. and {Lapenna}, E. and {Lanzoni}, B. and {Mucciarelli}, A. and {Dalessandro}, E. and {Pallanca}, C.},
        title = "{Searching in the dark: the dark mass content of the Milky Way globular clusters NGC288 and NGC6218}",
      journal = {\mnras},
         year = 2016,
        month = oct,
       volume = {462},
       number = {2},
        pages = {1937-1951},
          doi = {10.1093/mnras/stw1779},
archivePrefix = {arXiv},
       eprint = {1607.05612},
 primaryClass = {astro-ph.SR},
       adsurl = {https://ui.adsabs.harvard.edu/abs/2016MNRAS.462.1937S}
}

@ARTICLE{sollima2017,
       author = {{Sollima}, A. and {Dalessandro}, E. and {Beccari}, G. and {Pallanca}, C.},
        title = "{Testing multimass dynamical models of star clusters with real data: mass segregation in three Galactic globular clusters}",
      journal = {\mnras},
         year = 2017,
        month = feb,
       volume = {464},
       number = {4},
        pages = {3871-3881},
          doi = {10.1093/mnras/stw2628},
archivePrefix = {arXiv},
       eprint = {1610.02300},
 primaryClass = {astro-ph.GA},
       adsurl = {https://ui.adsabs.harvard.edu/abs/2017MNRAS.464.3871S}
}

@ARTICLE{stetson2019,
       author = {{Stetson}, P.~B. and {Pancino}, E. and {Zocchi}, A. and {Sanna}, N. and {Monelli}, M.},
        title = "{Homogeneous photometry - VII. Globular clusters in the Gaia era}",
      journal = {\mnras},
         year = 2019,
        month = may,
       volume = {485},
       number = {3},
        pages = {3042-3063},
          doi = {10.1093/mnras/stz585},
archivePrefix = {arXiv},
       eprint = {1902.09925},
 primaryClass = {astro-ph.SR},
       adsurl = {https://ui.adsabs.harvard.edu/abs/2019MNRAS.485.3042S}
}

@ARTICLE{tailo2020,
       author = {{Tailo}, M. and {Milone}, A.~P. and {Lagioia}, E.~P. and {D'Antona}, F. and {Marino}, A.~F. and {Vesperini}, E. and {Caloi}, V. and {Ventura}, P. and {Dondoglio}, E. and {Cordoni}, G.},
        title = "{Mass-loss along the red giant branch in 46 globular clusters and their multiple populations}",
      journal = {\mnras},
         year = 2020,
        month = nov,
       volume = {498},
       number = {4},
        pages = {5745-5771},
          doi = {10.1093/mnras/staa2639},
archivePrefix = {arXiv},
       eprint = {2009.01080},
 primaryClass = {astro-ph.SR},
       adsurl = {https://ui.adsabs.harvard.edu/abs/2020MNRAS.498.5745T}
}

@ARTICLE{torelli2019,
       author = {{Torelli}, M. and {Iannicola}, G. and {Stetson}, P.~B. and {Ferraro}, I. and {Bono}, G. and {Salaris}, M. and {Castellani}, M. and {Dall'Ora}, M. and {Fontana}, A. and {Monelli}, M. and {Pietrinferni}, A.},
        title = "{Horizontal branch morphology: A new photometric parametrization}",
      journal = {\aap},
         year = 2019,
        month = sep,
       volume = {629},
          eid = {A53},
        pages = {A53},
          doi = {10.1051/0004-6361/201935995},
archivePrefix = {arXiv},
       eprint = {1907.09568},
 primaryClass = {astro-ph.SR},
       adsurl = {https://ui.adsabs.harvard.edu/abs/2019A&A...629A..53T}
}

@ARTICLE{valcin2020,
       author = {{Valcin}, David and {Bernal}, Jos{\'e} Luis and {Jimenez}, Raul and {Verde}, Licia and {Wandelt}, Benjamin D.},
        title = "{Inferring the age of the universe with globular clusters}",
      journal = {\jcap},
         year = 2020,
        month = dec,
       volume = {2020},
       number = {12},
          eid = {002},
        pages = {002},
          doi = {10.1088/1475-7516/2020/12/002},
archivePrefix = {arXiv},
       eprint = {2007.06594},
 primaryClass = {astro-ph.CO},
       adsurl = {https://ui.adsabs.harvard.edu/abs/2020JCAP...12..002V}
}

@ARTICLE{gaiadr3,
       author = {{Gaia Collaboration} and {Vallenari}, A. and {Brown}, A.~G.~A. and {Prusti}, T. and {de Bruijne}, J.~H.~J. and {Arenou}, F. and {Babusiaux}, C. and {Biermann}, M. and {Creevey}, O.~L. and {Ducourant}, C. and {Evans}, D.~W. and {Eyer}, L. and {Guerra}, R. and {Hutton}, A. and {Jordi}, C. and {Klioner}, S.~A. and {Lammers}, U.~L. and {Lindegren}, L. and {Luri}, X. and {Mignard}, F. and {Panem}, C. and {Pourbaix}, D. and {Randich}, S. and {Sartoretti}, P. and {Soubiran}, C. and {Tanga}, P. and {Walton}, N.~A. and {Bailer-Jones}, C.~A.~L. and {Bastian}, U. and {Drimmel}, R. and {Jansen}, F. and {Katz}, D. and {Lattanzi}, M.~G. and {van Leeuwen}, F. and {Bakker}, J. and {Cacciari}, C. and {Casta{\~n}eda}, J. and {De Angeli}, F. and {Fabricius}, C. and {Fouesneau}, M. and {Fr{\'e}mat}, Y. and {Galluccio}, L. and {Guerrier}, A. and {Heiter}, U. and {Masana}, E. and {Messineo}, R. and {Mowlavi}, N. and {Nicolas}, C. and {Nienartowicz}, K. and {Pailler}, F. and {Panuzzo}, P. and {Riclet}, F. and {Roux}, W. and {Seabroke}, G.~M. and {Sordo}, R. and {Th{\'e}venin}, F. and {Gracia-Abril}, G. and {Portell}, J. and {Teyssier}, D. and {Altmann}, M. and {Andrae}, R. and {Audard}, M. and {Bellas-Velidis}, I. and {Benson}, K. and {Berthier}, J. and {Blomme}, R. and {Burgess}, P.~W. and {Busonero}, D. and {Busso}, G. and {C{\'a}novas}, H. and {Carry}, B. and {Cellino}, A. and {Cheek}, N. and {Clementini}, G. and {Damerdji}, Y. and {Davidson}, M. and {de Teodoro}, P. and {Nu{\~n}ez Campos}, M. and {Delchambre}, L. and {Dell'Oro}, A. and {Esquej}, P. and {Fern{\'a}ndez-Hern{\'a}ndez}, J. and {Fraile}, E. and {Garabato}, D. and {Garc{\'\i}a-Lario}, P. and {Gosset}, E. and {Haigron}, R. and {Halbwachs}, J. -L. and {Hambly}, N.~C. and {Harrison}, D.~L. and {Hern{\'a}ndez}, J. and {Hestroffer}, D. and {Hodgkin}, S.~T. and {Holl}, B. and {Jan{\ss}en}, K. and {Jevardat de Fombelle}, G. and {Jordan}, S. and {Krone-Martins}, A. and {Lanzafame}, A.~C. and {L{\"o}ffler}, W. and {Marchal}, O. and {Marrese}, P.~M. and {Moitinho}, A. and {Muinonen}, K. and {Osborne}, P. and {Pancino}, E. and {Pauwels}, T. and {Recio-Blanco}, A. and {Reyl{\'e}}, C. and {Riello}, M. and {Rimoldini}, L. and {Roegiers}, T. and {Rybizki}, J. and {Sarro}, L.~M. and {Siopis}, C. and {Smith}, M. and {Sozzetti}, A. and {Utrilla}, E. and {van Leeuwen}, M. and {Abbas}, U. and {{\'A}brah{\'a}m}, P. and {Abreu Aramburu}, A. and {Aerts}, C. and {Aguado}, J.~J. and {Ajaj}, M. and {Aldea-Montero}, F. and {Altavilla}, G. and {{\'A}lvarez}, M.~A. and {Alves}, J. and {Anders}, F. and {Anderson}, R.~I. and {Anglada Varela}, E. and {Antoja}, T. and {Baines}, D. and {Baker}, S.~G. and {Balaguer-N{\'u}{\~n}ez}, L. and {Balbinot}, E. and {Balog}, Z. and {Barache}, C. and {Barbato}, D. and {Barros}, M. and {Barstow}, M.~A. and {Bartolom{\'e}}, S. and {Bassilana}, J. -L. and {Bauchet}, N. and {Becciani}, U. and {Bellazzini}, M. and {Berihuete}, A. and {Bernet}, M. and {Bertone}, S. and {Bianchi}, L. and {Binnenfeld}, A. and {Blanco-Cuaresma}, S. and {Blazere}, A. and {Boch}, T. and {Bombrun}, A. and {Bossini}, D. and {Bouquillon}, S. and {Bragaglia}, A. and {Bramante}, L. and {Breedt}, E. and {Bressan}, A. and {Brouillet}, N. and {Brugaletta}, E. and {Bucciarelli}, B. and {Burlacu}, A. and {Butkevich}, A.~G. and {Buzzi}, R. and {Caffau}, E. and {Cancelliere}, R. and {Cantat-Gaudin}, T. and {Carballo}, R. and {Carlucci}, T. and {Carnerero}, M.~I. and {Carrasco}, J.~M. and {Casamiquela}, L. and {Castellani}, M. and {Castro-Ginard}, A. and {Chaoul}, L. and {Charlot}, P. and {Chemin}, L. and {Chiaramida}, V. and {Chiavassa}, A. and {Chornay}, N. and {Comoretto}, G. and {Contursi}, G. and {Cooper}, W.~J. and {Cornez}, T. and {Cowell}, S. and {Crifo}, F. and {Cropper}, M. and {Crosta}, M. and {Crowley}, C. and {Dafonte}, C. and {Dapergolas}, A. and {David}, M. and {David}, P. and {de Laverny}, P. and {De Luise}, F. and {De March}, R. and {De Ridder}, J. and {de Souza}, R. and {de Torres}, A. and {del Peloso}, E.~F. and {del Pozo}, E. and {Delbo}, M. and {Delgado}, A. and {Delisle}, J. -B. and {Demouchy}, C. and {Dharmawardena}, T.~E. and {Di Matteo}, P. and {Diakite}, S. and {Diener}, C. and {Distefano}, E. and {Dolding}, C. and {Edvardsson}, B. and {Enke}, H. and {Fabre}, C. and {Fabrizio}, M. and {Faigler}, S. and {Fedorets}, G. and {Fernique}, P. and {Fienga}, A. and {Figueras}, F. and {Fournier}, Y. and {Fouron}, C. and {Fragkoudi}, F. and {Gai}, M. and {Garcia-Gutierrez}, A. and {Garcia-Reinaldos}, M. and {Garc{\'\i}a-Torres}, M. and {Garofalo}, A. and {Gavel}, A. and {Gavras}, P. and {Gerlach}, E. and {Geyer}, R. and {Giacobbe}, P. and {Gilmore}, G. and {Girona}, S. and {Giuffrida}, G. and {Gomel}, R. and {Gomez}, A. and {Gonz{\'a}lez-N{\'u}{\~n}ez}, J. and {Gonz{\'a}lez-Santamar{\'\i}a}, I. and {Gonz{\'a}lez-Vidal}, J.~J. and {Granvik}, M. and {Guillout}, P. and {Guiraud}, J. and {Guti{\'e}rrez-S{\'a}nchez}, R. and {Guy}, L.~P. and {Hatzidimitriou}, D. and {Hauser}, M. and {Haywood}, M. and {Helmer}, A. and {Helmi}, A. and {Sarmiento}, M.~H. and {Hidalgo}, S.~L. and {Hilger}, T. and {H{\l}adczuk}, N. and {Hobbs}, D. and {Holland}, G. and {Huckle}, H.~E. and {Jardine}, K. and {Jasniewicz}, G. and {Jean-Antoine Piccolo}, A. and {Jim{\'e}nez-Arranz}, {\'O}. and {Jorissen}, A. and {Juaristi Campillo}, J. and {Julbe}, F. and {Karbevska}, L. and {Kervella}, P. and {Khanna}, S. and {Kontizas}, M. and {Kordopatis}, G. and {Korn}, A.~J. and {K{\'o}sp{\'a}l}, {\'A}. and {Kostrzewa-Rutkowska}, Z. and {Kruszy{\'n}ska}, K. and {Kun}, M. and {Laizeau}, P. and {Lambert}, S. and {Lanza}, A.~F. and {Lasne}, Y. and {Le Campion}, J. -F. and {Lebreton}, Y. and {Lebzelter}, T. and {Leccia}, S. and {Leclerc}, N. and {Lecoeur-Taibi}, I. and {Liao}, S. and {Licata}, E.~L. and {Lindstr{\o}m}, H.~E.~P. and {Lister}, T.~A. and {Livanou}, E. and {Lobel}, A. and {Lorca}, A. and {Loup}, C. and {Madrero Pardo}, P. and {Magdaleno Romeo}, A. and {Managau}, S. and {Mann}, R.~G. and {Manteiga}, M. and {Marchant}, J.~M. and {Marconi}, M. and {Marcos}, J. and {Marcos Santos}, M.~M.~S. and {Mar{\'\i}n Pina}, D. and {Marinoni}, S. and {Marocco}, F. and {Marshall}, D.~J. and {Martin Polo}, L. and {Mart{\'\i}n-Fleitas}, J.~M. and {Marton}, G. and {Mary}, N. and {Masip}, A. and {Massari}, D. and {Mastrobuono-Battisti}, A. and {Mazeh}, T. and {McMillan}, P.~J. and {Messina}, S. and {Michalik}, D. and {Millar}, N.~R. and {Mints}, A. and {Molina}, D. and {Molinaro}, R. and {Moln{\'a}r}, L. and {Monari}, G. and {Mongui{\'o}}, M. and {Montegriffo}, P. and {Montero}, A. and {Mor}, R. and {Mora}, A. and {Morbidelli}, R. and {Morel}, T. and {Morris}, D. and {Muraveva}, T. and {Murphy}, C.~P. and {Musella}, I. and {Nagy}, Z. and {Noval}, L. and {Oca{\~n}a}, F. and {Ogden}, A. and {Ordenovic}, C. and {Osinde}, J.~O. and {Pagani}, C. and {Pagano}, I. and {Palaversa}, L. and {Palicio}, P.~A. and {Pallas-Quintela}, L. and {Panahi}, A. and {Payne-Wardenaar}, S. and {Pe{\~n}alosa Esteller}, X. and {Penttil{\"a}}, A. and {Pichon}, B. and {Piersimoni}, A.~M. and {Pineau}, F. -X. and {Plachy}, E. and {Plum}, G. and {Poggio}, E. and {Pr{\v{s}}a}, A. and {Pulone}, L. and {Racero}, E. and {Ragaini}, S. and {Rainer}, M. and {Raiteri}, C.~M. and {Rambaux}, N. and {Ramos}, P. and {Ramos-Lerate}, M. and {Re Fiorentin}, P. and {Regibo}, S. and {Richards}, P.~J. and {Rios Diaz}, C. and {Ripepi}, V. and {Riva}, A. and {Rix}, H. -W. and {Rixon}, G. and {Robichon}, N. and {Robin}, A.~C. and {Robin}, C. and {Roelens}, M. and {Rogues}, H.~R.~O. and {Rohrbasser}, L. and {Romero-G{\'o}mez}, M. and {Rowell}, N. and {Royer}, F. and {Ruz Mieres}, D. and {Rybicki}, K.~A. and {Sadowski}, G. and {S{\'a}ez N{\'u}{\~n}ez}, A. and {Sagrist{\`a} Sell{\'e}s}, A. and {Sahlmann}, J. and {Salguero}, E. and {Samaras}, N. and {Sanchez Gimenez}, V. and {Sanna}, N. and {Santove{\~n}a}, R. and {Sarasso}, M. and {Schultheis}, M. and {Sciacca}, E. and {Segol}, M. and {Segovia}, J.~C. and {S{\'e}gransan}, D. and {Semeux}, D. and {Shahaf}, S. and {Siddiqui}, H.~I. and {Siebert}, A. and {Siltala}, L. and {Silvelo}, A. and {Slezak}, E. and {Slezak}, I. and {Smart}, R.~L. and {Snaith}, O.~N. and {Solano}, E. and {Solitro}, F. and {Souami}, D. and {Souchay}, J. and {Spagna}, A. and {Spina}, L. and {Spoto}, F. and {Steele}, I.~A. and {Steidelm{\"u}ller}, H. and {Stephenson}, C.~A. and {S{\"u}veges}, M. and {Surdej}, J. and {Szabados}, L. and {Szegedi-Elek}, E. and {Taris}, F. and {Taylor}, M.~B. and {Teixeira}, R. and {Tolomei}, L. and {Tonello}, N. and {Torra}, F. and {Torra}, J. and {Torralba Elipe}, G. and {Trabucchi}, M. and {Tsounis}, A.~T. and {Turon}, C. and {Ulla}, A. and {Unger}, N. and {Vaillant}, M.~V. and {van Dillen}, E. and {van Reeven}, W. and {Vanel}, O. and {Vecchiato}, A. and {Viala}, Y. and {Vicente}, D. and {Voutsinas}, S. and {Weiler}, M. and {Wevers}, T. and {Wyrzykowski}, {\L}. and {Yoldas}, A. and {Yvard}, P. and {Zhao}, H. and {Zorec}, J. and {Zucker}, S. and {Zwitter}, T.},
        title = "{Gaia Data Release 3. Summary of the content and survey properties}",
      journal = {\aap},
         year = 2023,
        month = jun,
       volume = {674},
          eid = {A1},
        pages = {A1},
          doi = {10.1051/0004-6361/202243940},
archivePrefix = {arXiv},
       eprint = {2208.00211},
 primaryClass = {astro-ph.GA},
       adsurl = {https://ui.adsabs.harvard.edu/abs/2023A&A...674A...1G}
}

@ARTICLE{vandenberg2013,
       author = {{VandenBerg}, Don A. and {Brogaard}, K. and {Leaman}, R. and {Casagrande}, L.},
        title = "{The Ages of 55 Globular Clusters as Determined Using an Improved $\Delta V^{HB}_{TO}$ Method along with Color-Magnitude Diagram Constraints, and Their Implications for Broader Issues}",
      journal = {\apj},
         year = 2013,
        month = oct,
       volume = {775},
       number = {2},
          eid = {134},
        pages = {134},
          doi = {10.1088/0004-637X/775/2/134},
archivePrefix = {arXiv},
       eprint = {1308.2257},
 primaryClass = {astro-ph.GA},
       adsurl = {https://ui.adsabs.harvard.edu/abs/2013ApJ...775..134V}
}

@ARTICLE{vasiliev2021,
       author = {{Vasiliev}, Eugene and {Baumgardt}, Holger},
        title = "{Gaia EDR3 view on galactic globular clusters}",
      journal = {\mnras},
         year = 2021,
        month = aug,
       volume = {505},
       number = {4},
        pages = {5978-6002},
          doi = {10.1093/mnras/stab1475},
archivePrefix = {arXiv},
       eprint = {2102.09568},
 primaryClass = {astro-ph.GA},
       adsurl = {https://ui.adsabs.harvard.edu/abs/2021MNRAS.505.5978V}
}

@ARTICLE{viaux2013,
       author = {{Viaux}, N. and {Catelan}, M. and {Stetson}, P.~B. and {Raffelt}, G.~G. and {Redondo}, J. and {Valcarce}, A.~A.~R. and {Weiss}, A.},
        title = "{Particle-physics constraints from the globular cluster M5: neutrino dipole moments}",
      journal = {\aap},
         year = 2013,
        month = oct,
       volume = {558},
          eid = {A12},
        pages = {A12},
          doi = {10.1051/0004-6361/201322004},
archivePrefix = {arXiv},
       eprint = {1308.4627},
 primaryClass = {astro-ph.SR},
       adsurl = {https://ui.adsabs.harvard.edu/abs/2013A&A...558A..12V}
}

@ARTICLE{vitral2021,
       author = {{Vitral}, Eduardo},
        title = "{BALRoGO: Bayesian Astrometric Likelihood Recovery of Galactic Objects - Global properties of over one hundred globular clusters with Gaia EDR3}",
      journal = {\mnras},
         year = 2021,
        month = jun,
       volume = {504},
       number = {1},
        pages = {1355-1369},
          doi = {10.1093/mnras/stab947},
archivePrefix = {arXiv},
       eprint = {2102.04841},
 primaryClass = {astro-ph.GA},
       adsurl = {https://ui.adsabs.harvard.edu/abs/2021MNRAS.504.1355V}
}

@ARTICLE{wise,
       author = {{Wright}, Edward L. and {Eisenhardt}, Peter R.~M. and {Mainzer}, Amy K. and {Ressler}, Michael E. and {Cutri}, Roc M. and {Jarrett}, Thomas and {Kirkpatrick}, J. Davy and {Padgett}, Deborah and {McMillan}, Robert S. and {Skrutskie}, Michael and {Stanford}, S.~A. and {Cohen}, Martin and {Walker}, Russell G. and {Mather}, John C. and {Leisawitz}, David and {Gautier}, Thomas N., III and {McLean}, Ian and {Benford}, Dominic and {Lonsdale}, Carol J. and {Blain}, Andrew and {Mendez}, Bryan and {Irace}, William R. and {Duval}, Valerie and {Liu}, Fengchuan and {Royer}, Don and {Heinrichsen}, Ingolf and {Howard}, Joan and {Shannon}, Mark and {Kendall}, Martha and {Walsh}, Amy L. and {Larsen}, Mark and {Cardon}, Joel G. and {Schick}, Scott and {Schwalm}, Mark and {Abid}, Mohamed and {Fabinsky}, Beth and {Naes}, Larry and {Tsai}, Chao-Wei},
        title = "{The Wide-field Infrared Survey Explorer (WISE): Mission Description and Initial On-orbit Performance}",
      journal = {\aj},
         year = 2010,
        month = dec,
       volume = {140},
       number = {6},
        pages = {1868-1881},
          doi = {10.1088/0004-6256/140/6/1868},
archivePrefix = {arXiv},
       eprint = {1008.0031},
 primaryClass = {astro-ph.IM},
       adsurl = {https://ui.adsabs.harvard.edu/abs/2010AJ....140.1868W}
}

@ARTICLE{ying2025,
       author = {{Ying}, Jiaqi (Martin) and {Chaboyer}, Brian and {Boylan-Kolchin}, Michael and {Weisz}, Daniel R. and {Goebel-Bain}, Rowan},
        title = "{The Absolute Age of Milky Way Globular Clusters}",
      journal = {\apj},
         year = 2025,
        month = jul,
       volume = {987},
       number = {1},
          eid = {52},
        pages = {52},
          doi = {10.3847/1538-4357/add471},
archivePrefix = {arXiv},
       eprint = {2505.02969},
 primaryClass = {astro-ph.GA},
       adsurl = {https://ui.adsabs.harvard.edu/abs/2025ApJ...987...52Y}
}

@ARTICLE{zaritsky2002,
       author = {{Zaritsky}, Dennis and {Harris}, Jason and {Thompson}, Ian B. and {Grebel}, Eva K. and {Massey}, Philip},
        title = "{The Magellanic Clouds Photometric Survey: The Small Magellanic Cloud Stellar Catalog and Extinction Map}",
      journal = {\aj},
         year = 2002,
        month = feb,
       volume = {123},
       number = {2},
        pages = {855-872},
          doi = {10.1086/338437},
archivePrefix = {arXiv},
       eprint = {astro-ph/0110665},
 primaryClass = {astro-ph},
       adsurl = {https://ui.adsabs.harvard.edu/abs/2002AJ....123..855Z}
}

@ARTICLE{zloczewski2012,
       author = {{Zloczewski}, K. and {Kaluzny}, J. and {Rozyczka}, M. and {Krzeminski}, W. and {Mazur}, B.},
        title = "{A Proper Motion Study of the Globular Clusters M4, M12, M22, NGC 3201, NGC 6362 and NGC 6752}",
      journal = {\actaa},
         year = 2012,
        month = dec,
       volume = {62},
       number = {4},
        pages = {357-375},
          doi = {10.48550/arXiv.1301.1198},
archivePrefix = {arXiv},
       eprint = {1301.1198},
 primaryClass = {astro-ph.SR},
       adsurl = {https://ui.adsabs.harvard.edu/abs/2012AcA....62..357Z}
}

\clearpage
\appendix

\section{Rejected data sets}
\label{rejected}

We do not use some available data sets for clusters under consideration (some were considered in \citetalias{ngc5904}--\citetalias{ngc288}) due to the following reasons.

The {\it HST} ACS photometry by \citet{sarajedini2007} has been replaced by \citetalias{nardiello2018}.
The {\it HST} Wide Field and Planetary Camera 2 (WFPC2) data sets \citep{piotto2002} for NGC\,362, NGC\,1904, NGC\,4372, NGC\,5904, NGC\,6205, and NGC\,6218 show large 
differences between photometry from different CCD chips used.
The HB, AGB, and RGB are saturated in the Parallel-Field Catalogues of the {\it HST} UV Legacy Survey of Galactic Globular Clusters \citep{simioni2018} for NGC\,5904 and NGC\,6218
as well as in the {\it HST} WFPC2 data set by \citet{layden2005} for NGC\,5904.
The {\it HST} photometry by \citet{cohen1997} for NGC\,6205 is of unacceptable quality due to significant photometric errors and contamination by non-members.
The $JK_s$ photometry obtained by \citet{cohen2015} with Infrared Side Port Imager mounted on the 4-m Blanco telescope at CTIO for NGC\,288 and NGC\,362 
is not precisely referred to the 2MASS system (as evident from our isochrone fitting) and, hence, has a poorly defined system.
The IR photometry in the 3.6-$\mu$m filter of the {\it Spitzer} Space Telescope Infrared Array Camera obtained by \citet{sage} within 
the Surveying the Agents of Galaxy Evolution in the Tidally-Disrupted, Low-Metallicity Small Magellanic Cloud (SAGE) for NGC\,362 is related only to
stars brighter than TO and not modeled by BaSTI.
The Str\"omgren $vby$ photometry from the Isaac Newton Telescope -- Wide Field Camera \citep{savino2018} for NGC\,6205 shows unacceptable quality in the observations of 
valuable bright stars, such as those on the HB, AGB, and RGB.

Some data sets have been included into the current version of the \citetalias{stetson2019} data sets and, hence, are not used by us separately:
the $VI$ photometry by \citet{rosenberg2000a,rosenberg2000b} for all the clusters,
$BV$ photometry with the Wide Field Imager mounted on the 2.2-m telescope, ESO, La Silla for NGC\,288 and NGC\,6218 \citep{sollima2016},
$UBVRI$ photometry by \citet{viaux2013} for NGC\,5904,
$VI$ photometry for NGC\,4372 by \citet{kacharov2014} obtained with the Wide Field Imager at the 2.2-m MPG/ESO telescope at La Silla,
and others.

We do not use any data set represented only by a fiducial sequence (i.e., without data for individual stars), as this does not allow for cross-identification, which is essential to our approach.
Such data sets used in \citetalias{ngc5904}--\citetalias{ngc288} are 
the $BV$ photometry obtained for NGC\,288 with the 4-m and 0.9-m telescopes at CTIO \citep{bolte1992};
the $JK$ photometry obtained for NGC\,288 and NGC\,6205 with the 3.6-m Canada--France--Hawaii Telescope (CFHT) \citep{davidge1995,davidge1997};
the $ugriz$ photometry obtained for NGC\,6205 with the MegaCam wide-field imager on CFHT \citep{clem2008};
the $BV$ photometry obtained for NGC\,6205 with the 1.23-m telescope at the German-Spanish Astronomical Center, Calar Alto, Spain \citep{paltrinieri1998};
the $JK$ photometry obtained for NGC\,5904 and NGC\,6205 with the WIRCam imager on CFHT \citep{brasseur2010}.

Also, we do not use the promising $ugriz$ photometry obtained by \citet{shanks2015} with the Very Large Telescope ATLAS survey for NGC\,288, 
as it is not precisely calibrated to the SDSS system (as evident from our isochrone fitting) and therefore has a poorly defined photometric system.

\section{Filters used and cleaning of the data sets}
\label{lambda}

\begin{table}
\def\baselinestretch{1}\normalsize\scriptsize
\caption[]{The adopted effective wavelength $\lambda_\mathrm{eff}$ (nm) for the filters under consideration, data set numbers (see text),
and typical photometric uncertainty cut (mag) applied (slightly varying depending on cluster and data set).
We relax the photometry cuts by 0.03 mag for distant NGC\,1904, while tight them by 0.02 mag for highly contaminated NGC\,362.
}
\label{filters}
\[
\begin{tabular}{lrlc}
\hline
\noalign{\smallskip}
Filter & $\lambda_\mathrm{eff}$ & Data set numbers & Cut \\
\hline
\noalign{\smallskip}
{\it HST}/WFC3 $F275W$             & 285 & \ref{filterhst}                                              & 0.08 \\
{\it HST}/WFC3 $F336W$             & 340 & \ref{filterhst}                                              & 0.08 \\
Str\"omgren $u$                    & 349 & \ref{filtergrundahl}                                         & 0.08 \\
SDSS $u_\mathrm{SDSS}$             & 360 & \ref{filtersdss}                                             & 0.10 \\
Johnson--Kron--Cousins $U$         & 366 & \ref{filterstetson}, \ref{filtermcps}, \ref{filterkravtsov}  & 0.10 \\
Str\"omgren $v$                    & 414 & \ref{filtergrundahl}                                         & 0.07 \\
{\it HST}/WFC3 $F438W$             & 438 & \ref{filterhst}                                              & 0.07  \\
Johnson--Kron--Cousins $B$         & 452 & \ref{filterstetson}, \ref{filternarloch}, \ref{filtermcps}, \ref{filterhargis}, \ref{filterzloczewski}, \ref{filterkravtsov}, \ref{filterrey}   & 0.08 \\
Str\"omgren $b$                    & 467 & \ref{filtergrundahl}, \ref{filterlee}                        & 0.06  \\
SDSS $g_\mathrm{SDSS}$             & 471 & \ref{filtersdss}                                             & 0.07  \\  
DES DR2 DECam $g_\mathrm{DECam}$   & 481 & \ref{filterdes}                                              & 0.02  \\
PS1 $g_\mathrm{PS1}$               & 496 & \ref{filterps1}                                              & 0.08  \\  
{\it Gaia} DR3 $G_\mathrm{BP}$     & 505 & \ref{filtergaia}                                             & 0.10  \\
SMSS $g_\mathrm{SMSS}$             & 514 & \ref{filtersmss}                                             & 0.10  \\
Str\"omgren $y$                    & 548 & \ref{filtergrundahl}, \ref{filterlee}                        & 0.06  \\
Johnson--Kron--Cousins $V$         & 552 & \ref{filterstetson}, \ref{filternarloch}, \ref{filtermcps}, \ref{filterhargis}, \ref{filterzloczewski}, \ref{filterkravtsov}, \ref{filterrey}, \ref{filterbellazzini}  & 0.08 \\
{\it HST}/ACS $F606W$              & 599 & \ref{filterhst}                                              & 0.06  \\
{\it Gaia} DR3 $G$                 & 604 & \ref{filtergaia}                                             & 0.03  \\
SMSS $r_\mathrm{SMSS}$             & 615 & \ref{filtersmss}                                             & 0.10  \\
SDSS $r_\mathrm{SDSS}$             & 621 & \ref{filtersdss}                                             & 0.07  \\
PS1 $r_\mathrm{PS1}$               & 621 & \ref{filterps1}                                              & 0.07  \\
DES DR2 DECam $r_\mathrm{DECam}$   & 644 & \ref{filterdes}                                              & 0.02   \\
SDSS $i_\mathrm{SDSS}$             & 743 & \ref{filtersdss}                                             & 0.07  \\
PS1 $i_\mathrm{PS1}$               & 752 & \ref{filterps1}                                              & 0.07  \\
{\it Gaia} DR3 $G_\mathrm{RP}$     & 770 & \ref{filtergaia}                                             & 0.10  \\
SMSS $i_\mathrm{SMSS}$             & 776 & \ref{filtersmss}                                             & 0.10  \\
DES DR2 DECam $i_\mathrm{DECam}$   & 784 & \ref{filterdes}                                              & 0.02    \\
{\it HST}/ACS $F814W$              & 807 & \ref{filterhst}                                              & 0.06  \\
Johnson--Kron--Cousins $I$         & 807 & \ref{filterstetson}, \ref{filtermcps}, \ref{filterbellazzini}, \ref{filterhargis}   & 0.08  \\
PS1 $z_\mathrm{PS1}$               & 867  & \ref{filterps1}                                              & 0.08  \\
SDSS $z_\mathrm{SDSS}$             & 885  & \ref{filtersdss}                                             & 0.08 \\
SMSS $z_\mathrm{SMSS}$             & 913  & \ref{filtersmss}                                             & 0.10 \\
DES DR2 DECam $z_\mathrm{DECam}$   & 927  & \ref{filterdes}                                              & 0.04  \\
PS1 $y_\mathrm{PS1}$               & 971  & \ref{filterps1}                                              & 0.10  \\
VISTA $Y_\mathrm{VISTA}$           & 1021 & \ref{filtervista}                                            & 0.10  \\
UKIDSS $Y_\mathrm{UKIDSS}$         & 1031 & \ref{filterukidss}                                           & 0.10  \\
UKIDSS $J_\mathrm{UKIDSS}$         & 1248 & \ref{filterukidss}                                           & 0.12  \\
VISTA $J_\mathrm{VISTA}$           & 1254 & \ref{filtervista}                                            & 0.12  \\
VISTA $K_\mathrm{VISTA}$           & 2149 & \ref{filtervista}                                            & 0.15  \\
2MASS $K_\mathrm{2MASS}$           & 2176 & \ref{filtercoppola}                                          & 0.05  \\
UKIDSS $K_\mathrm{UKIDSS}$         & 2201 & \ref{filterukidss}                                           & 0.15  \\
{\it WISE} $W1$                    & 3317 & \ref{filterwise}                                             & 0.15  \\
\hline
\end{tabular}
\]
\end{table}

Table~\ref{filters} presents the effective wavelength $\lambda_\mathrm{eff}$ in nm for the filters used, their correspondence with the data sets, and typical photometric uncertainty cut level.
This cut level is equal to $3\,\sigma$ of the average photometric uncertainty $\sigma$ stated by the authors of the data sets.
We eliminate stars with inaccurate photometry as those with a photometric uncertainty larger than the cut level. 
As an exception, we increase the cut level to 0.15\,mag for the {\it WISE} $W1$ filter for better representation of the TO and bright MS stars.

As in our previous papers, to clean the data sets, we follow the recommendations of their authors to select single star-like objects with reliable photometry.
The {\it HST} WFC3 and ACS data sets are cleaned by selecting stars with $|{\tt sharp}|<0.15$, quality fit $>0.9$, and membership probability $>0.9$ or $-1$ (this probability is discussed in \ref{details}).
For cleaning of the \citetalias{stetson2019}, \citetalias{grundahl1999}, SDSS, and other data sets with the stated $\chi$ and {\tt sharp} parameters, we select stars with $\chi<3$ and $|{\tt sharp}|<0.3$.

We leave {\it Gaia} stars with \verb|astrometric_excess_noise|$<1$ ($\epsilon i<1$);
a renormalised unit weight error not exceeding $1.4$ (\verb|RUWE|$<1.4$); 
and a corrected excess factor \verb"phot_bp_rp_excess_factor" (i.e. \verb"E(BP/RP)Corr") between $-0.14$ and $0.14$ (see \citealt{riello2021}).

We have to eliminate the brightest stars ($V<14.5$ mag) from the data set of \citet{zloczewski2012} as well as such stars ($Y_\mathrm{VISTA}<12.5$, $J_\mathrm{VISTA}<12$ or $K_\mathrm{VISTA}<11.5$) from the VISTA data sets due to their unacceptable systematics (e.g. see Fig.~\ref{ngc4372_dr}).

\section{Some more details of the study}
\label{details}

\begin{figure*}
\centering
\includegraphics{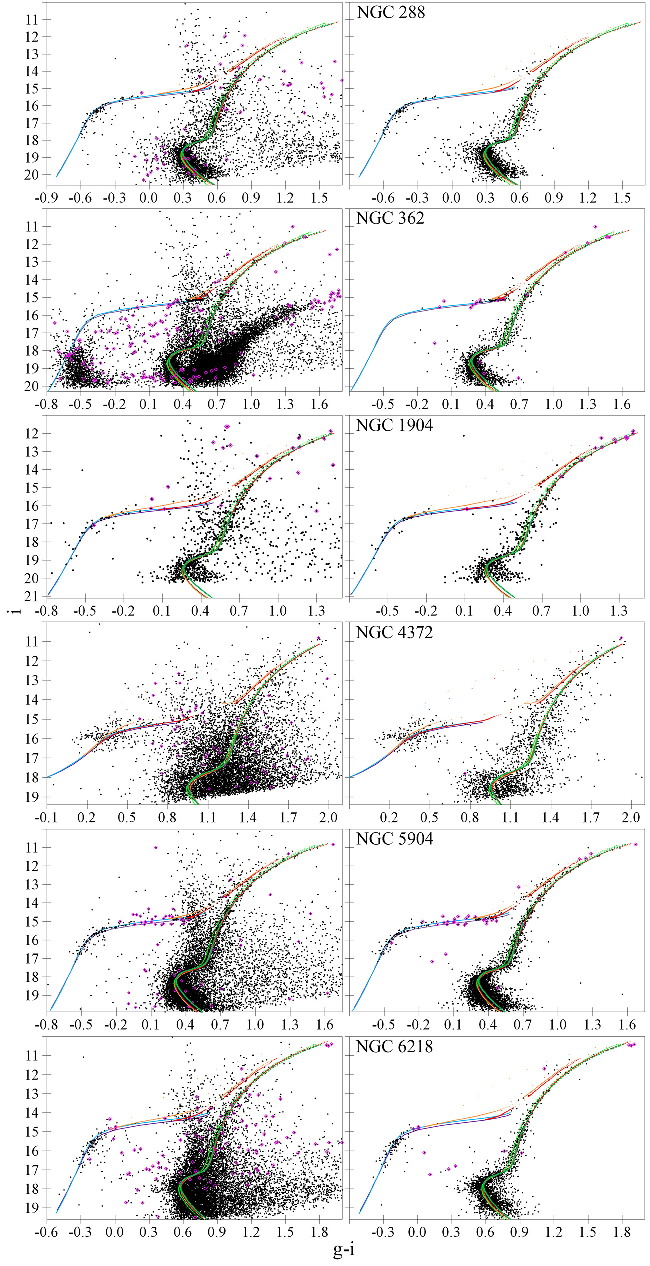}
\caption{The SMSS $g_\mathrm{SMSS}-i_\mathrm{SMSS}$ versus $i_\mathrm{SMSS}$ CMDs for six clusters before (left) and after (right) selection of the cluster members using the {\it Gaia} parallaxes and proper motions. DR is not corrected.
Variable stars are shown by the magenta diamonds.
The initial CMD for NGC\,362 is strongly contaminated by the Small Magellanic Cloud.
The isochrones for a primordial $Y\approx0.25$ from BaSTI (red), BaSTI ZAHB (purple), and DSED (green), isochrones for $Y=0.275$ from BaSTI (orange) and BaSTI ZAHB (blue), as well as isochrones for $Y=0.33$ from DSED (luminous green) are calculated with the best-fitting parameters.
}
\label{smss_init}
\end{figure*}

Fig.~\ref{smss_init} presents some examples of CMDs before and after the selection of cluster members when the remaining data set cleaning is applied.
The crucial importance of this selection is evident.

\begin{figure}
\centering
\includegraphics{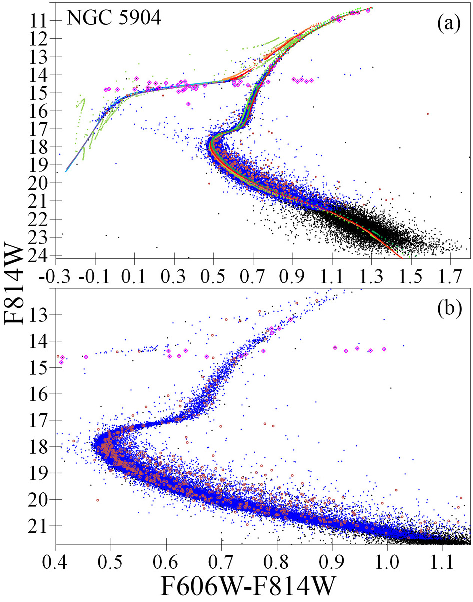}
\caption{(a) $F606W-F814W$ versus $F814W$ CMD for the \citetalias{nardiello2018} stars of NGC\,5904 with good photometry.
Stars with membership probability $>90\%$ -- blue symbols, with membership probability $<90\%$ -- brown symbols, with undefined membership probability $=-1$ -- black symbols.
Variable stars are shown by the magenta diamonds.
The isochrones for a primordial $Y\approx0.25$ from BaSTI (red), BaSTI ZAHB (purple), DSED (green), and DSED HB/AGB track (light green), isochrones for $Y=0.275$ from BaSTI (orange) and BaSTI ZAHB (blue), as well as isochrones for $Y=0.33$ from DSED (luminous green) are calculated with the best-fitting parameters.
(b) Central part of the same CMD.
}
\label{ngc5904_nardiello_f606wf814w_members_init}
\end{figure}

\begin{figure}
\centering
\includegraphics{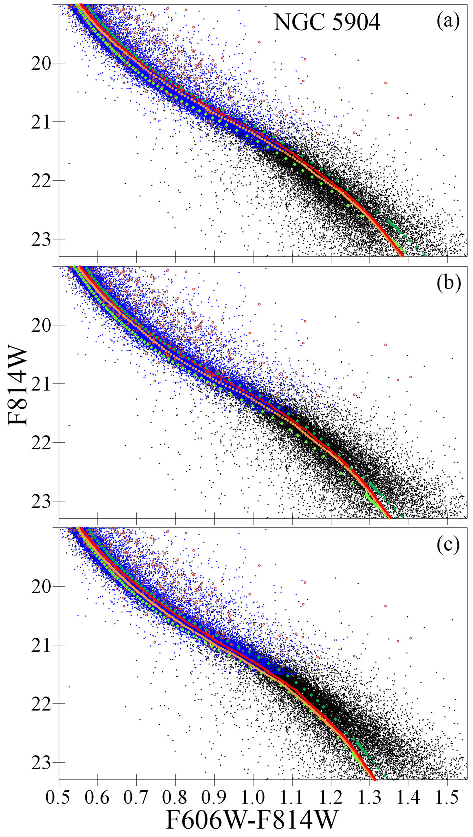}
\caption{The MS part of the $F606W-F814W$ versus $F814W$ CMD for the \citetalias{nardiello2018} stars of NGC\,5904 with good photometry.
Stars with membership probability $>90\%$ -- blue symbols, with membership probability $<90\%$ -- brown symbols, with undefined membership probability $=-1$ -- black symbols.
The isochrones for a primordial $Y\approx0.25$ from BaSTI (red) and DSED (green), isochrones for $Y=0.275$ from BaSTI (orange) as well as isochrones for $Y=0.33$ from DSED (luminous green) are calculated with the best-fitting parameters with (a) [Fe/H]$=-1.1$ and $-1.2$ for the BaSTI and DSED isochrones, respectively, (b) the finally accepted best estimates [Fe/H]$=-1.2$ and $-1.3$ for the BaSTI and DSED isochrones, respectively, and (c) [Fe/H]$=-1.3$ and $-1.4$ for the BaSTI and DSED isochrones, respectively.
}
\label{ngc5904_nardiello_f606wf814w_metal}
\end{figure}

The determination of the best-fit [Fe/H] from the \citetalias{nardiello2018} CMDs and its relation to the membership probability in the \citetalias{nardiello2018} data sets requires the following comments.
\citetalias{nardiello2018} data sets cover only few central arcminutes of the cluster fields where very likely cluster members (\citetalias{nardiello2018} membership probability $>90\%$) dominate: from >97\% among all stars for slightly contaminated clusters such as NGC\,5904 down to about 80\% for strongly contaminated ones such as NGC\,362. This is evident from Fig.~\ref{ngc5904_nardiello_f606wf814w_members_init} where very likely members, less likely members (\citetalias{nardiello2018} membership probability $<90\%$), and stars with indeterminate membership (membership probability $=-1$) are shown by different colours in a typical \citetalias{nardiello2018} CMD for NGC\,5904. It is seen that the stars with indeterminate membership probability are mainly faint MS stars. We assume that, similar to the bright MS stars, cluster members dominate among the faint MS stars with indeterminate membership probability. Therefore, we retain them in the data sets in contrast to the eliminated less likely members. These faint MS stars with indeterminate membership probability are important for our determination of [Fe/H] as evident from Fig.~\ref{ngc5904_nardiello_f606wf814w_metal} for the same \citetalias{nardiello2018} CMD for NGC\,5904 best-fitted by isochrones with different [Fe/H] (and other best-fitted parameters changed respectively).

It is worth noting that [Fe/H] can be best determined from fitting of the faint MS or bright RGB stars. The former are used for the determination of the best-fit [Fe/H] from the \citetalias{nardiello2018} CMDs, since their bright RGBs are sparsely populated, while the latter are used for that from the {\it Gaia}, \citetalias{stetson2019}, and other data sets, since their faint MSs are incomplete.


\begin{figure*}
\includegraphics{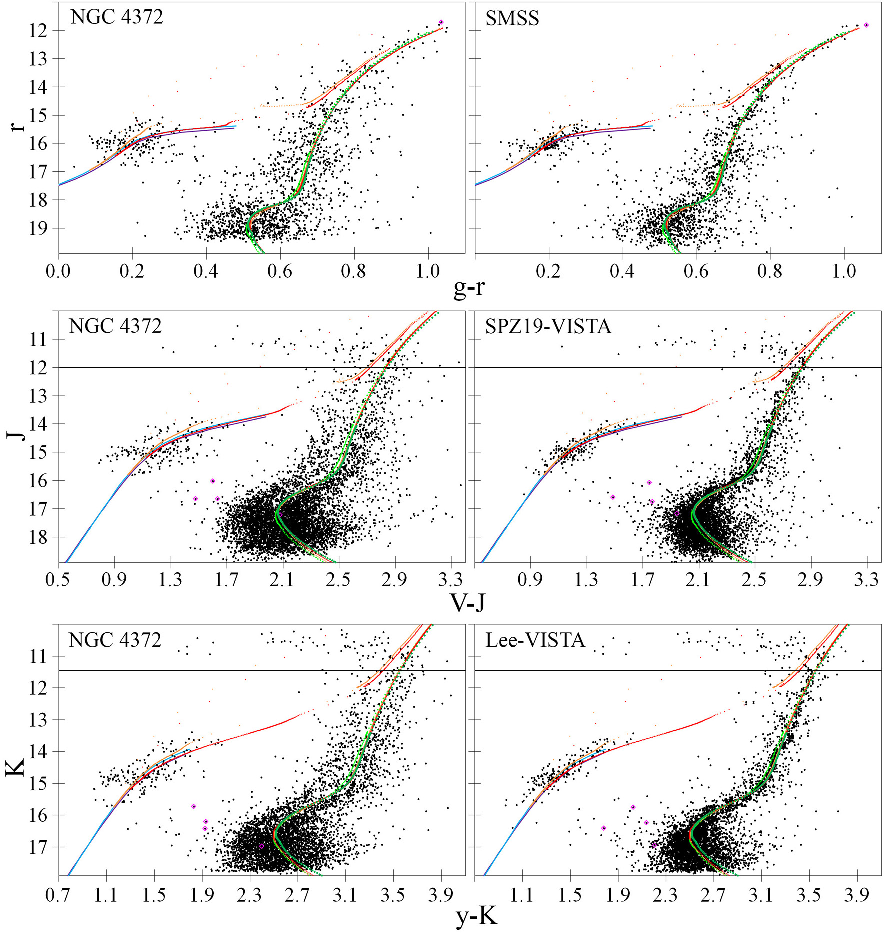}
\caption{The same as Fig.\ref{gaia} but for the NGC\,4372 SMSS $g-r$ versus $r$, \citetalias{stetson2019}--VISTA $V-J$ versus $J$, and Lee--VISTA $y-K$ versus $K$ CMDs
before (left column) and after (right column) our DR correction.
The saturated stars with the VISTA photometry at $J_\mathrm{VISTA}<12$ or $K_\mathrm{VISTA}<11.5$ are separated by the horizontal line and not used in the DR map construction.
}
\label{ngc4372_dr}
\end{figure*}

DR requires the following comments.
Fig.~\ref{ngc4372_dr} presents some examples of CMDs for NGC\,4372 before and after applying our DR corrections.

\begin{figure*}
\includegraphics{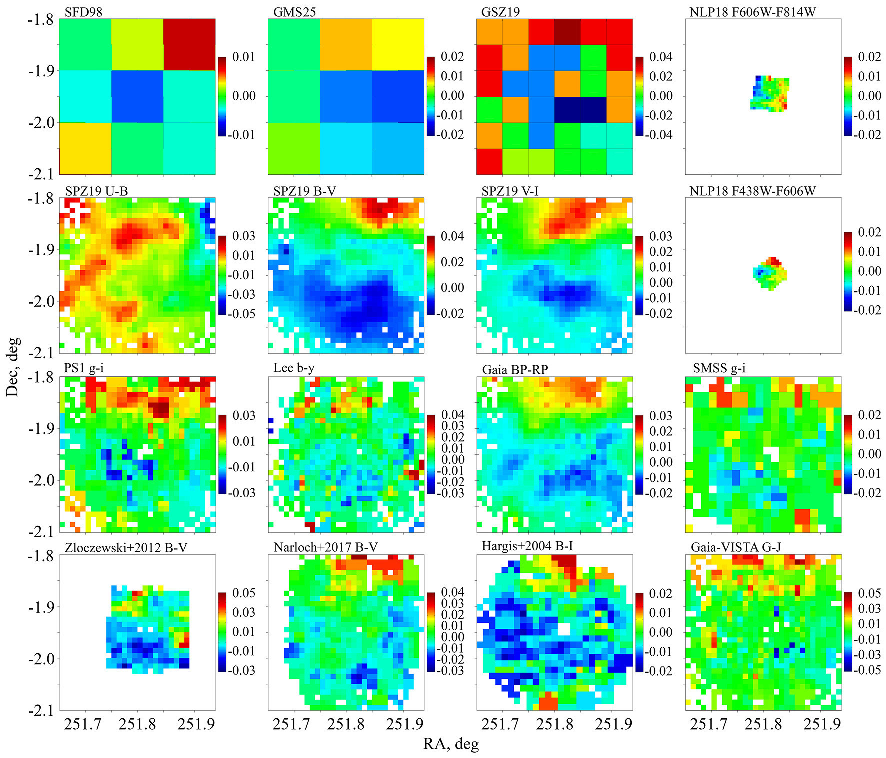}
\caption{DR maps derived from \citetalias{sfd98}, \citetalias{gms2025}, and \citetalias{green2019} and various CMDs for the same NGC\,6218 field.
All the maps are converted into $\Delta E(B-V)$ using the \citetalias{ccm89} extinction law with $R_\mathrm{V}=3.1$.
The white areas have no estimates.
}
\label{ngc6218_drmap}
\end{figure*}

DR maps derived from \citetalias{sfd98}, \citetalias{gms2025}, \citetalias{green2019}, and various CMDs for the NGC\,6218 field are presented in Fig.~\ref{ngc6218_drmap}.
These DR maps show slight differences in their DR values due to systematics in the data and, consequently, we have to display them on different scales.
A similar pattern is seen for \citetalias{sfd98}, \citetalias{gms2025}, \citetalias{green2019}, {\it Gaia}, \citetalias{stetson2019} $B-V$ and $V-I$ and some other maps:
the highest and lowest reddenings occur at the upper edge and the center of the field, respectively.
The same pattern is seen in the DR map presented by \citet{jang2022} in their figure 5 and \citet{pancino2024} in their figure A3.
However, this similarity is not so sharp as in the NGC\,4372 DR maps. 
For example, \citetalias{green2019} report DR variations four times greater than those in \citetalias{sfd98}, likely due to different methods employed in the creation of these maps.
Moreover, some disagreement is seen between the DR maps for different data sets and even for different CMDs/colours of the same data set 
(e.g. \citetalias{stetson2019} $U-B$ versus $B-V$ or $V-I$).
Similar to figure 5 in \citetalias{ngc5024} for NGC\,7099, Fig.~\ref{ngc6218_drmap} illustrates that the DR for NGC\,6218 is only marginally greater than the systematic uncertainties 
in the data sets.
The DR in the fields of the remaining clusters is at a level of other systematics.


Table~\ref{cmds} presents the best solutions of our isochrone fitting for two models and some key CMDs.

\begin{table*}
\def\baselinestretch{1}\normalsize\small
\caption{The results of our isochrone fitting for two models and some key CMDs. \\
In all the CMDs, the colour is the abscissa, the reddening is the colour excess, and the magnitude in the redder filter is the ordinate, 
except the Lee's data set for NGC\,5904 where the ordinate is the $V$ magnitude.
Each derived reddening is followed by corresponding $E(B-V)$, given in parentheses and calculated using extinction coefficients from 
\citet{casagrande2014,casagrande2018a,casagrande2018b} or \citetalias{ccm89} with $R_\mathrm{V}=3.1$.
Age is in Gyr, cluster distance from the Sun $R$ is in kpc.
The complete table is available online.
}
\label{cmds}
\[
\begin{tabular}{lcccccccc} 
\hline
\noalign{\smallskip}
   & \multicolumn{4}{c}{BaSTI} & \multicolumn{4}{c}{DSED} \\
Data set and colour                       & [Fe/H] & Age  & $R$  & Reddening       & [Fe/H] & Age  & $R$  & Reddening   \\
\hline
\noalign{\smallskip}
   & \multicolumn{8}{c}{NGC\,288} \\
\noalign{\smallskip}
\citetalias{nardiello2018} $F606W-F814W$  & $-1.2$ & 13.0 & 8.80 & $0.010$ [0.010]     & $-1.3$ & 13.0 & 8.80 & $0.029$ [0.029] \\ 
{\it Gaia} $G_\mathrm{BP}-G_\mathrm{RP}$  & $-1.2$ & 13.5 & 8.75 & $0.015$ [0.011]     & $-1.3$ & 13.0 & 8.80 & $0.062$ [0.044] \\ 
\citetalias{stetson2019} $B-I$            & $-1.2$ & 13.0 & 9.00 & $-0.028$ [$-0.013$] & $-1.3$ & 12.5 & 9.10 & $0.037$ [0.017] \\ 
Lee Str\"omgren $b-y$                     & $-1.2$ & 13.0 & 8.80 & $0.012$ [0.018]     & $-1.4$ & 12.5 & 8.80 & $0.041$ [0.060] \\ 
\ldots & \ldots & \ldots & \ldots & \ldots  & \ldots  & \ldots  & \ldots  & \ldots \\
\noalign{\smallskip}
   & \multicolumn{8}{c}{NGC\,362} \\
\citetalias{nardiello2018} $F606W-F814W$  & $-1.2$ & 10.5 & 9.00 & $0.010$ [0.010]     & $-1.3$ & 11.0 & 8.85 & $0.026$ [0.026] \\ 
{\it Gaia} $G_\mathrm{BP}-G_\mathrm{RP}$  & $-1.2$ & 10.5 & 8.81 & $0.038$ [0.027]     & $-1.3$ & 10.5 & 8.80 & $0.083$ [0.059] \\ 
\citetalias{stetson2019} $B-I$            & $-1.2$ & 10.5 & 9.00 & $0.013$ [0.006]     & $-1.3$ & 10.5 & 8.90 & $0.076$ [0.035] \\ 
Lee Str\"omgren $b-y$                     & $-1.2$ & 10.0 & 9.20 & $0.007$ [$0.010$] & $-1.4$ & 10.5 & 8.95 & $0.036$ [0.052] \\
\ldots & \ldots & \ldots & \ldots & \ldots  & \ldots  & \ldots  & \ldots  & \ldots \\
\noalign{\smallskip}
   & \multicolumn{8}{c}{NGC\,1904} \\
{\it Gaia} $G_\mathrm{BP}-G_\mathrm{RP}$  & $-1.6$ & 13.5 & 12.50 & $0.042$ [0.030]  & $-1.7$ & 13.0 & 12.70 & $0.084$ [0.060]  \\ 
\citetalias{stetson2019} $B-I$            & $-1.6$ & 14.0 & 12.90 & $0.017$ [0.008]  & $-1.6$ & 13.0 & 12.70 & $0.071$ [0.033] \\ 
Lee Str\"omgren $b-y$                     & $-1.5$ & 13.0 & 12.50 & $0.011$ [0.016]  & $-1.7$ & 13.0 & 12.40 & $0.036$ [0.053] \\ 
\ldots & \ldots & \ldots & \ldots & \ldots  & \ldots  & \ldots  & \ldots  & \ldots \\
\noalign{\smallskip}
   & \multicolumn{8}{c}{NGC\,4372} \\
{\it Gaia} $G_\mathrm{BP}-G_\mathrm{RP}$  & $-2.2$ & 12.0 & 5.20 & $0.728$ [0.520] & $-2.2$ & 12.5 & 5.20 & $0.746$ [0.533]  \\ 
\citetalias{stetson2019} $B-I$            & $-2.3$ & 12.5 & 5.15 & $1.186$ [0.549] & $-2.3$ & 13.5 & 5.00 & $1.197$ [0.554] \\ 
Lee Str\"omgren $b-y$                     & $-2.2$ & 12.5 & 5.00 & $0.392$ [0.570] & $-2.3$ & 12.5 & 5.05 & $0.405$ [0.589] \\ 
\ldots & \ldots & \ldots & \ldots & \ldots  & \ldots  & \ldots  & \ldots  & \ldots \\
\noalign{\smallskip}
   & \multicolumn{8}{c}{NGC\,5904} \\
\citetalias{nardiello2018} $F606W-F814W$  & $-1.2$ & 11.5 & 7.25 & $0.021$ [0.020] & $-1.3$ & 12.0 & 7.05 & $0.041$ [0.040] \\ 
{\it Gaia} $G_\mathrm{BP}-G_\mathrm{RP}$  & $-1.3$ & 11.5 & 7.27 & $0.059$ [0.042] & $-1.4$ & 11.5 & 7.25 & $0.105$ [0.075]  \\ 
\citetalias{stetson2019} $B-I$            & $-1.3$ & 11.5 & 7.46 & $0.048$ [0.022] & $-1.3$ & 11.5 & 7.25 & $0.086$ [0.040] \\ 
Lee Str\"omgren $b-y$                     & $-1.3$ & 11.0 & 7.30 & $0.036$ [0.052] & $-1.4$ & 11.5 & 7.05 & $0.061$ [0.088] \\ 
\ldots & \ldots & \ldots & \ldots & \ldots  & \ldots  & \ldots  & \ldots  & \ldots \\
\noalign{\smallskip}
   & \multicolumn{8}{c}{NGC\,6205} \\
\citetalias{nardiello2018} $F606W-F814W$  & $-1.5$ & 12.5 & 7.41 & $0.010$ [0.010] & $-1.7$ & 13.0 & 7.38 & $0.035$ [0.034] \\ 
{\it Gaia} $G_\mathrm{BP}-G_\mathrm{RP}$  & $-1.6$ & 12.5 & 7.40 & $0.035$ [0.025] & $-1.6$ & 12.0 & 7.40 & $0.074$ [0.053] \\ 
\citetalias{stetson2019} $B-I$            & $-1.6$ & 13.5 & 7.40 & $0.017$ [0.008] & $-1.6$ & 13.0 & 7.39 & $0.054$ [0.025] \\ 
Lee Str\"omgren $b-y$                     & $-1.4$ & 13.0 & 7.25 & $0.012$ [0.017] & $-1.6$ & 13.0 & 7.15 & $0.040$ [0.058] \\ 
\ldots & \ldots & \ldots & \ldots & \ldots  & \ldots  & \ldots  & \ldots  & \ldots \\
\noalign{\smallskip}
   & \multicolumn{8}{c}{NGC\,6218} \\
\citetalias{nardiello2018} $F606W-F814W$  & $-1.2$ & 13.0 & 5.05 & $0.190$ [0.183] & $-1.3$ & 13.0 & 5.06 & $0.209$ [0.202] \\ 
{\it Gaia} $G_\mathrm{BP}-G_\mathrm{RP}$  & $-1.2$ & 13.5 & 4.80 & $0.276$ [0.197] & $-1.3$ & 12.5 & 4.93 & $0.319$ [0.228] \\ 
\citetalias{stetson2019} $B-I$            & $-1.2$ & 14.0 & 4.90 & $0.378$ [0.175] & $-1.3$ & 12.5 & 5.03 & $0.458$ [0.212] \\ 
Lee Str\"omgren $b-y$                     & $-1.2$ & 13.0 & 4.85 & $0.146$ [0.212] & $-1.4$ & 12.5 & 4.90 & $0.175$ [0.255] \\ 
\ldots & \ldots & \ldots & \ldots & \ldots  & \ldots  & \ldots  & \ldots  & \ldots \\
\hline
\end{tabular}
\]
\end{table*}
%


\begin{figure}
\includegraphics{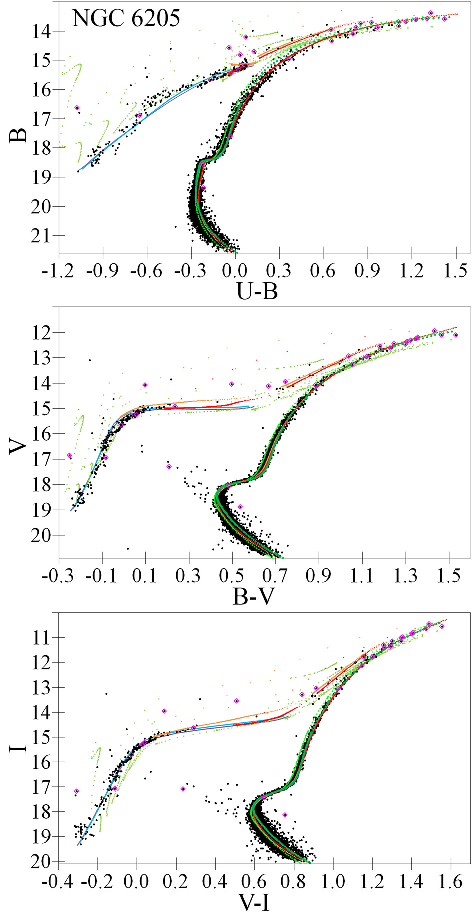}
\caption{$U-B$ versus $B$, $B-V$ versus $V$, and $V-I$ versus $I$ CMDs for the NGC\,6205 {\it Gaia} DR3 members from the \citetalias{stetson2019} data set.
The isochrones for a primordial $Y\approx0.25$ from BaSTI (red), BaSTI ZAHB (purple), and DSED (green), DSED HB/AGB tracks (light green), 
isochrones for $Y=0.275$ from BaSTI (orange) and BaSTI ZAHB (blue),  
as well as isochrones for $Y=0.33$ from DSED (luminous green) are calculated with the best-fitting parameters.
Variable stars are shown by the magenta diamonds.
}
\label{ngc6205_stetson}
\end{figure}

\begin{figure}
\includegraphics{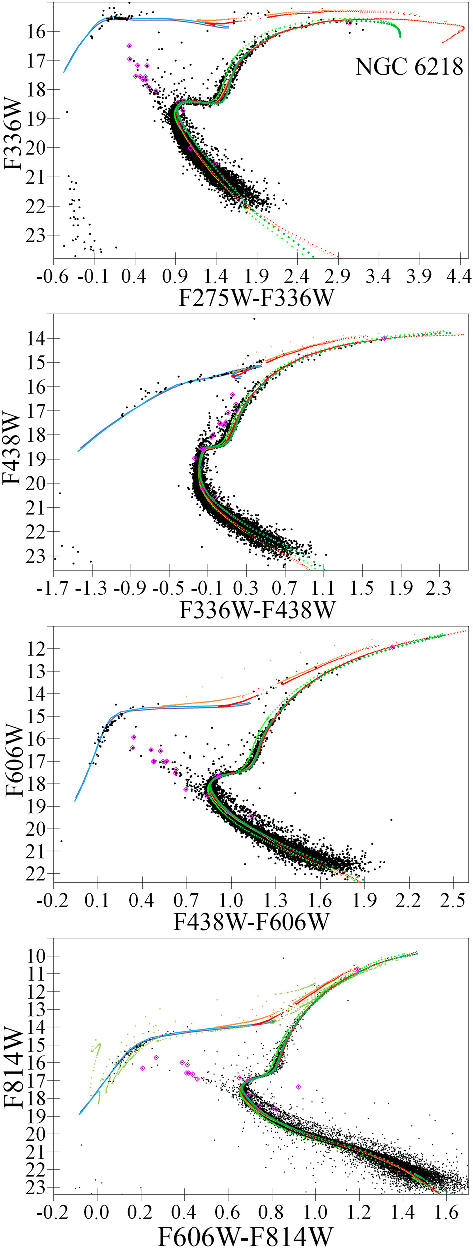}
\caption{$F275W-F336W$ versus $F336W$, $F336W-F438W$ versus $F438W$, $F438W-F606W$ versus $F606W$, and $F606W-F814W$ versus $F814W$ CMDs for the \citetalias{nardiello2018} stars of NGC\,6218.
The isochrones for a primordial $Y\approx0.25$ from BaSTI (red), BaSTI ZAHB (purple), DSED (green), and DSED HB/AGB track (light green), isochrones for $Y=0.275$ from BaSTI (orange) and BaSTI ZAHB (blue), as well as isochrones for $Y=0.33$ from DSED (luminous green) are calculated with the best-fitting parameters.
Variable stars are shown by the magenta diamonds.
}
\label{ngc6218_nardiello}
\end{figure}

\begin{table*}
\def\baselinestretch{1}\normalsize\normalsize
\caption{The same as Table~\ref{estimates} but for several CMDs of the same data set.
`Average' is the average for these CMD estimates, while uncertainty is half the range of the estimates.\\
}
\label{cmdtocmdtable}
\[
\begin{tabular}{lcccccccc} 
\hline
\noalign{\smallskip}
   & \multicolumn{4}{c}{BaSTI} & \multicolumn{4}{c}{DSED} \\
Data set and colour                       & [Fe/H] & Age  & $R$  & Reddening       & [Fe/H] & Age  & $R$  & Reddening   \\
\hline
\noalign{\smallskip}
\noalign{\smallskip}
   & \multicolumn{8}{c}{NGC\,6205} \\
\citetalias{stetson2019} $U-B$ & $-1.5$  & 12.5 & 7.70 & $-0.011$ [$-0.015$] & $-1.6$  & 12.0 & 7.60 & $0.023$ [$0.030$] \\ 
\citetalias{stetson2019} $B-V$ & $-1.6$  & 13.5 & 7.40 &  $0.008$ [$0.008$]  & $-1.6$  & 12.5 & 7.40 & $0.033$ [$0.033$] \\ 
\citetalias{stetson2019} $V-I$ & $-1.5$  & 13.0 & 7.50 & $0.000$ [$0.000$]   & $-1.6$  & 13.0 & 7.40 & $0.030$ [$0.024$] \\ 
\hline
Average                        & $-1.53$ & 13.0 & 7.53 & [$-0.003$]        & $-1.60$ & 12.5 & 7.47 & [$0.029$] \\
Uncertainty                    & $0.05$  & 0.50 & 0.15 & [$0.012$]	       & $0.00$  & 0.50  & 0.10 & [$0.005$] \\
\hline
\noalign{\smallskip}
   & \multicolumn{8}{c}{NGC\,6218} \\
\citetalias{nardiello2018} $F275W-F336W$ & $-1.2$  & 12.0 & 5.00 & $0.176$ [$0.195$] & $-1.4$  & 12.5 & 5.00 & $0.194$ [$0.215$] \\ 
\citetalias{nardiello2018} $F336W-F438W$ & $-1.2$  & 12.5 & 5.00 & $0.192$ [$0.190$] & $-1.4$  & 12.5 & 5.00 & $0.212$ [$0.210$] \\ 
\citetalias{nardiello2018} $F438W-F606W$ & $-1.2$  & 13.5 & 5.00 & $0.216$ [$0.170$] & $-1.3$  & 13.0 & 4.90 & $0.265$ [$0.209$] \\ 
\citetalias{nardiello2018} $F606W-F814W$ & $-1.2$  & 13.0 & 5.05 & $0.190$ [$0.183$] & $-1.3$  & 13.0 & 5.06 & $0.209$ [$0.202$] \\ 
\hline
Average                                  & $-1.20$ & 12.8  & 5.01 & [$0.185$]       & $-1.35$ & 12.8    & 4.99 & [$0.209$] \\
Uncertainty                              & $0.00$  & 0.75  & 0.03 & [$0.013$]       & $0.05$  & 0.50    & 0.08 & [$0.007$]  \\
\hline
\end{tabular}
\]
\end{table*}

Figs~\ref{ngc6205_stetson} and \ref{ngc6218_nardiello} present examples of several CMDs of the same data set with the UV and optical filters. A reliable fitting of all the CMDs is evident.
Similarly to Table~\ref{estimates}, Table~\ref{cmdtocmdtable} presents the best solutions for these CMDs as well as the average estimates of the parameters and their uncertainties as half the range of the parameter estimates. It is seen that the average estimates agree with the final ones in Table~\ref{estimates}, while the uncertainties are within our estimates of total uncertainties.

\section{Adjustment}
\label{adjustment}


\begin{figure}
\includegraphics{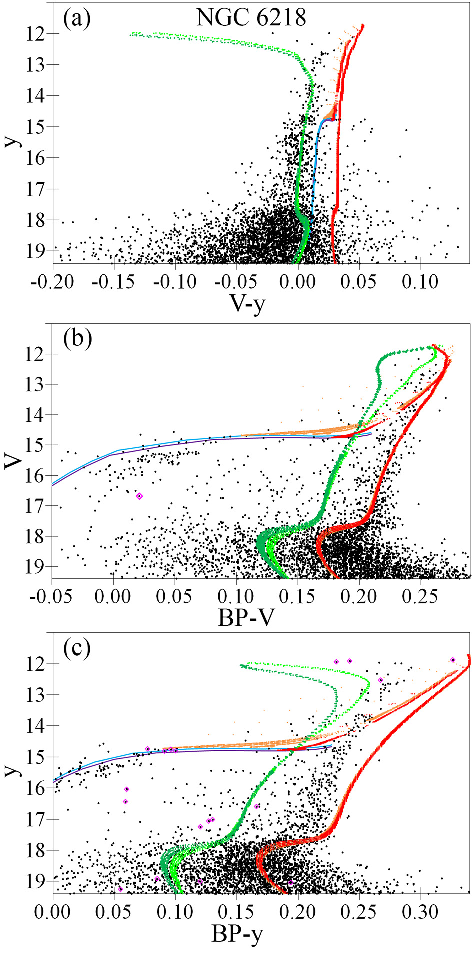}
\caption{(a) \citetalias{stetson2019}-Lee $V-y$ versus $y$, (b) {\it Gaia}-\citetalias{stetson2019} $G_\mathrm{BP}-V$ versus $V$, and (c) {\it Gaia}-Lee $G_\mathrm{BP}-y$ versus $y$ CMDs for cluster members of NGC\,6218.
The isochrones for a primordial $Y\approx0.25$ from BaSTI (red), BaSTI ZAHB (purple), and DSED (green), as well as for $Y=0.275$ from BaSTI (orange) and BaSTI ZAHB (blue) are calculated with the best-fitting [Fe/H], $R$, and reddening from Table~\ref{cmds} and age between 12.5 and 14 Gyr.
Variable stars are shown by the magenta diamonds.
}
\label{ngc6218_adjustment}
\end{figure}

We adjust data sets with similar filters (e.g. \citetalias{stetson2019} $V$ versus Lee Str\"omgren $y$ versus {\it Gaia} $G_\mathrm{BP}$, other pairs can be found in Table~\ref{filters}) following our approach described in section 6 of \citetalias{ngc6205} and in \citetalias{ngc288}.

Such similar filters provide CMDs in which the position of stars and isochrones is almost independent of the cluster parameters. Hence, any colour shift of an isochrone w.r.t. a bulk of the stars is due to a systematic error of a data set or an isochrone used (the latter may be, for example, due to a wrong colour--$T_\mathrm{eff}$ relation). 
An example of such CMDs for the \citetalias{stetson2019}, Lee, and {\it Gaia} similar filters for NGC\,6218 is shown in Fig.~\ref{ngc6218_adjustment}. The isochrones of the same model for various age and $Y$ almost coincide. A reasonable variation of reddening, $R$, and [Fe/H] also changes the isochrone colour negligibly. 
Fig.~\ref{ngc6218_adjustment} shows that the MS and HB have a large scatter of the stars, while the RGB is appropriate to analyse the colour shifts.

Such CMDs present relative colours shifts for `data set--data set--isochrone' triples (e.g. a nearly zero \citetalias{stetson2019}--Lee $V-y$ colour of the star bulk w.r.t. the DSED isochrone in Fig.~\ref{ngc6218_adjustment} (a)), but not those of a data set or an isochrone separately. Therefore, we cannot use such diagrams to clarify the systematic errors of a data set or an isochrone. However, we can adjust a series of data sets with similar filters by calculating constant colour corrections to the `data set -- isochrone pairs' in order to make the colours of all the data sets in the series closer to each other. For such a calculation we impose an additional condition: an average colour of all the data sets in the series w.r.t. an isochrone is fixed. This condition fixes the derived reddenings and extinctions. It is worth noting that since we correct only colours and since these corrections are rather small (within a few hundredths of a magnitude), all the derived estimates of [Fe/H], $R$, and age are also fixed.
In the case of Fig.~\ref{ngc6218_adjustment}, the corrections of $+0.013$, $+0.013$, and $-0.026$\,mag would minimise, respectively, the \citetalias{stetson2019}, Lee, and {\it Gaia} colour shifts w.r.t. the DSED isochrone, while the corrections of $+0.01$, $-0.02$, and $+0.01$\,mag would do the same w.r.t. the BaSTI isochrone.
This adjustment suppresses systematic differences between data sets for the same cluster and fitted by the same model.
As a result, this reduces the scatter of extinctions derived from the data sets around an average extinction for the model--cluster pair.
Accordingly, this leads to a better determination of empirical extinction law and a higher precision of the extinction-to-reddening ratio $R_\mathrm{V}$.

\section{Supplemental material: some CMDs for the clusters}
\label{addcmds}

We present some CMDs with our best isochrone fitting as a supplemental material.

\end{document}